\documentclass{article}
\usepackage{graphicx} 

\usepackage{amsmath}
\usepackage{amssymb} 
\usepackage{verbatim} 
\usepackage{caption}
\usepackage{subcaption}

\usepackage{tikz}

\usepackage{amsthm}

\newtheorem{theorem}{Theorem}[section]
\newtheorem{lemma}[theorem]{Lemma}
\newtheorem{proposition}[theorem]{Proposition}

\theoremstyle{definition}
\newtheorem{assumption}{Assumption} 
\newtheorem{definition}[theorem]{Definition}

\theoremstyle{remark}

\usepackage[
natbibapa
]{apacite} 
\usepackage[hyphens]{url}
\usetikzlibrary{positioning, arrows.meta, decorations.pathreplacing}
\usepackage{pifont} 
\usepackage{array}  
\usepackage{graphicx}
\usepackage{subcaption}
\usepackage{booktabs}
\usepackage{tabularx}
\usepackage{siunitx} 
\usepackage{etoolbox} 
\usepackage{makecell} 

\newcommand{\diff}[2]{\frac{d #1}{d #2}}

\newcommand{\TABLE}[3]{%
	\caption{#1} #2 \par \vspace{2pt} {\footnotesize #3}
}
\newcommand{\up}{\rule{0pt}{3ex}}    
\newcommand{\down}{\rule[-1.5ex]{0pt}{0pt}} 

\newcommand{\Halmos}{\hfill$\square$}

\begin{document}
\title{Distributionally Robust Contract Design with Deferred Inspection}
\author{Halil I. Bayrak, Martin Bichler}

\date{January 2026}

\maketitle

\section*{Abstract}
We study a robust contract design problem with deferred inspection, in which a principal allocates a scarce resource to an agent, observes the agent's realized outcome ex post at negligible cost, and conditions transfers on this information through rewards. The principal faces ambiguity about the agent's value distribution and seeks to maximize worst-case expected revenue subject to incentive compatibility and limited liability. In contrast to existing work on inspection mechanisms, which relies on common-prior assumptions, we adopt a distributionally robust approach based on moment information.
Our main contribution is a complete characterization of the robust contract design problem with a single agent. When the ambiguity set is defined by the first moment, we identify a robustly optimal contract with a concave allocation rule and a linear payment rule. We further show that robustness does not uniquely pin down transfers: we construct a Pareto robustly optimal contract that preserves the same allocation while extracting maximal feasible payments from all types, yielding strictly higher expected revenue under non-worst-case distributions.
We also derive structural results for multi-agent extensions. For ambiguity sets defined by the first $N$ moments, we show that robust optimality requires aggregate payments to be lower bounded by a multi-dimensional polynomial of degree $N$. However, unlike the single-agent case, robust multi-agent mechanisms are substantially more complex: dominant-strategy incentive compatibility becomes necessary, simple monotone mechanisms are no longer tractable, and worst-case distributions may involve correlated types or degenerate to a Dirac distribution at the mean. These results highlight a sharp contrast between robust contract design and robust multi-agent mechanism design with inspection.

\newpage
\tableofcontents

\newpage

\section{{Introduction}}

\noindent Mechanism design deals with the design of incentive-compatible mechanisms to achieve desirable outcomes in the presence of self-interested participants. This line of research has led to landmark papers in auction theory and related fields ($e.g.$, \citet{hurwicz1973design}, \citet{myerson1983mechanism}). Traditionally, it is assumed that the designer cannot observe the participants' valuations ex-ante or ex-post. However, in some applications, the designer might actually be able to do so. Related problems are referred to as mechanism design with inspection ($e.g.$, \citet{ben2014optimal}).

Our motivation is inspired by recent capacity allocation mechanisms used by large platforms such as Amazon's Fulfillment by Amazon (FBA) program \citep{AmazonFulfillment}.
Sellers bid for additional storage capacity and pay a reservation fee that is partially rebated based on realized sales, which the platform observes ex post at negligible cost. This setting naturally combines deposits, inspection, and rewards, and illustrates how ex-post information can be leveraged to improve revenue.

While the Amazon example illustrates the economic forces behind inspection and rewards in a platform setting, the remainder of the paper focuses on a single-agent contract design problem, which admits a complete and tractable characterization under distributional ambiguity.
Imagine a Manufacturer (Principal) launching a new product (e.g., a new smartphone or fashion line) and dealing with a specific Retailer (Agent) in a new territory. Because the product is new or the territory is untested, the Manufacturer does not know the demand distribution ($\mathbb{P}$). They only know the support (e.g., demand is between 0 and 1000 units) or the first moment (average demand in similar regions). The Retailer has better local market intelligence. They privately observe a signal $\nu$ indicating the likely demand (valuation) for the product. The Manufacturer must decide how much inventory ($x$) to allocate to this Retailer. Inventory is scarce or costly to produce. The Retailer reports a demand forecast (bid) and pays a deposit for the inventory. The Manufacturer allocates inventory based on this report. Modern Point-of-Sale (POS) systems allow the Manufacturer to observe the Retailer's actual sales ex-post at effectively zero cost. If the Retailer actually sells the inventory (proving the high demand they claimed), the Manufacturer provides a sales rebate or retroactive discount (the reward). If the Retailer over-reported demand to hoard inventory but fails to sell it, they receive no rebate, effectively losing their deposit on the unsold units.
We use this setting as a motivating example and focus primarily on the corresponding single-agent contract design problem, which we fully characterize under distributional ambiguity. Multi-agent extensions are discussed later and are shown to exhibit substantially higher complexity.

\subsection{Prior Work}
In a seminal paper, \citet{alaei2024optimal} study mechanism design with deferred inspection and rewards under a common prior. In their model, a seller allocates an indivisible good, observes agents' realized values ex post at
no cost, and conditions transfers on this information through rewards. They show that inspection substantially increases revenue relative to classical auctions and characterize the optimal truthful mechanism under full distributional knowledge. The authors show that the optimal mechanism increases the principal's payoff by around 50\% compared to the classical optimal auction that does not use the ex-post information. 
Although these results are compelling, the assumption of a common prior is strong. The Wilson doctrine argues against reliance on precise distributional assumptions \citep{Wilson_1987}, motivating robust mechanism design.
Accordingly, we assume that the agent's type distribution is unknown but belongs to a commonly known ambiguity set characterized by moment information, and we study the problem from the perspective of an ambiguity-averse designer maximizing worst-case expected revenue.

Our analysis builds on and extends the robust selling framework of \citet{carrasco2018optimal}. As in their work, moment-based ambiguity leads to a polynomial characterization of worst-case revenue.
However, the introduction of deferred inspection and rewards fundamentally alters feasibility: the optimal polynomial no longer uniquely determines the mechanism, and robust optimality may be achieved by infinitely many contracts. We also study extensions to settings with multiple agents and derive important structural insights for these problems.

\subsection{Contributions}
Our main contribution is a complete characterization of the single-agent robust contract design problem with deferred inspection when the seller knows only the first moment of the buyer's value distribution. {From an optimization perspective, deferred inspection enlarges the feasible region of the designer’s max–min problem by decoupling allocation and information rents, which fundamentally alters the geometry of the robust optimization problem.} We show that robust optimality can be achieved by a mechanism with a concave allocation rule and a linear payment rule, and that robustness does not uniquely determine transfers. In particular, there exist infinitely many robustly optimal mechanisms that share the same allocation rule but differ in their payment rules. Among them, we identify a mechanism that extracts the maximum feasible payment from each type and prove that it is Pareto robustly optimal. This focus on the first moment is strategically chosen: it allows us to derive intuitive linear contracts that are easy to implement, avoiding the complex high-degree polynomial structures that arise when robustness is required over higher-order moments.

More generally, we provide a structural simplification of robust contract design under moment-based ambiguity. Extending the results of \citet{carrasco2018optimal} to settings with deferred inspection and
rewards, we show that when the first $N$ moments are known, robust optimality requires aggregate payments to be lower bounded by a polynomial of degree~$N$. Unlike in standard robust selling problems, this polynomial characterization no longer uniquely determines the mechanism, reflecting the enlarged feasibility induced by inspection and rewards.

Finally, we study extensions to settings with multiple agents and derive important structural insights. For ambiguity sets with symmetric first moments, we characterize worst-case distributions in two regimes. With two agents, worst-case distributions typically involve correlated types and are characterized by solutions to nonlinear optimization problems. With three or more agents, the worst-case distribution collapses to a Dirac mass at the symmetric mean. In all cases, robustly optimal multi-agent mechanisms are substantially more complex than in the single-agent setting, must induce dominant-strategy equilibria under ambiguity, and often rely on randomized allocations. As a result, the simple and tractable characterization obtained for the single-agent case does not extend to multi-agent settings. We therefore study their properties using numerical analyses.

{Methodologically, our analysis reduces a robust mechanism design problem to a sequence of tractable optimization problems with polynomial structure, revealing when robustness leads to linear contracts and when it necessarily induces high-dimensional nonlinear constraints.}

\subsection{Illustration}\label{sec: illustration}

Conceptually, our mechanism proceeds in four distinct steps (see Figure \ref{fig:timeline}):

\begin{enumerate}
	\item \textbf{Reporting:} The buyer observes their private type $\nu \in [0,1]$ and submits a report $\hat{\nu}$. This report serves as a binding bid, and the agent pays a deposit equal to this report immediately. 
	
	\item \textbf{Allocation:} The designer allocates the good according to an allocation rule $x(\hat{\nu})$, where $x$ represents the probability of receiving the item.
	
	\item \textbf{Deferred Inspection:} After allocation, the designer observes the agent's true type $\nu$ with certainty and without cost. Crucially, as established in the model by \cite{alaei2024optimal}, this revelation is {independent of the allocation}. The seller observes the true type regardless of whether the agent received the good or not.
	
	\item \textbf{Reward and Net Payment:} Based on the inspected true type, the designer may provide a reward $r(\nu)$. Consequently, the final net payment $p(\nu)$ extracted from the agent is the initial deposit minus the ex-post reward: $p(\nu) = \hat{\nu} - r(\nu)$. 
\end{enumerate}

\begin{figure}[h]
	\centering
	\resizebox{\textwidth}{!}{
	\begin{tikzpicture}[
		node distance=1.6cm,
		auto,
		>=Latex,
		thick,
		timeline/.style={->, ultra thick, gray!80},
		event/.style={circle, fill=black, inner sep=1.5pt},
		label/.style={text width=2.5cm, align=center, font=\small}
		]
		
		\draw[timeline] (0,0) -- (12,0);
		
		\node[event] (t0) at (0,0) {};
		\node[event] (t1) at (3,0) {};
		\node[event] (t2) at (6,0) {};
		\node[event] (t3) at (9,0) {};
		\node[event] (t4) at (12,0) {};
		
		\node[label, above=0.2cm of t0] (l0) {\textbf{Type Realization}\\ Buyer observes private type $\nu$};
		\node[label, above=0.2cm of t1] (l1) {\textbf{Reporting}\\ Buyer reports $\hat{\nu}$ and pays deposit equal to report};
		\node[label, above=0.2cm of t2] (l2) {\textbf{Allocation}\\ Seller allocates $x(\hat{\nu})$ based on report};
		\node[label, above=0.2cm of t3] (l3) {\textbf{Deferred Inspection}\\ Seller observes true $\nu$ with certainty};
		\node[label, above=0.2cm of t4] (l4) {\textbf{Reward Phase}\\ Seller pays reward $r(\nu)$ if $\hat{\nu} = \nu$};
		
		\draw[decoration={brace, mirror, amplitude=10pt}, decorate] (0,-0.5) -- (2.9,-0.5) node[midway, below=15pt] {\textit{Ex-Ante}};
		\draw[decoration={brace, mirror, amplitude=10pt}, decorate] (3.1,-0.5) -- (8.9,-0.5) node[midway, below=15pt] {\textit{Interim (Allocation)}};
		\draw[decoration={brace, mirror, amplitude=10pt}, decorate] (9.1,-0.5) -- (12,-0.5) node[midway, below=15pt] {\textit{Ex-Post (Verification)}};
		
		
	\end{tikzpicture}
	}
	\caption{Timeline of the Deferred Inspection Mechanism. The net payment is calculated as $p(\nu) = \hat{\nu} - r(\nu)$.}
	\label{fig:timeline}
\end{figure}

It is important to distinguish our model from ``verification-with-penalties.'' We operate under limited liability where the seller cannot impose arbitrary fines. The seller can reward truthful reporting but cannot extract more than the initial deposit if a misreport is detected. Therefore, the mechanism is strictly a {verification-with-only-rewards} system: the agent is incentivized via potential reimbursement (rewards) rather than the threat of ex-post fines exceeding their bid.

\begin{figure}
	\centering
	\includegraphics[scale=0.7]{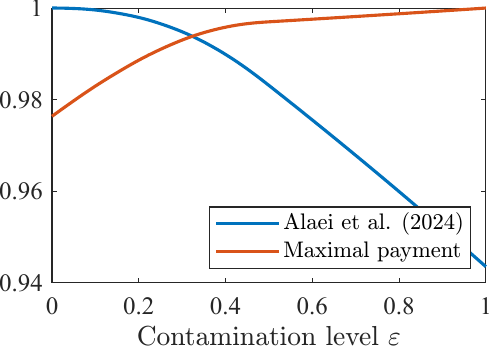}
	\caption{Expected payoffs of the mechanism from \citet{alaei2024optimal} and the robustly optimal mechanism with maximal payments.}
	\label{fig: experimentIntro}
\end{figure}

Let us now illustrate the expected payoff of our distributionally robust mechanism for this model. For this, we report a numerical experiment comparing the optimal mechanism from \citet{alaei2024optimal} tailored to a specific distribution and our robustly optimal mechanism that maximizes the worst-case payoff across a set of distributions. For a detailed discussion, see the end of Section \ref{sec: one agent}. In this experiment, we assume that the true distribution of the types is a convex combination of the uniform distribution and a worst-case distribution that we identify. Figure \ref{fig: experimentIntro} reports the performances of two mechanisms, where a performance of one means that the mechanism achieves the same expected payoff as the optimal mechanism tailored to the true distribution. At one extreme point, when the true distribution is exactly the uniform ($\varepsilon=0$), the mechanism characterized by \citet{alaei2024optimal} is optimal, whereas our robustly optimal mechanism (maximal payment) becomes optimal when the misspecification is at its most severe ($\varepsilon=1$). As apparent from the figure, the robustly optimal mechanism with maximal payments consistently performs close to the true optimal solution and outperforms the mechanism from \citet{alaei2024optimal} when $\varepsilon \geq 0.324$. 

\subsection{Organization of the paper}

In the remainder of the paper, we first review the most relevant work in Section \ref{sec: literature}. In Section \ref{sec: one agent}, we first introduce our model for the seller {who faces a single agent and wants to maximize her worst-case expected payoff when she only knows the first $N$ moments of the distribution. We illustrate how these results lead to characterization of the robustly optimal solution when $N=1$. In Section \ref{sec: multi-agentCase}, we show that the structural results of the single-agent case extend to the multi-agent setting, but the characterization of the optimal solution is much more complex. For the case of $N=1$, we characterize the worst-case distributions and present numerical experiments to analyze the robustly optimal mechanisms. We then conclude in Section \ref{sec: conclusion}. All proofs are relegated to the Electronic Companion.
}
\section{Related Literature}\label{sec: literature}

We introduce distributional robustness for mechanisms to the setting of \citet{alaei2024optimal}, which is related to both auction design and allocation with verification literature. In a nutshell, \citet{alaei2024optimal} introduce costless inspection and rewards to the auction setting and report the optimal mechanism, which significantly differs from the classical one and results in a higher revenue. A similar approach, without inspection, was taken by \citet{hansen1985auctions}, who considered auctions with contingent payments and noted that non-cash bidding mediums with contingent pricing features can yield higher revenue. See \citet{skrzypacz2013auctions} for an overview of this literature. 

The exploration of allocation with costly verification originated with \citet{townsend1979optimal}, who delved into a principal-agent model involving monetary transfers. \citet{ben2014optimal} expanded this analysis to accommodate multiple agents but excluded monetary transfers. They show that a simple threshold mechanism optimally solves the problem. Their work stimulated extensive subsequent research (see \cite{mylovanov2017optimal}, \cite{li2020mechanism}, \cite{bayrak2025distributionally}). \citet{alaei2024optimal} contribute to this literature by reintroducing monetary transfers while assuming inspections are not costly. Unlike earlier studies, their approach allows for a continuous optimal allocation function rather than a threshold mechanism. We demonstrate that this observation remains valid in the presence of distributional ambiguity,
{there are infinitely many robustly optimal mechanisms one can choose from, and the single-agent mechanism does not extend easily to the multi-agent case due to the dominant strategy incentive compatibility requirement. There have been other extensions of the problem from \cite{alaei2024optimal} that are not related to robust optimization. Here, we note the most relevant ones. \cite{belloni2025approximately} consider the problem of allocating multiple units to buyers who arrive online and may have multi-unit demand. As in \cite{alaei2024optimal}, the seller can inspect and reward the agents after the allocation, but the inspection is limited to agents who are allocated. \cite{li2025mechanism} consider a multi-round interaction between a seller and a buyer. The seller can inspect the reported type post-allocation at no cost, but neither holds deposits nor returns any rewards. Instead, she punishes by having no interaction in the future.}

This paper also contributes to the literature on (distributionally) robust mechanism design. \citet{carroll2019robustness} surveys this literature and notes that the study of robustness may lead to novel mechanisms or help us explain real-life market interactions. Essentially, the robust mechanism design literature diverges from classical studies, which model uncertainty through random variables governed by known probability distributions, and considers non-stochastic distributional uncertainty.
The closest study to our setting from the robust mechanism design literature is \cite{carrasco2018optimal}. They consider the mechanism design problem of the seller who faces a single buyer and has only partial information about the buyer's value distribution, $i.e.$, the first $N$ moments. They assume the seller to be a worst-case maximizer and show that the robustly optimal mechanism is a randomization over posted price mechanisms. The optimal allocation rule translates as the distribution of the posted prices, and the payment rule is identified through a non-negative monotonic hull of a polynomial of degree $N$.
We extend their setting by introducing deferred inspection, rewards, {and multiple agents. We first show that their simplification result for the seller's problem remains valid in our multi-agent setting. Here, we compare our results from the single-agent setting, which unfortunately do not entirely extend to the multi-agent one.} When the first moment is known, (randomized) posted price mechanisms are no longer robustly optimal, but there is an optimal payment rule confirming their results by being linear (polynomial of degree 1). However, the optimal allocation rule is in stark contrast to theirs, and we can identify another payment rule that {is not a polynomial} and results in a higher expected payoff for non-worst-case distributions. Hence, thanks to deferred inspection and rewards, the seller can improve her worst-case payoff by deviating from posted price mechanisms {and polynomial payment rules}.
For further examples of robust studies in auction design and allocation with costly verification settings, see \citet{bergemann2005robust}, \citet{bandi2014optimal}, and \citet{bayrak2025distributionally}. However, to the best of our knowledge, we are the first to derive closed-form optimal mechanisms for auction design with deferred inspection and rewards under distributional ambiguity. 

{Lastly, we mention some extensions of \cite{carrasco2018optimal} that are relevant. \cite{chen2024screening} consider the same single-agent setting with generic convex ambiguity sets and generalize the results from \cite{carrasco2018optimal} using duality theory. \cite{wang2024minimax} sticks with the moment ambiguity set and solves the problem from the shoes of a regret-minimizing seller. \cite{bachrach2022distributional} restrict the ambiguity set to contain only the first moment and the bounds while extending to the multi-agent setting. They assume independently and identically distributed types for the buyers and find that the robustly optimal mechanism is a second-price auction with randomized reserve prices. Although their setting is quite different from ours (no inspection and rewards), interestingly, they also establish that the case of two buyers gives rise to different results compared to the case of three or more buyers.}

\section{Robust Contract Design} \label{sec: one agent}

The robust contract design problem discussed in this section captures one agent's dominant-strategy incentives and serves as a building block for DSIC mechanisms with multiple agents.

\subsection{The Model}

A seller (she) with a single indivisible object faces a risk-neutral buyer (he). The seller's cost is normalized to zero, 
{whereas the buyer's valuation, which is bounded above, is normalized to the set of values $[0,1]$. The buyer learns his valuation (type) privately. Hence, from the perspective of the seller, it can be modeled as a random variable $\nu \in [0,1]$ generated by some probability distribution $\mathbb{P}$. This paper considers the environment where the seller has limited information on this distribution, $i.e.,$ first $N$ moments that are given by a vector $\mathbf{k}=(k_1,\ldots,k_N)$:
	\begin{equation}\label{eq: ambiguity set}
		\mathcal{P}_{\mathbf{k}} = \{\mathbb{P} \in \mathcal{P}_0([0,1]): \int_0^1 \nu^i d \mathbb{P}(\nu) = k_i \; \forall i \in [N]\},
	\end{equation}
	where $\mathcal{P}_0([0,1])$ denotes the family of all probability distributions on $[0,1]$, and $[N]$ denotes the set $\{1,\ldots,N\}$ for any $N \in \mathbb{N}$.
}
We will consider the seller to be ambiguity-averse in the sense that she wants to maximize her worst-case payoff over the admissible distributions.
{As is common in the robust mechanism design literature, the problem can be posed as a zero-sum game between the designer and an adversary who chooses the distribution. To solve this, we need to identify a robustly optimal mechanism and a worst-case distribution pair that constitutes a Nash equilibrium.
}

The ambiguity set above also appears in \cite{carrasco2018optimal} with one small difference. That is, the last moment information is not exact but only serves as an upper bound. In their setting, where the buyer's valuation set is unbounded, this small difference plays a crucial role in establishing the existence of an equilibrium in the zero-sum game. When the set of types is bounded, the strategy set of the adversary given in \eqref{eq: ambiguity set} is compact, ensuring equilibrium existence. So, we can consider any available moment information to be exact without any issues. To avoid degenerate and trivial cases, we assume the following conditions on $\mathbf{k}$, which are also given in \cite{carrasco2018optimal}.

\begin{assumption}\label{assumption: N moments}
	We consider $\mathbf{k} \in (0,1)^N$ such that for some $\varepsilon>0$, we have $\mathcal{P}_{\hat{\mathbf{k}}} \neq \emptyset$ for all $\hat{\mathbf{k}} \in (0,1)^N$ satisfying $|\hat{\mathbf{k}} - \mathbf{k}| < \varepsilon$.
\end{assumption}

The difference from \cite{carrasco2018optimal} is that the seller can costlessly inspect the buyer's type after the allocation has been made. Observing the true type, the seller may reward the buyer if his bid matches his true type, but cannot impose a penalty. Under these limitations, the seller wants to design a revenue-maximizing auction, in which she secures the bid of the buyer as a deposit, allocates the good according to an allocation rule $x: \mathbb{R} \mapsto [0,1]$ and rewards the agent according to a reward rule $r: \mathbb{R} \times \mathbb{R} \mapsto \mathbb{R}_+$ if the inspection reveals truthful reporting. 

In this setting, a variant of the revelation principle is proven by \citet{alaei2024optimal}, allowing us to focus on truthful mechanisms. As we are only interested in the truthful equilibria, we would not reward the untruthful bids, $r(\nu,\hat{\nu})=0$ for all $\nu \neq \hat{\nu}$. 
Hence, we can simplify the notation by substituting $r(\nu, \hat \nu)$ by $r(\nu)$. 
The following optimization problem formalizes the seller's problem.
\begingroup
\setlength{\abovedisplayskip}{2pt} 
\setlength{\belowdisplayskip}{2pt}
\begin{align}
	{z^*} =\sup_{x,r} \;&\inf_{\mathbb{P} \in \mathcal{P}_{\mathbf{k}}} \mathbb{E}_{\mathbb{P}}[\nu-r(\nu)]  \nonumber\\
	\text{s.t.} \;&\nu x(\nu)- \nu +r(\nu) \geq \nu x(\hat{\nu}) - \hat{\nu} &\forall \nu,\hat{\nu} \in [0,1], \label{IC}\tag{IC}\\
	&\nu x(\nu)- \nu +r(\nu) \geq 0 &\forall \nu \in [0,1], \label{IR}\tag{IR}\\
	&x(\nu) \leq 1 &\forall \nu \in [0,1], \nonumber\\
	&x(\nu), r(\nu) \geq 0 &\forall \nu \in [0,1], \nonumber
\end{align}
\endgroup
where {$z^*$ denotes the optimal objective value. Here,} \eqref{IC} constraints ensure that the agent with type $\nu$ does not benefit from deviating to any other type $\hat{\nu}$, \eqref{IR} constraints ensure participation of all types, and the remaining constraints ensure the feasibility of the allocation and the reward rules.

Throughout the paper, we distinguish between the \emph{deposit} paid at the reporting stage and the \emph{net payment} collected by the seller after inspection. Formally, the agent pays a deposit equal to the reported type $\hat{\nu}$, and after inspection receives a reward $r(\nu)$ if the report is truthful. We denote by $p(\nu)=\nu-r(\nu)$ the resulting \emph{net payment}.
All optimization problems in what follows are written in terms of the net payment $p(\nu)$.
We restrict attention to mechanisms in which the deposit equals the reported type. Allowing arbitrary deposit rules would trivialize the problem, as the seller could extract the entire surplus ex ante and rebate it after inspection. To rule out such degenerate mechanisms and focus on economically meaningful contracts, we normalize deposits to equal reported values and work directly with net payments.

We let $p(\nu)=\nu-r(\nu)$ and work with the following equivalent formulation:
\begingroup
\setlength{\abovedisplayskip}{2pt} 
\setlength{\belowdisplayskip}{2pt}
\begin{align}
	{z^*} = \sup_{x,p} \;&\inf_{\mathbb{P} \in \mathcal{P}_{\mathbf{k}}} \mathbb{E}_{\mathbb{P}}[p(\nu)] \label{eq: MDP} \tag{MDP}\\
	\text{s.t.} \;&p(\nu) \leq \nu \big( x(\nu)- x(\hat{\nu}) \big) + \hat{\nu} &\forall \hat{\nu}, \nu \in [0,1], \tag{\ref{IC}}\\
	&p(\nu) \leq \nu x(\nu) &\forall \nu \in [0,1], \tag{\ref{IR}}\\
	&x(\nu) \in [0,1] &\forall \nu \in [0,1]. \nonumber
\end{align} 
\endgroup
We refer to the above mechanism design problem as \eqref{eq: MDP}.
{Under the incentive constraints \emph{(IC)}, any feasible mechanism can be transformed into one with
	a weakly increasing allocation rule without decreasing the seller's worst-case expected revenue.
	Hence, monotonicity of $x(\cdot)$ can be imposed without loss of generality.
}
This result is formally stated in Proposition \ref{prop: resultsAlaei} below.

\begin{figure}
	\centering
	\begin{tikzpicture}[x=0.75pt,y=0.75pt,yscale=-1,xscale=1]
		\node[anchor=south west,inner sep=0] at (0,0) {\includegraphics[width=0.85\textwidth]{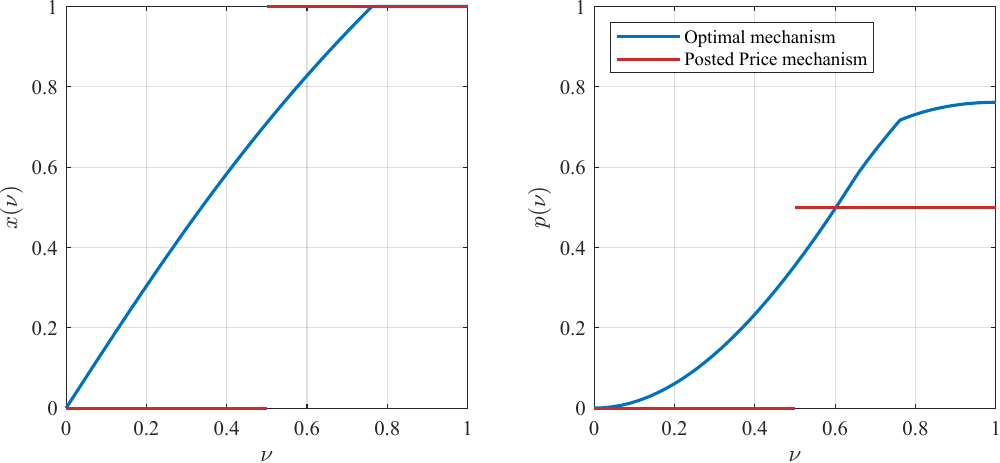}};
	\end{tikzpicture}
	\caption{Optimal mechanism from \citet{alaei2024optimal} compared to the optimal Posted Price mechanism when $\nu  \sim \mathcal{U}[0,1]$.}
	\label{fig: comparisonToPosted}
\end{figure}

If we substitute the term $\hat{\nu}$ on the right-hand-side of \eqref{IC} with $p(\hat{\nu})$, the constraints of the above formulation exactly match those of the well-known {contract} design problem, where one seeks a pair of \eqref{IC} \& \eqref{IR} allocation and payment rules. 
{When $\mathcal{P}_{\mathbf{k}}$ is a singleton,}
we know that the optimal solution is a Posted price mechanism {(see \cite{myerson1981optimal})}, so the question is if the seller can extract more revenue when he can also inspect and reward. The answer is indeed affirmative as \citet{alaei2024optimal} show via an example, when the prior is the uniform distribution over $[0,1]$, that the optimal {mechanism} with deferred inspection and rewards yields an expected revenue around 0.38, whereas the optimal Posted price mechanism yields 0.25. Hence, securing the bids as deposits, inspecting the value derived from the allocation, and returning a reward to the truthful reports greatly helps the auctioneer to increase the revenue. For ease of comparison, Figure \ref{fig: comparisonToPosted} illustrates the allocation and payment rules of these two mechanisms when the agent's type is uniformly distributed.

We next present the results of \citet{alaei2024optimal} that remain valid in the face of ambiguity.

\begin{proposition}[\citet{alaei2024optimal}]\label{prop: resultsAlaei}
	Given an ambiguity set $\mathcal{P}_{\mathbf{k}}$, there exists a robustly optimal deferred-inspection auction $(x,p)$ that satisfies
	\begin{itemize}
		\item[(i)] $p(\nu)=\inf_{\hat \nu \leq \nu} \{ \nu(x(\nu)-x(\hat \nu)) +\hat \nu \}$ for all $\nu \in [0,1]$,
		\item[(ii)] $x$ is weakly increasing with $x(0)=0$ and $x(1)=1$, and it is concave.
	\end{itemize}
\end{proposition}

Note that the above result appears as a necessary condition for optimality in \citet{alaei2024optimal}, whereas robust optimality may be achieved without it. This is because \citet{alaei2024optimal} assume their prior distribution to be everywhere strictly positive, whereas, our ambiguity set also includes discrete or mixed distributions. Nevertheless, the proof of \citet{alaei2024optimal} can be deployed in our setting to show the existence of a robustly optimal mechanism satisfying the same conditions.

Given any allocation rule $x(\nu)$ weakly increasing and concave, the optimal value of $p(\nu)$, $i.e.$ the maximum feasible payment that can be extracted from $\nu$, can be found using the formula in Proposition \ref{prop: resultsAlaei}. This requires finding the type $\hat \nu$ that satisfies the first order condition for the function $-\nu x(\hat \nu) +\hat \nu$.
Hence, the optimal mechanism could be rather complex for application as the agents should be able to calculate first-order conditions to check its incentive compatibility. 

As it is evident from Figure \ref{fig: comparisonToPosted}, deferred inspection with rewards enlarges the set of feasible solutions of the auction design problem, which renders simple mechanisms such as Posted price obsolete if one can embrace the complexity that comes with it. However, the optimal mechanism from \citet{alaei2024optimal} and the optimal Posted price mechanism are in stark difference, which may suggest the existence of other mechanisms that can balance performance and simplicity. {Our results affirm this observation {for the one agent case}.
	
	{In the rest of {Section \ref{sec: one agent}}, we will illustrate that there can be infinitely many robustly optimal mechanisms that yield the same expected revenue in the worst-case but perform quite differently under non-worst distributions that also belong to the same ambiguity set. In such a case, it is reasonable to search for mechanisms that promise more than worst-case optimality. The concept of Pareto robust optimality from the robust optimization literature requires stronger conditions, ensuring that one cannot improve the solution with respect to some admissible distribution without worsening it under another. {Pareto robust optimality provides a refinement criterion that is particularly relevant in robust optimization problems with non-unique worst-case solutions, allowing the designer to select mechanisms that preserve worst-case guarantees while improving performance under non-adversarial scenarios.} Its formal definition, presented below, is adjusted to fit the problem of interest. See \cite{iancu2014pareto} for a more general treatment.
		
		\begin{definition}
			A mechanism $(x',p')$ that is feasible in \eqref{eq: MDP} \textit{weakly Pareto robustly dominates} another mechanism $(x,p)$ if, for all $\mathbb{P} \in \mathcal{P}_{\mathbf{k}}$,
			\begin{equation}
				\mathbb{E}_{\mathbb{P}}[p'(\nu)] \geq \mathbb{E}_{\mathbb{P}}[p(\nu)]. \label{eq: PRO definition}
			\end{equation}
			Moreover, if \eqref{eq: PRO definition} strictly holds for some $\mathbb{P} \in \mathcal{P}_{\mathbf{k}}$, then $(x',p')$ \textit{Pareto robustly dominates} $(x,p)$. A mechanism is called \textit{Pareto robustly optimal} if it solves \eqref{eq: MDP}, and there does not exist another feasible mechanism that Pareto robustly dominates it.
		\end{definition}
	}
	
	\subsection{Mechanisms with Polynomial Payment Rules}
	
	We will now argue that the seller who is interested in maximizing her worst-case expected revenue under 
	{the ambiguity set \eqref{eq: ambiguity set}} can focus on mechanisms that admit {polynomial} payment rules.
	{In contrast to classical robust selling problems, the polynomial obtained from duality arguments characterizes only a lower bound on aggregate payments. Deferred inspection breaks the tight coupling between this polynomial and a unique feasible mechanism.}
	{Note that when the seller only knows the expectation, this payment rule simplifies to a linear function.} 
	This would simplify our search for a robustly optimal mechanism and may also return an easy-to-use mechanism. 
	{A similar result is also proven by \cite{carrasco2018optimal}. They show that the seller's problem can be reduced to finding the coefficients of a degree $N$ polynomial, which in turn determines the robustly optimal payment and allocation rules. We show that these results partially extend to our setting, allowing us to focus on polynomial payment rules. However, the enlarged set of feasible mechanisms in the presence of deferred inspection and rewards leads to two distinctions compared to \cite{carrasco2018optimal}. First, the robustly optimal polynomial payment rule may not uniquely determine a corresponding feasible allocation rule, which prevents us from using the solution approach of \cite{carrasco2018optimal}. Second, our robust mechanism design problem may admit infinitely many solutions, not restricted to the set of mechanisms with polynomial payment rules. Hence, in the next section, we identify two different robustly optimal mechanisms: one that employs a polynomial payment rule, and the other that is Pareto robustly optimal.
	}
	
	{
		\begin{lemma}[\cite{carrasco2018optimal}]\label{lem: polynomialPayment}
			Let $p(.)$ be 
			bounded and $\mathbb{P}^* \in \mathcal{P}_{\mathbf{k}}$ be such that
			\begin{equation*}
				\int_0^1 p(\nu) d \mathbb{P}^*(\nu) = \min_{\mathbb{P} \in \mathcal{P}_{\mathbf{k}}} \int_0^1 p(\nu) d \mathbb{P}(\nu) \geq 0.
			\end{equation*}
			Then, there is $\boldsymbol{\lambda}=(\lambda_0,\lambda_1,\ldots,\lambda_N)$ satisfying:
			\begin{itemize}
				\item[(i)] $p(\nu) \geq \sum_{i=0}^N \lambda_i \nu^i$ for all $\nu \in [0,1]$ with equality $\mathbb{P}^*$ almost surely;
				\item[(ii)] $\lambda_0 \leq 0$ for $p(.)$ feasible in \eqref{eq: MDP};
				\item[(iii)] the seller's revenue is $\int_0^1 p(\nu) d \mathbb{P}^*(\nu) = \sum_{i=0}^N \lambda_i k_i$. 
			\end{itemize}
		\end{lemma}
		
		Lemma \ref{lem: polynomialPayment} {above is a clipped version of what \cite{carrasco2018optimal} presents, because we omit the results that are not relevant for our setting and keep only the assumptions that are needed for the remaining results. For instance, \cite{carrasco2018optimal} state the above result for $p(.)$ weakly increasing, which assumption is then leveraged for showing that $p(\nu)$ is higher than the monotonic hull of $\sum_{i=0}^N \lambda_i \nu^i$. Although one can also prove that a monotone robustly optimal $p(.)$ exists in our setting, the results we are interested in do not require it. Focusing our attention on what is relevant will simplify our arguments for the multi-agent case. In fact, we extend Lemma \ref{lem: polynomialPayment} to $J \geq 1$ many agents and present its proof for the sake of completeness (see Lemma \ref{lem: polynomialPayment-multiAgent}). 
			
			Another difference from \cite{carrasco2018optimal}} is in part $(ii)$, which originally deduces negativity of first and last non-zero polynomial coefficients. The proof for this part utilizes the optimality of posted price mechanisms when $\mathcal{P}_{\mathbf{k}}$ is a singleton, which no longer holds in the presence of deferred inspection and rewards. Instead, we can only deduce that $\lambda_0$ is non-positive, which follows from part $(i)$ of Lemma \ref{lem: polynomialPayment}, and the \eqref{IR} constraint for type $0$, $p(0) \leq 0$.
		
		Lemma \ref{lem: polynomialPayment} {states the following. Any payment rule, which is interesting ($i.e.$, feasible and leads to a non-negative expected worst-case payoff),} is bounded below by a polynomial of degree $N$. 
		We next present the model where the seller is restricted to polynomial payment rules:
		\begingroup
		\setlength{\abovedisplayskip}{2pt} 
		\setlength{\belowdisplayskip}{2pt}
		\begin{align}
			\max_{x,\boldsymbol{\lambda}} \; &\sum_{i=0}^N \lambda_i k_i \; \label{eq: MDP lambda} \tag{MDP$_{\lambda}$}\\
			\text{s.t.} \;&\sum_{i=0}^N \lambda_i \nu^i \leq \inf_{\hat{\nu} \leq \nu} \{ \nu ( x(\nu)-x(\hat{\nu}) ) + \hat{\nu} \} &\forall \nu \in [0,1], \nonumber\\
			&x(\nu) \in [0,1] &\forall \nu \in [0,1]. \nonumber
		\end{align}
		\endgroup
		The analog of the above model presented in \cite{carrasco2018optimal} is much simpler, as it is only over $\boldsymbol{\lambda}$, maximizes the same objective, and only requires the uniquely determined allocation to be within $[0,1]$. This is not the case in our setting, as the optimal choice of $\boldsymbol{\lambda}$ can be coupled with different allocation rules. Nevertheless, our designer can also solve a simpler problem, \eqref{eq: MDP lambda}, to find her optimal worst-case expected revenue and a robustly optimal polynomial payment rule.
		
		\begin{proposition}\label{prop: equivalence of MDPs}
			A mechanism $(x^*,p^*)$ solves \eqref{eq: MDP} if and only if there exists some $\boldsymbol{\lambda}^*$ such that $(x^*, \boldsymbol{\lambda}^*)$ solves \eqref{eq: MDP lambda} and $p^*(\nu)$ satisfies the following set of inequalities 
			\begin{equation*}
				\begin{aligned}
					p^*(\nu) &\geq \sum_{i=0}^N \lambda_i^* \nu^i &\forall \nu \in [0,1],\\
					p^*(\nu) &\leq \inf_{\hat{\nu} \leq \nu} \{ \nu ( x^*(\nu)-x^*(\hat{\nu}) ) + \hat{\nu} \} &\forall \nu \in [0,1].		 
				\end{aligned}
			\end{equation*}
			Moreover, $z^* = \sum_{i=0}^N \lambda_i^* k_i$.
		\end{proposition}
		
		{In our setting, Proposition \ref{prop: equivalence of MDPs} follows from Proposition \ref{prop: equivalence of MDPs-multiAgent}, which extends the analog result from \cite{carrasco2018optimal} to any number of agents.}
		
		Proposition \ref{prop: equivalence of MDPs} is a looser version of what \cite{carrasco2018optimal} presents, but it still helps in simplifying the seller's problem. We observe that a robustly optimal mechanism with inspection and rewards can still follow a polynomial function; however, this is not a one-to-one relationship, as in \cite{carrasco2018optimal}. That is, identification of the optimal polynomial function no longer completely pins down a unique optimal solution. As a result, we are not able to utilize the solution approach of \cite{carrasco2018optimal} to identify the optimal mechanism. Still, studying the numerical solutions of \eqref{eq: MDP lambda} points in the direction of a possibly infinite set of robustly optimal mechanisms.
	}
	
	\subsection{First Moment}
	
	In this section, we consider the ambiguity set $\mathcal{P}_{\mu}$ that specifies only the first moment, $k_1 = \mu$. Accordingly, we set $N=1$ in the polynomial framework established in the previous section, which naturally leads us to search for robustly optimal mechanisms with linear payment rules. We present a worst-case distribution, a robustly optimal mechanism with a linear payment rule, and a Pareto robustly optimal mechanism for a subset of values $\mu \in (0,1)$. The results for the remaining values of $\mu$ are relegated to the Electronic Companion.
	
	\begin{lemma}\label{lem:upperBounds}
		For any $\mu \in (0,1)$, the two-point distribution that assigns probability $\alpha=\mu/2$ to type $1$ and probability $1-\alpha$ to type ${\nu}' = \frac{\mu}{2-\mu}$ is contained in the ambiguity set $\mathcal{P}_{\mu}$, under which the seller's optimal revenue is $\frac{ \mu}{2- \mu}$. Hence, $z^* \leq \frac{ \mu}{2- \mu}$.
	\end{lemma}
	
	Lemma \ref{lem:upperBounds} points to a two-point distribution contained in $\mathcal{P}_{\mu}$, under which the optimal mechanism yields an expected value of $\mu/(2-\mu)$, which is strictly less than $\mu$ for any $\mu \in (0,1)$. Note that limiting the strategy set of the adversary would only increase the equilibrium value of the zero-sum game. Hence, Lemma \ref{lem:upperBounds} presents an upper bound on the value of the seller's max-min problem.
	
	The two-point distribution in Lemma \ref{lem:upperBounds} leaves the principal indifferent between the following two mechanisms and any convex combination of them. The first mechanism 
	allocates the good to both types with probability $1$ and asks for a payment of ${\nu}'= \mu/(2- \mu)$ from either type. The second one allocates the good to type 1 with probability 1, to type ${\nu}'$ with ${\nu}'$, and asks for payment amounts $1$ and ${\nu'}^2$, respectively. The latter mechanism leaves both types with exactly zero information rent, whereas under the first mechanism, type $1$ receives strictly positive information rent. Nevertheless, their expected revenues coincide under the given two-point distribution, leaving the seller indifferent. This is because, in the first mechanism, the expected gains from allocating the good to type $\nu'$ with probability $1$ and the expected loss from the information rent given to type $1$ break even.
	
	Our next result presents a robustly optimal mechanism with a linear payment rule that attains the upper bound given in Lemma \ref{lem:upperBounds} for a subset of $\mu$ values. We will illustrate this mechanism for different $\mu$ values and argue that the payment rule can be improved at the cost of complexity. Our final result in this section presents the payment rule that extracts the maximum feasible payment from each type under the same allocation rule, thereby achieving Pareto robust optimality.
	
	\begin{theorem}\label{thrm: twoPointOptimalMechanism} 
		{Define $\mu' = \frac{2z'}{z'+1}$ where $z'$ is the optimal objective value of the following problem: 
			\begin{equation*}
				\begin{aligned}
					z' = \max_{t \in [0,1]} \; & z(t) \\
					\text{s.t.} \;&z(t) = \frac{(3t^2-2t) + t\sqrt{2(t^4+2t^2-6t+3)}}{2-t^2}, \\ 
					& t^4+2t^2-6t+3 \geq 0. 
				\end{aligned}
			\end{equation*}
			If $\mu \in [\mu',1]$,} then the following mechanism is robustly optimal with a worst-case expected payoff of $z^* = \mu/(2- \mu)$, \begin{equation*}\label{eq:linearPaymentMechanism} \begin{aligned} 
				&p^*(\nu) = \lambda_1 \nu+ \lambda_0 &\forall \nu \in [0,1],\\
				&x^*(\nu) = 
				\begin{cases} 
					\frac{\lambda_1}{2\mu^2}\nu &\text{if } \nu \in [0, \underline \nu),\\ 
					\lambda_1+ \frac{\lambda_0}{\nu} &\text{if } \nu \in [\underline \nu, \nu^\circ),\\ 
					\nu+\lambda_1-2z^* &\text{if } \nu \in [\nu^\circ, \overline{\nu}),\\ 
					1 &\text{if } \nu \in [\overline{\nu},1],
				\end{cases}
			\end{aligned} 
		\end{equation*} 
		where $\lambda_1 = 2(z^*/\mu)^2$, $\lambda_0 =-{z^*}^2$, $\underline \nu = \mu^2$, $\nu^\circ = z^*$, and $\overline{\nu}= \lambda_1+\lambda_0$. Moreover, $x^*$ is continuous and weakly increasing.
	\end{theorem}
	
	Theorem \ref{thrm: twoPointOptimalMechanism} describes a robustly optimal mechanism for 
	{the case $\mu \in [\mu',1]$, where $\mu'=2z'/(z'+1)$ and $z'$ is the optimal objective value of a nonlinear maximization problem. The nonlinear model only serves to find the minimum value of $\mu$, for which the given mechanism is feasible. Finding the exact value of $z'$ requires solving a high-degree polynomial and results in $\mu' = \frac{2z'}{z'+1} \approx 0.315$. The plots of the nonlinear functions are given in Figure \ref{fig: supportingFigureForMu'}. For the case $\mu < \mu'$, the mechanism in Theorem \ref{thrm: twoPointOptimalMechanism} becomes infeasible, as it no longer satisfies \eqref{IC} constraints for a subset of types. In such a case, the adversary can do better by deviating from the two-point distribution given in Lemma \ref{lem:upperBounds} to a three or four-point distribution, and the upper bound $\mu/(2-\mu)$ is no longer within reach for the designer. 
		We treat the case of $\mu \in (0,\mu')$ in Section \ref{sec: smallMu} and discuss when it is possible to identify the solution in closed form and propose an approximation mechanism for the remaining cases.
		
		\begin{figure}
			\centering
			\includegraphics[scale=0.7]{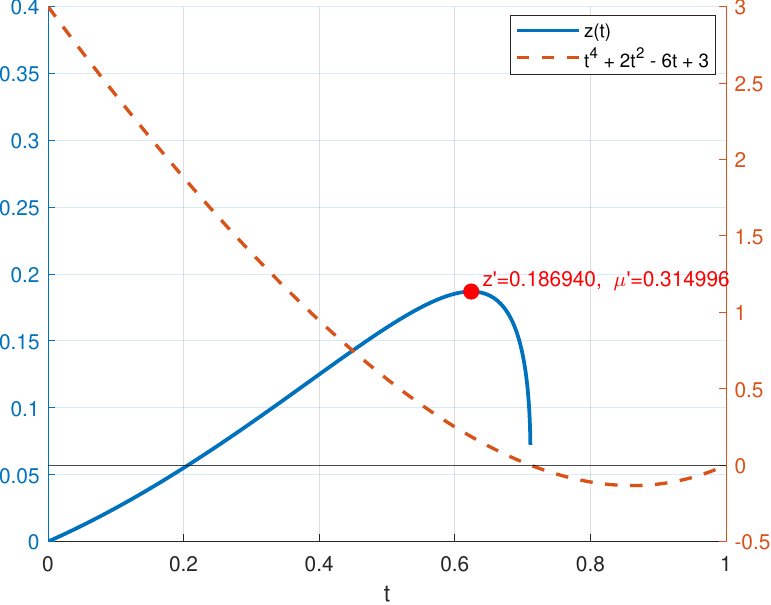}
			\caption{The plots of the nonlinear functions that determine $\mu'$.}
			\label{fig: supportingFigureForMu'}
		\end{figure}
		
		\begin{figure}
			\centering
			\includegraphics[scale=0.8]{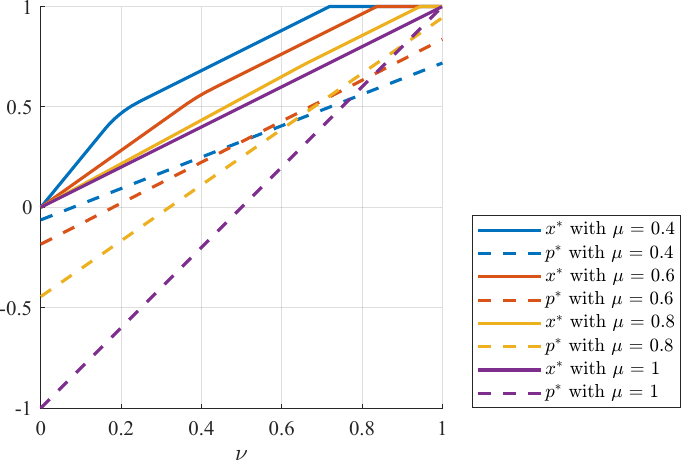}
			\caption{A robustly optimal mechanism with a linear payment rule for different $\mu$ values.}
			\label{fig: optimalTwoPointMechanisms}
		\end{figure}
		
		Figure \ref{fig: optimalTwoPointMechanisms} illustrates the mechanism $(x^*,p^*)$ for different values of $\mu \in [\mu',1)$. The first thing to be noticed in Figure \ref{fig: optimalTwoPointMechanisms} is that there are types that make negative net payments, $i.e.$ receive money from the principal. One can omit these negative payments and use a {Clipped} linear payment rule instead, which is equal to $p^*$ whenever $p^*$ is non-negative and equal to zero when $p^*$ is negative. 
		
		{It is easy to see that, {under the Clipped linear payment rule,} the principal's worst-case expected payoff remains the same, as we do not lower the payment for any type. For feasibility, notice that the constraints on the payment rule, \eqref{IC} and \eqref{IR}, only impose upper bounds that are always non-negative when the allocation rule is weakly increasing. As the proposed  allocation rule, $x^*$ is weakly increasing, we can substitute the negative parts of the linear payment rule with zero without disturbing feasibility.}		
		Then, the {Clipped} linear payment rule is feasible, and it would improve the expected payoff under some non-worst-case distributions in $\mathcal{P}_{\mu}$. {Since these two mechanisms, $x^*$ coupled with the linear and {Clipped} linear payment rules, are distinct, and we have a convex feasibility set, any convex combination of these two payment rules is feasible and guarantee the same worst-case expected revenue. In other words, we have infinitely many robustly optimal mechanisms. {It is clear that the Clipped linear payment rule is superior to the linear one, but can we do better?} Our next result presents the payment rule that extracts the maximum payment from each type under the fixed allocation rule $x^*$, for which we also prove Pareto robust optimality.}
		
		\begin{theorem}\label{thrm: twoPointMaximalPayment}
			If $\mu \in [\mu',1]$, then the following payment rule, $p_m^*$ extracts the maximum feasible payment from each type under $x^*$, and $(x^*,p_m^*)$ is {Pareto} robustly optimal:
			\begin{equation*}
				\begin{aligned}
					&p_m^*(\nu) = \begin{cases}
						\frac{\lambda_1}{2\mu^2}\nu^2 &\text{if } \nu \in [0, \underline \nu),\\
						\lambda_1 \nu+ \lambda_0 &\text{if } \nu \in [\underline \nu, \nu^\circ),\\
						\nu(\nu +\lambda_1 -2z^*) &\text{if } \nu \in [\nu^\circ, \nu^\star],\\
						\nu^2-2z^*(\nu -\sqrt{\nu}) &\text{if } \nu \in (\nu^\star, \overline{\nu}),\\
						\nu(1-\lambda_1)+2z^*\sqrt{\nu} &\text{if } \nu \in [\overline{\nu},1],
					\end{cases}
				\end{aligned}
			\end{equation*}
			where $\nu^\star = \mu^2(2-\mu)^2$.
		\end{theorem}
		
		\begin{figure}
			\centering
			\includegraphics[scale=0.8]{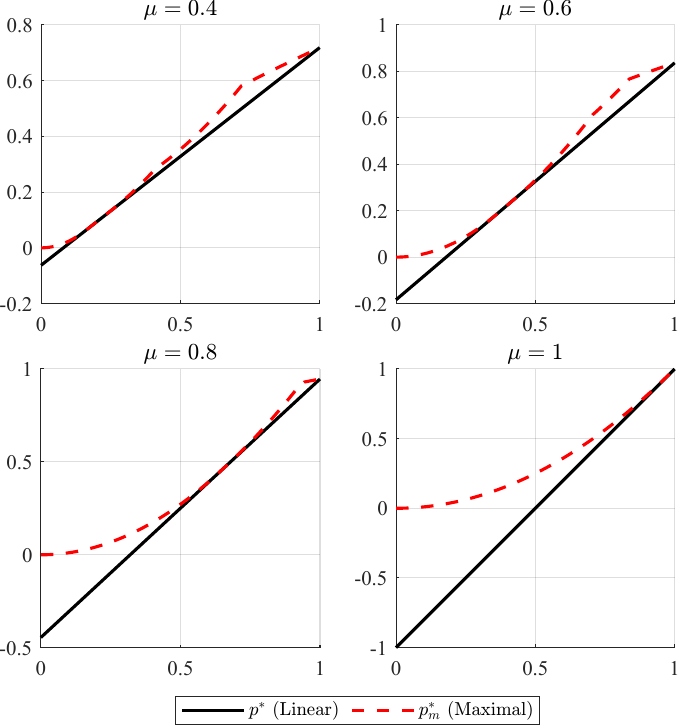}
			\caption{Plots of the linear and the maximal payment rules for different $\mu$ values.}
			\label{fig: maximalTwoPointMechanisms}
		\end{figure}
		
		Figure \ref{fig: maximalTwoPointMechanisms} illustrates $p_m^*$ and $p^*$ for different $\mu$ values. We see that $p_m^*$ is mostly non-linear but strictly improves the payment extracted from some types. 
		It leaves any type $\nu \leq \nu^\star$ with zero information rent, so they do not gain anything by participating in the mechanism. All other types, $\nu >\nu^*$, receive a positive information rent, and their payments leave them indifferent between truthfully reporting $\nu$ and posing themselves as $z^*\sqrt{\nu}$. {More importantly, the characterization of $(x^*,p_m^*)$, which Pareto robustly dominates $(x^*,p^*)$, underlines the difference between our setting and that of \cite{carrasco2018optimal}. That is, inclusion of deferred inspection and rewards requires a more detailed analysis of robustly optimal solutions, as a polynomial of degree $N$ payment rule no longer uniquely determines the optimal solution. Moreover, to achieve Pareto robust optimality, one may need to use piecewise functions with higher-degree polynomials and radicals.}

		\textbf{Example.} When $\nu$ is uniformly distributed over $[0,1]$, the optimal mechanism from \citet{alaei2024optimal} (see Figure \ref{fig: comparisonToPosted}) generates an expected revenue of $0.3808$, and the optimal posted price mechanism generates $0.25$. Assume that the prior is the same, but we use the robustly optimal mechanism for $\mathcal{P}_\mu$ when $\mu = 0.5$. Then, the mechanism with the linear payment rule (Theorem \ref{thrm: twoPointOptimalMechanism}) generates an expected revenue of $\int_0^1 (\alpha \nu + \beta) \text{d} \nu = \int_0^1 (8 \nu/9 -1/9) \text{d} \nu \approx 0.3333$, whereas the Pareto robustly optimal mechanism (Theorem \ref{thrm: twoPointMaximalPayment}) generates $0.3716$. Hence, the linear payment rule brings simplicity at the cost of performance, but it still outperforms the optimal posted price mechanism. On the other hand, if complexity is not an issue, then using the robustly optimal mechanism with the maximal payment rule instead of the optimal mechanism tailored to a specific distribution ensures near-optimal performance guarantees and requires only the support and expected value information, thereby ensuring robustness against possible errors in the data.
		
		\textbf{Numerical experiment.} We also report a numerical experiment to compare the performances of three mechanisms, namely the optimal mechanism from \citet{alaei2024optimal}, our two robustly optimal mechanisms with linear and maximal payment rules, when the type distribution is contaminated with the worst-case distribution that we identified in Lemma \ref{lem:upperBounds}. For this experiment, we consider a discrete type set to be able to use linear programming.  The type set consists of $100$ equidistant points on $[0,1]$. We let $\mathbb P^N$ be the nominal distribution that puts $1/100$ probability to each point in the type set, and we let $\mathbb P^W$ be the worst-case distribution with the same expected value $1/2$. We will calculate the true type distribution as $\mathbb P^\varepsilon = (1-\varepsilon)\mathbb P^N + \varepsilon \mathbb P^W$ where $\varepsilon \in [0,1]$ is the contamination level,
		and report the performances of the three mechanisms across different contamination levels. The performance of a mechanism refers to its expected payoff divided by the maximum achievable payoff under the unique prior $\mathbb P^\varepsilon$ without ambiguity. {In Figure \ref{fig: experiment}, the vertical axis represents performance relative to the optimal prior-dependent mechanism, not absolute expected payoff. For instance, when $\epsilon=1$, the underlying distribution coincides with the worst-case distribution used to design our robust mechanisms. In this case, the robustly optimal mechanisms coincide with the optimal prior-dependent mechanism, which explains why their relative performance equals~1.}
		
		\begin{figure}
			\centering
			\includegraphics[scale=0.7]{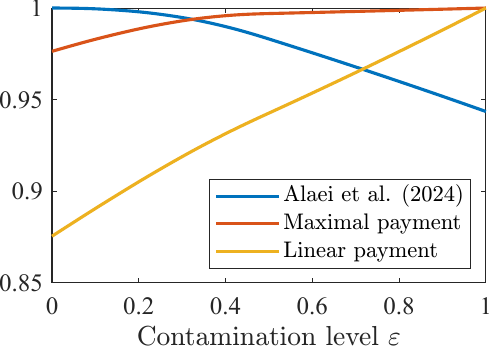}
			\caption{Performances of three mechanisms compared to the mechanism that is optimal under the unique prior $\mathbb{P}^\varepsilon$.}
			\label{fig: experiment}
		\end{figure}
		
		Figure \ref{fig: experiment} visualizes the relative performances of the three mechanisms. As the figure shows, the Pareto robustly optimal mechanism with maximal payments consistently performs close to the true optimal solution and outperforms the nominal mechanism when the contamination level exceeds $0.324$. For contamination levels beyond $0.716$, the nominal mechanism performs worse than both of our robustly optimal mechanisms. 
		Therefore, in addition to achieving the worst-case optimal payoffs, our robustly optimal mechanisms also perform favorably compared to optimal mechanisms designed for a misspecified distribution.

		{In summary, if a mechanism designer lacks data on buyer valuations, they should implement the Linear Rule for simplicity if subsidies are permitted. If subsidies are impossible, they should implement the Clipped Linear Rule. If the platform has high computational power and wants to maximize revenue, they should implement the Maximal Payment Rule.}	{ In Table \ref{table: comparison of mechanisms}, we compare and summarize the properties of the mechanisms that solve the one-agent case under different assumptions. In the following sections, we demonstrate that the polynomial structure of simple, robustly optimal mechanisms extends beyond the case of a single agent. Nevertheless, the extension is not straightforward, and one needs to treat the two-agent case separately.}
		
\begin{table}[ht]
	\TABLE
	{Comparison of our robustly optimal mechanisms and the mechanism from \cite{alaei2024optimal}\label{table: comparison of mechanisms}}
	{\begin{tabularx}{\textwidth}{@{} l c c c c @{}}
			\hline\up
			Mechanism & \makecell{Robust\\Optimality} & \makecell{Rent for\\Low Types} & Pros & Cons \\ \hline\up
			\cite{alaei2024optimal} & \ding{55} & zero & \makecell{High\\expected\\revenue\\if the prior\\is known} & \makecell{Robustly\\suboptimal\\ \& complex\\payment rule} \\ 
			Linear Payment & \ding{51} & \makecell{high with\\subsidy} & \makecell{Robustness \\ \& simplicity} & \makecell{Pareto\\robustly\\dominated} \\ 
			Maximal Payment & \ding{51} & zero & \makecell{Pareto\\robust\\optimality} & \makecell{Complex\\payment rule} \down\\ \hline
	\end{tabularx}}
	{
	}
\end{table}

\section{Symmetrical Multiple Agents}\label{sec: multi-agentCase}

{
	In this section, we extend our results to the case where there are $J>1$ symmetric agents. Note that \cite{alaei2024optimal}, when working on the multi-agent setting, assume that the types are identically and independently distributed. Our symmetry assumption, which enforces our ambiguity set to have the same bounds and moments for each agent, is not as restrictive. We will see that the agents' types do not need to be independently distributed under the worst-case distribution.
	
	First, we extend our notation in a way that the case of $J =1$ formulates the one-agent model. We use $\boldsymbol{\nu} \in [0,1]^J$ to denote the vector of agents' types, $(\nu_1,\ldots,\nu_J)$, which can also be denoted as $(\nu_j,\boldsymbol{\nu}_{-j})$ for all $j \in [J]$, where $\boldsymbol{\nu}_{-j}$ is the vector obtained from $\boldsymbol{\nu}$ by removing the type of agent $j$. Then, the ambiguity set with the first $N$ moments, $\mathbf K \in (0,1)^{N \times J}$, can be written as follows:
	\begin{equation}\label{eq: ambiguitySetGeneralized}
			\mathcal{P}_{\mathbf K} = \{ \mathbb{P} \in \mathcal{P}_0([0,1]^J): \; \int_0^1 \nu_j^i d \mathbb{P}(\boldsymbol{\nu}) = K_{ij}, \forall j \in [J], i \in [N]\}.
	\end{equation}
	For symmetrical agents, we have $K_{ij} = K_i$ for all $j \in [J]$ and $i \in [N]$. We solve the problem under the following assumption to avoid degenerate and trivial cases.
	\begin{assumption}\label{assumption: N moments-multiAgent}
		We consider $\mathbf K \in (0,1)^{N \times J}$ such that for some $\varepsilon>0$, we have $\mathcal{P}_{\hat{\mathbf{K}}} \neq \emptyset$ for all $\hat{\mathbf{K}} \in (0,1)^{N \times J}$ satisfying $|\hat{\mathbf{K}} - \mathbf{K}| < \varepsilon$.
	\end{assumption}
	Note that Assumption \ref{assumption: N moments-multiAgent} reduces to Assumption \ref{assumption: N moments} when $J=1$. Our model generalizes to:
	\begingroup
	\setlength{\abovedisplayskip}{2pt} 
	\setlength{\belowdisplayskip}{2pt}
	\begin{align}
		{z^*} = \sup_{x \geq 0,p} \;&\inf_{\mathbb{P} \in \mathcal{P}_{\mathbf{K}}} \mathbb{E}_{\mathbb{P}}[\sum_{j \in |J|} p_j(\boldsymbol{\nu})] \label{eq: MDPmultiAgent} \tag{MDP$^J$}\\
		\text{s.t.} \;&p_j(\boldsymbol{\nu}) \leq \nu_j \big( x_j(\boldsymbol{\nu})- x_j(\hat{\nu},\boldsymbol{\nu}_{-j}) \big) + \hat{\nu}  &\forall \hat{\nu} \in [0,1], \forall j \in [J], \forall \boldsymbol{\nu} \in [0,1]^J, \label{DSIC} \tag{DS-IC}\\
		&p_j(\boldsymbol{\nu}) \leq \nu_j x_j(\boldsymbol{\nu}) &\forall j \in [J], \forall \boldsymbol{\nu} \in [0,1]^J, \label{EPIR} \tag{EP-IR}\\
		&\sum_{j \in [J]} x_j(\boldsymbol{\nu}) \leq 1 &\forall \boldsymbol{\nu} \in [0,1]^J,\nonumber
	\end{align} 
	\endgroup
	where \eqref{DSIC} constraints ensure Dominant Strategy Incentive Compatibility, and \eqref{EPIR} ensure Ex-post Individual Rationality. When a mechanism is \eqref{DSIC}, any agent $j \in [J]$ is always better off reporting his true type for any vector of remaining agents' types, $\boldsymbol{\nu}_{-j} \in [0,1]^{J-1}$. As there is distributional ambiguity, we need to find a mechanism that admits a Dominant Strategy Equilibrium instead of a Bayesian one. This is essential for our multi-agent mechanism to be robust. The same reason also demands \eqref{EPIR}, $i.e.$, any agent $j$ is always better off participating in the mechanism for any vector of remaining agents' types, $\boldsymbol{\nu}_{-j}$. 
	{Under distributional ambiguity, Bayesian incentive compatibility would depend on beliefs over other agents' types, which are themselves ill-defined.
		\eqref{DSIC} and \eqref{EPIR} avoid this dependence and is therefore necessary for robust implementation.}
	
	When extending their results to the multiple agents, \cite{alaei2024optimal} are able to work on a reduced formulation with expected allocation and payment rules thanks to their assumption of independently distributed types, and Bayesian \eqref{IC} and Ex-ante \eqref{IR} constraints. Our constraints, \eqref{DSIC} and \eqref{EPIR}, prevent us from doing the same. Instead, we extend the results of \cite{carrasco2018optimal} to the multi-agent setting.
	
	\begin{lemma}\label{lem: polynomialPayment-multiAgent}
		For $J \in \mathbb{Z}_+$ many agents, let $\sum_{j \in [J]} p_j(.)$ be 
		bounded and $\mathbb{P}^* \in \mathcal{P}_{\mathbf{K}}$ solve $\min_{\mathbb{P} \in \mathcal{P}_{\mathbf{K}}} \int_{[0,1]^J} \sum_{j \in [J]} p_j(\boldsymbol{\nu}) d \mathbb{P}(\boldsymbol{\nu}) \geq 0$.
		Then, there is $\lambda_0$ and $\lambda_{ij}$ for all $i \in [N]$ and $j \in [J]$ satisfying:
		\begin{itemize}
			\item[(i)] $\sum_{j \in [J]} p_j(\boldsymbol{\nu}) \geq \sum_{i \in [N]} \sum_{j \in [J]} \lambda_{ij} \nu_j^i + \lambda_0$ for all $\boldsymbol{\nu} \in [0,1]^J$ with equality $\mathbb{P}^*$ almost surely;
			\item[(ii)] $\lambda_0 \leq 0$ for $p(.)$ feasible in \eqref{eq: MDPmultiAgent};
			\item[(iii)] the seller's revenue is $\int_{[0,1]^J} \sum_{j \in [J]} p_j(\boldsymbol{\nu}) d \mathbb{P}^*(\nu) = \sum_{i \in [N]} \sum_{j \in [J]} \lambda_{ij} K_{ij} + \lambda_0$. 
		\end{itemize}
	\end{lemma}
	
	Lemma \ref{lem: polynomialPayment-multiAgent} shows that there exists a multi-dimensional polynomial that yields the designer's robustly optimal payoff, which can be found by solving the following formulation.
	\begingroup
	\setlength{\abovedisplayskip}{2pt} 
	\setlength{\belowdisplayskip}{2pt}
	\begin{align}
		\max_{x \geq 0,\boldsymbol{\lambda}} \; &\sum_{i \in [N]} \sum_{j \in [J]} \lambda_{ij} K_{ij} +\lambda_0 \label{eq: MDP lambda-multiAgent} \tag{MDP$_{\lambda}^J$}\\
		\text{s.t.} \; &\sum_{i \in [N]} \sum_{j \in [J]} \lambda_{ij} \nu_j^i +\lambda_0 \leq \sum_{j \in [J]} \inf_{\hat{\nu} \leq \nu_j} \{ \nu_j ( x_j(\boldsymbol{\nu})-x_j(\hat{\nu},\boldsymbol{\nu}_{-j}) ) + \hat{\nu} \} &\forall \boldsymbol{\nu} \in [0,1]^J, \nonumber\\
		&\sum_{j \in [J]} x_j(\boldsymbol{\nu}) \leq 1 &\forall \boldsymbol{\nu} \in [0,1]^J. \nonumber
	\end{align}
	\endgroup
	
	Our next result affirms that the designer can restrict attention to payment rules whose \textit{aggregate} value is equal to a multi-dimensional polynomial.
	
	\begin{proposition}\label{prop: equivalence of MDPs-multiAgent}
		For $J \in \mathbb{Z}_+$ many agents, a mechanism $(x^*,p^*)$ solves \eqref{eq: MDPmultiAgent} if and only if there exists some $\lambda_0 \in \mathbb{R}$ and $\boldsymbol{\lambda}^* \in \mathbb{R}^{N \times J}$ such that $(x^*, \lambda_0^*, \boldsymbol{\lambda}^*)$ solves \eqref{eq: MDP lambda-multiAgent} and $p^*(.)$ satisfies the following set of inequalities for all $\boldsymbol{\nu} \in [0,1]^J$:
		\begin{equation*}
			\begin{aligned}
				&\sum_{j \in [J]} p_j^*(\boldsymbol{\nu}) \geq \sum_{i \in [N]} \sum_{j \in [J]} \lambda_{ij}^* \nu_j^i +\lambda_0^*, \\
				&p_j^*(\boldsymbol{\nu}) \leq \inf_{\hat{\nu} \leq \nu_j} \{ \nu_j ( x_j^*(\boldsymbol{\nu})-x_j^*(\hat{\nu}, \boldsymbol{\nu}_{-j}) ) + \hat{\nu} \}, \; \forall j \leq J.		 
			\end{aligned}
		\end{equation*}
		Moreover, $z^* = \sum_{i \in [N]} \sum_{j \in [J]} \lambda_{ij}^* K_{ij} +\lambda_0^*$.
	\end{proposition}
	We deduce the multi-dimensional polynomial by leveraging KKT conditions for the adversary's problem, which lower bounds the aggregate payment rule and provides no further insight into the individual payments from each agent. In the following subsections, we will see that the simple structure of the aggregate payment rule does not necessarily make the mechanism more practical, which also complicates the characterization of the robustly optimal mechanism.
}

{Proposition~5 characterizes the structure of aggregate payments only. It does not imply that individual payment or allocation rules admit a similarly simple form, which explains the analytical complexity of robustly optimal multi-agent mechanisms. The multi-dimensional polynomial is obtained by leveraging KKT conditions for the adversary's problem and provides only a lower bound on aggregate payments, offering no further insight into the structure of individual payments. As we show in the following subsections, the simplicity of the aggregate payment rule does not necessarily translate into practical or easily interpretable mechanisms.}

\subsection{The Case of Two Agents \& First Moment}\label{sec:2agents}

{In this section, we analyze our problem when there are two symmetrical agents, and the ambiguity set contains only the first-moment information. Let $\mathcal{P}_{\boldsymbol{\mu}}$ denote the ambiguity set from \eqref{eq: ambiguitySetGeneralized} with only the first moment information. We focus on the case $\mu_j = \mu$ for all $j \in [J]$ for some $\mu \in (0,1)$ and show that the adversary's best option is to use a distribution, under which the agents' types are correlated. This worst-case distribution cannot be written in closed form using the problem parameters, as it requires solving a complex nonlinear optimization problem.
	
	\begin{lemma}\label{lem: upperbounds-2agents}
		For any $\mu \in (0,1)$, the distribution that assigns probability $\beta/ r$ to profile $(r \nu^*, r \nu^*)$, probability $\beta$ to $(\nu^*, r \nu^*)$ and $(r \nu^*, \nu^*)$, probability $\beta r$ to $(\nu^*,\nu^*)$ and probability $\beta \nu^*$ to $(1,r \nu^*)$ and $(r \nu^*,1)$ is contained in the ambiguity set $\mathcal{P}_{\boldsymbol{\mu}}$, where $r, \nu^*, \beta$ solve the following:
		\begingroup
		\setlength{\abovedisplayskip}{2pt} 
		\setlength{\belowdisplayskip}{2pt}
		\begin{align}
			\min_{r, \nu^*, \beta} \;& 2 \beta \nu^* (r^2 + 3r/2 + \nu^* + 1) \nonumber\\
			\text{s.t.} \; &\beta = 1/(2\nu^*+ r + 1/r + 2), \label{eq: 2agentDist-equal to one}\\
			&r{\nu^*}^2 + (2r +3 -2\mu) \nu^* - \mu(r + 1/r + 2) = 0, \label{eq: 2agentDist-expectation}\\
			& r, \nu^*, \beta \in [0,1]. \label{eq: 2agentDist-nonnegativity}
		\end{align}
		\endgroup
		Letting $f(\mu)$ denote the optimal objective value of the above problem for all $\mu \in (0,1)$, seller's optimal revenue, $z^*$, is bounded above by $\min \{f(\mu), \mu\}$.
	\end{lemma}
	
	\begin{figure}
		\centering		
		\tikzset{every picture/.style={line width=0.75pt}} 
		
		\resizebox{0.55\textwidth}{!}{
			\begin{tikzpicture}[x=0.75pt,y=0.75pt,yscale=-1,xscale=1]
				
				\draw  (177,290.85) -- (459.62,290.85)(205.26,36.49) -- (205.26,319.11) (452.62,285.85) -- (459.62,290.85) -- (452.62,295.85) (200.26,43.49) -- (205.26,36.49) -- (210.26,43.49)  ;
				\draw  [dash pattern={on 0.84pt off 2.51pt}]  (202.29,62.63) -- (461.98,62.63) ;
				\draw  [dash pattern={on 0.84pt off 2.51pt}]  (432.47,40.03) -- (432.47,296.35) ;
				\draw  [dash pattern={on 0.84pt off 2.51pt}]  (202.29,146.94) -- (461.98,146.94) ;
				\draw  [dash pattern={on 0.84pt off 2.51pt}]  (202.29,248.96) -- (461.98,248.96) ;
				\draw  [dash pattern={on 0.84pt off 2.51pt}]  (349,40.03) -- (349,296.35) ;
				\draw  [dash pattern={on 0.84pt off 2.51pt}]  (246.98,40.03) -- (246.98,296.35) ;
				\draw  [fill={rgb, 255:red, 0; green, 0; blue, 0 }  ,fill opacity=1 ] (244.28,249.05) .. controls (244.28,247.6) and (245.45,246.43) .. (246.9,246.43) .. controls (248.34,246.43) and (249.51,247.6) .. (249.51,249.05) .. controls (249.51,250.49) and (248.34,251.66) .. (246.9,251.66) .. controls (245.45,251.66) and (244.28,250.49) .. (244.28,249.05) -- cycle ;
				\draw  [fill={rgb, 255:red, 0; green, 0; blue, 0 }  ,fill opacity=1 ] (346.3,249.05) .. controls (346.3,247.6) and (347.47,246.43) .. (348.92,246.43) .. controls (350.36,246.43) and (351.53,247.6) .. (351.53,249.05) .. controls (351.53,250.49) and (350.36,251.66) .. (348.92,251.66) .. controls (347.47,251.66) and (346.3,250.49) .. (346.3,249.05) -- cycle ;
				\draw  [fill={rgb, 255:red, 0; green, 0; blue, 0 }  ,fill opacity=1 ] (429.77,249.05) .. controls (429.77,247.6) and (430.94,246.43) .. (432.39,246.43) .. controls (433.83,246.43) and (435,247.6) .. (435,249.05) .. controls (435,250.49) and (433.83,251.66) .. (432.39,251.66) .. controls (430.94,251.66) and (429.77,250.49) .. (429.77,249.05) -- cycle ;
				\draw  [fill={rgb, 255:red, 0; green, 0; blue, 0 }  ,fill opacity=1 ] (346.3,147.03) .. controls (346.3,145.58) and (347.47,144.41) .. (348.92,144.41) .. controls (350.36,144.41) and (351.53,145.58) .. (351.53,147.03) .. controls (351.53,148.47) and (350.36,149.64) .. (348.92,149.64) .. controls (347.47,149.64) and (346.3,148.47) .. (346.3,147.03) -- cycle ;
				\draw  [fill={rgb, 255:red, 0; green, 0; blue, 0 }  ,fill opacity=1 ] (244.28,147.03) .. controls (244.28,145.58) and (245.45,144.41) .. (246.9,144.41) .. controls (248.34,144.41) and (249.51,145.58) .. (249.51,147.03) .. controls (249.51,148.47) and (248.34,149.64) .. (246.9,149.64) .. controls (245.45,149.64) and (244.28,148.47) .. (244.28,147.03) -- cycle ;
				\draw  [fill={rgb, 255:red, 0; green, 0; blue, 0 }  ,fill opacity=1 ] (244.28,62.71) .. controls (244.28,61.27) and (245.45,60.1) .. (246.9,60.1) .. controls (248.34,60.1) and (249.51,61.27) .. (249.51,62.71) .. controls (249.51,64.16) and (248.34,65.33) .. (246.9,65.33) .. controls (245.45,65.33) and (244.28,64.16) .. (244.28,62.71) -- cycle ;
				\draw    (247,147.47) -- (247,201.47) ;
				\draw [shift={(247,203.47)}, rotate = 270] [color={rgb, 255:red, 0; green, 0; blue, 0 }  ][line width=0.75]    (10.93,-3.29) .. controls (6.95,-1.4) and (3.31,-0.3) .. (0,0) .. controls (3.31,0.3) and (6.95,1.4) .. (10.93,3.29)   ;
				\draw    (247,62.47) -- (247,116.47) ;
				\draw [shift={(247,118.47)}, rotate = 270] [color={rgb, 255:red, 0; green, 0; blue, 0 }  ][line width=0.75]    (10.93,-3.29) .. controls (6.95,-1.4) and (3.31,-0.3) .. (0,0) .. controls (3.31,0.3) and (6.95,1.4) .. (10.93,3.29)   ;
				\draw    (348.92,147.03) -- (348.92,201.03) ;
				\draw [shift={(348.92,203.03)}, rotate = 270] [color={rgb, 255:red, 0; green, 0; blue, 0 }  ][line width=0.75]    (10.93,-3.29) .. controls (6.95,-1.4) and (3.31,-0.3) .. (0,0) .. controls (3.31,0.3) and (6.95,1.4) .. (10.93,3.29)   ;
				\draw    (348.92,147.03) -- (295.13,147.03) ;
				\draw [shift={(293.13,147.03)}, rotate = 360] [color={rgb, 255:red, 0; green, 0; blue, 0 }  ][line width=0.75]    (10.93,-3.29) .. controls (6.95,-1.4) and (3.31,-0.3) .. (0,0) .. controls (3.31,0.3) and (6.95,1.4) .. (10.93,3.29)   ;
				\draw    (348.92,249.05) -- (295.13,249.05) ;
				\draw [shift={(293.13,249.05)}, rotate = 360] [color={rgb, 255:red, 0; green, 0; blue, 0 }  ][line width=0.75]    (10.93,-3.29) .. controls (6.95,-1.4) and (3.31,-0.3) .. (0,0) .. controls (3.31,0.3) and (6.95,1.4) .. (10.93,3.29)   ;
				\draw    (435,249.05) -- (381.22,249.05) ;
				\draw [shift={(379.22,249.05)}, rotate = 360] [color={rgb, 255:red, 0; green, 0; blue, 0 }  ][line width=0.75]    (10.93,-3.29) .. controls (6.95,-1.4) and (3.31,-0.3) .. (0,0) .. controls (3.31,0.3) and (6.95,1.4) .. (10.93,3.29)   ;
				
				\draw (182.02,20.67) node [anchor=north west][inner sep=0.75pt]    {$\nu _{2}$};
				\draw (460.57,298.79) node [anchor=north west][inner sep=0.75pt]    {$\nu _{1}$};
				\draw (184.86,54.18) node [anchor=north west][inner sep=0.75pt]    {$1$};
				\draw (428.16,300.64) node [anchor=north west][inner sep=0.75pt]    {$1$};
				\draw (184.47,138.34) node [anchor=north west][inner sep=0.75pt]    {$\nu ^{*}$};
				\draw (181.1,240.36) node [anchor=north west][inner sep=0.75pt]    {$r\nu ^{*}$};
				\draw (344.29,297.95) node [anchor=north west][inner sep=0.75pt]    {$\nu ^{*}$};
				\draw (238.43,297.95) node [anchor=north west][inner sep=0.75pt]    {$r\nu ^{*}$};
				\draw (249.76,227.87) node [anchor=north west][inner sep=0.75pt]    {$\beta /r$};
				\draw (353.12,227.87) node [anchor=north west][inner sep=0.75pt]    {$\beta $};
				\draw (251.1,127.54) node [anchor=north west][inner sep=0.75pt]    {$\beta $};
				\draw (351.88,127.54) node [anchor=north west][inner sep=0.75pt]    {$\beta r$};
				\draw (435.57,226.87) node [anchor=north west][inner sep=0.75pt]    {$\beta \nu ^{*}$};
				\draw (250.08,41.38) node [anchor=north west][inner sep=0.75pt]    {$\beta \nu ^{*}$};
				
			\end{tikzpicture}
		}
		
		\caption{Probability mass distribution given in Lemma \ref{lem: upperbounds-2agents}, and the binding \eqref{IC} constraints illustrated by arrows that point to the payoff equivalent scenario with the false report.}
		\label{fig: dist-2agent}
	\end{figure}
	
	Figure \ref{fig: dist-2agent} illustrates the probability distribution from Lemma \ref{lem: upperbounds-2agents} over the set of scenarios $[0,1]^2$, which is a
	potential worst-case distribution, whose parameters can be calculated by solving a constrained nonlinear optimization problem. The complex optimization problem actually formulates feasibility conditions of the distribution as constraints and minimizes the expected payoff of the corresponding optimal mechanism. Hence, by definition, the value of $f(\mu)$ upper bounds the optimal worst-case payoff of the designer. The scenarios in Figure \ref{fig: dist-2agent} seem to be chosen so that several \eqref{IC} constraints bind without putting much probability into the scenarios with high value types. As a result, the agents' types are not independently distributed. Moreover, the types and the probability masses are chosen in a way that would leave the mechanism designer indifferent between allocating an agent with two different types in \textit{\eqref{IC}-adjacent scenarios}. Here, by \eqref{IC}-adjacency, we refer to scenarios that appear in the same binding \eqref{IC} constraint. For instance, in Figure \ref{fig: dist-2agent}, scenarios $(r \nu^*, r\nu^*)$ and $(\nu^*, r\nu^*)$ would be \eqref{IC}-adjacent, which is also highlighted by an arrow.
	
	\begin{figure}
		\centering
		\includegraphics[scale=0.45]{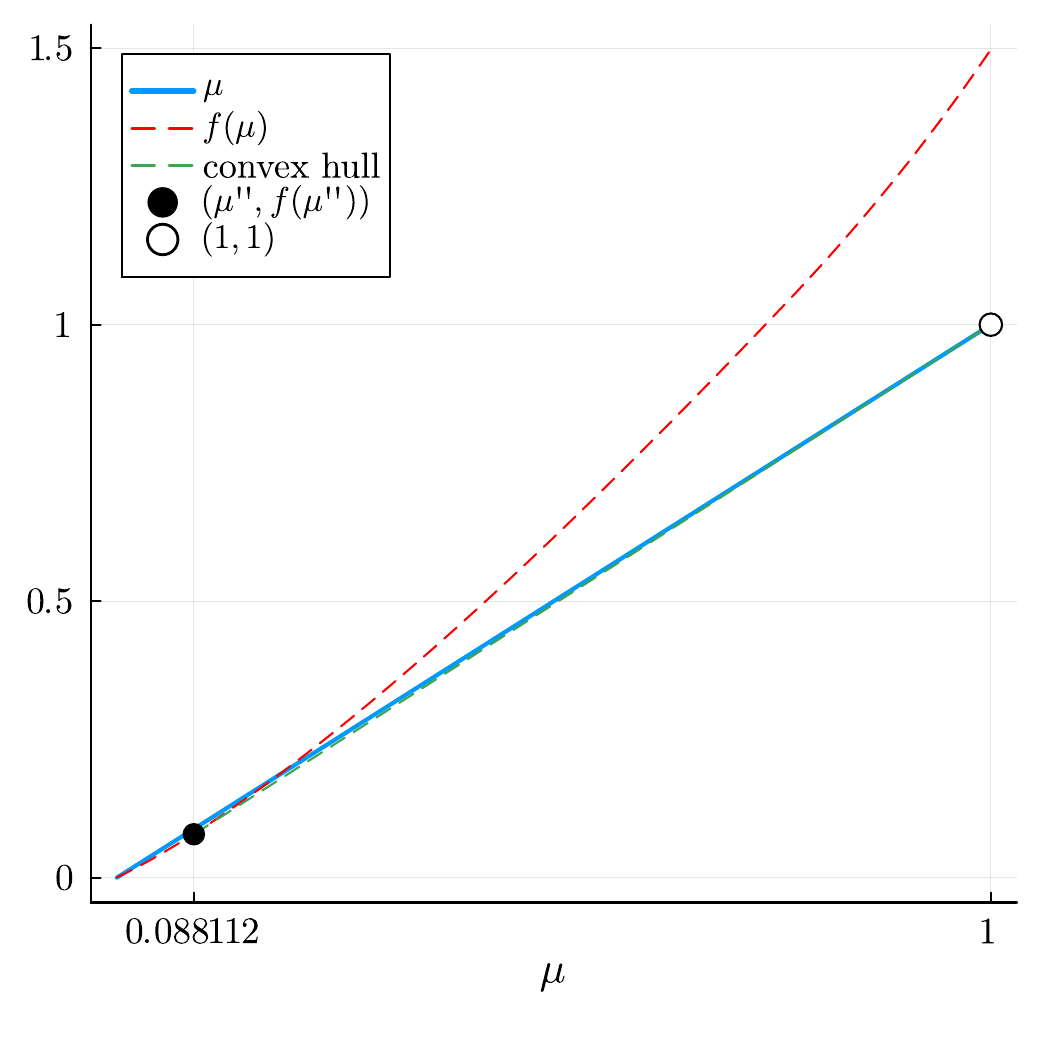}
		\caption{The upper bounds given in Lemma \ref{lem: upperbounds-2agents}, and their convex hull.}
		\label{fig: worstCasePayoff2agents}
	\end{figure}
	
	The distribution in Figure \ref{fig: dist-2agent} do not always coincide the minimizer from the set $\mathcal{P}_{\mu}$ for all $\mu \in (0,1)$ and might be even worst than the simple Dirac distribution at $(\mu,\mu)$, which leads to another upper bound $z^* \leq \mu$. In Figure \ref{fig: worstCasePayoff2agents}, we illustrate the upper bounds from Lemma \ref{lem: upperbounds-2agents} and their convex hull. When $\mu \in [\mu'',1]$, the adversary would prefer using a convex combination of the distribution from Lemma \ref{lem: upperbounds-2agents} evaluated at $\mu''$ and the Dirac distribution at $(1,1)$. Notice that, compared to the one-agent case, the upper bounds are much closer to $\mu$ values. This is because the mechanism designer can use lotteries to subtract more payment from multiple agents. 
	
	In order to prove that the convex hull represents the set of worst-case distributions for $\mu \in [\mu'',1]$, one needs to formulate a feasible mechanism yielding an expected payoff of $f(\mu)$ for all $\mathbb{P} \in \mathcal{P}_{\mu}$, which is not as easy as in the one agent case. Due to Proposition \ref{prop: equivalence of MDPs-multiAgent}, robust optimality requires the \textit{aggregate} payment rule to be always higher than a \textit{hyperplane} (multi-dimensional polynomial of degree one), whose coefficients we can deduce using the upper bounds given in Lemma \ref{lem: upperbounds-2agents}. However, the simple aggregate payment rule does not provide guidance on how to choose the individual payment and allocation rules. Numerical experiments reveal that even symmetric mechanisms with monotone individual rules are not straightforward to interpret. Hence, robustness does not necessarily lead to easy-to-use mechanisms in multi-agent settings. The good news is that a robustly optimal mechanism for the case of $\mu_j = \mu''$ for all $j \in [J]$ is also optimal when $\mu \in (\mu'',1)$. Figure \ref{fig: 2agentsMechanism} illustrates such a mechanism.
	
	\begin{figure}
		\centering
		\begin{subfigure}[t]{0.49\textwidth}
			\centering
			\includegraphics[width=\textwidth, trim=0 45 0 18, clip]{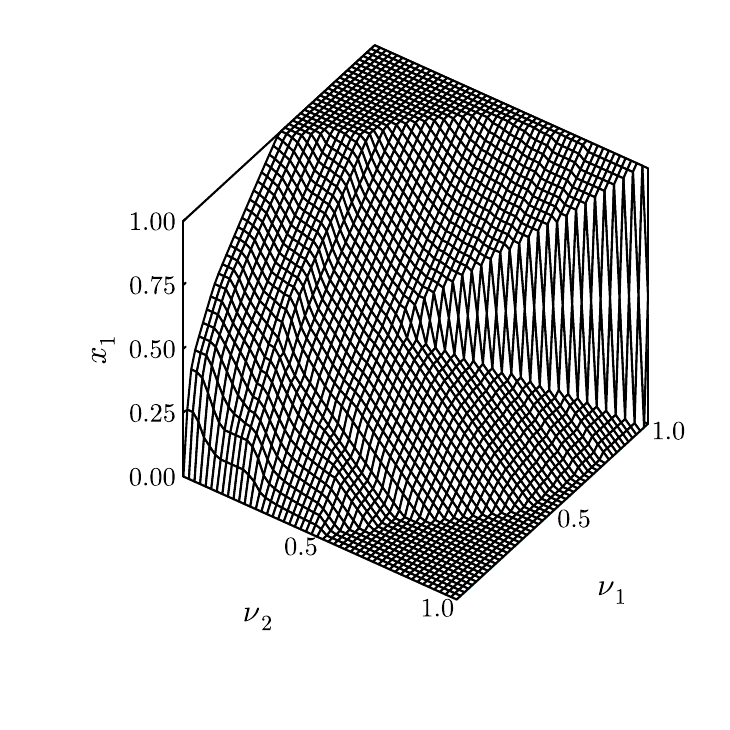}
			\caption{The allocation rule for agent 1.}
			\label{fig: 2agentsAllocationRule}
		\end{subfigure}
		\hfill
		\begin{subfigure}[t]{0.49\textwidth}
			\centering			
			\includegraphics[width=\textwidth, trim=0 45 0 18, clip]{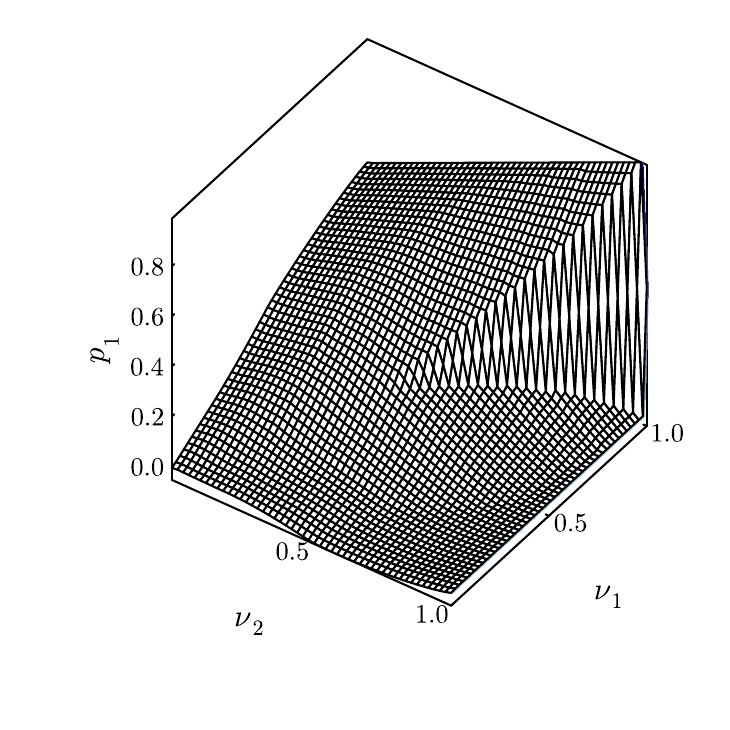}
			\caption{The payment rule for agent 1.}
			\label{fig: 2agentsPaymentRule}
		\end{subfigure}
		\caption{A symmetric mechanism that is robustly optimal for $\mu \in [\mu'',1]$ and has a hyperplane as an aggregate payment.}
		\label{fig: 2agentsMechanism}
	\end{figure}
	
	In order to numerically solve for the mechanism in Figure \ref{fig: 2agentsMechanism}, we consider a discrete type set and use linear programming. The type set is $50$ equidistant points from $[0,1]$, which is denoted by $\mathcal{T}^2$. The model contains two payment rules, $p$ whose aggregate equals a hyperplane, and $p_m$ whose aggregate is always bigger than the aggregate of $p$. The objective is to maximize the sum of all $p_m$ across all agents and types. We illustrate $p_m$ in Section \ref{sec: maximalPaymentSection}. The constraints of the model ensure that $p$ and $p_m$ are \eqref{DSIC}, \eqref{EPIR}, and the sum of $x$ over agents is upper bounded by $1$ for all scenarios $\boldsymbol{\nu} \in [0,1]^2$. Moreover, we constrain that aggregate $p$ equals $\sum_{j \in [J]} \lambda_1^* \nu_j + \lambda_0^*$, where:
	\begin{equation*}
		2 \lambda_1^* \mu'' + \lambda_0^* = f(\mu''), \quad \& \quad 2 \lambda_1^* + \lambda_0^* = 1.
	\end{equation*}
	Notice that we solve for a symmetric solution by letting $\lambda_{1j}^* = \lambda_1^*$ for all $j \in \{1,2\}$.
	The first equation of $\lambda_1^*$ and $\lambda_0^*$ above is chosen specifically so that the resulting mechanism would be robustly optimal, $i.e.$, yield an expected payoff matching the upper bound given in Lemma \ref{lem: upperbounds-2agents} for all $\mu \in [\mu'',1]$. The second one is due to our observations from the numerical results, $i.e.$, it should extract a payment of $1$ when both agents are of type $1$. Lastly, we include symmetry and monotonicity constraints. The symmetry constraints are written for $x$, $p$, and $p_m$ separately, and they enforce the following relation: $x_1(\nu_1,\nu_2) = x_2(\nu_2,\nu_1)$ for all $(\nu_1,\nu_2) \in \mathcal{T}^2$. The monotonicity constraints are only written for $x$ and $p$. We enforce $x_1$ to be weakly increasing in $\nu_1$ and weakly decreasing in $\nu_2$, and $p_1$ to be weakly increasing in $\nu_1$.
	As we have already noted, there are infinitely many robustly optimal mechanisms; therefore, we include the symmetry and monotonicity constraints to arrive at a mechanism that is easy to interpret. Note that enforcing $p_1$ to be also weakly decreasing in $\nu_2$ leads to infeasibility.
	
	In Figure \ref{fig: 2agentsAllocationRule}, we see that $x_1$ is no longer concave as in the one-agent case. Moreover, for fixed $\nu_1$ small enough, one may need high-order polynomials of $\nu_2$ to write it in closed form. Also, for high values of $\nu_1$, we expect $x_1$ to be discontinuous at $\nu_1 = \nu_2$. Analyzing Figure \ref{fig: 2agentsPaymentRule}, we see the same discontinuity issue for $p_1$ together with the negative payments made to low types as in the single-agent case. Other than that, $p_1$ seems like a piecewise function with nonlinear parts, whose thresholds depend on $\nu_2$ in a complex manner. 
	
	To summarize, the simplicity of robustly optimal mechanisms from the one-agent case does not carry over to the setting with 2 agents. The problem of finding the worst-case distribution becomes more complex, and it leads to correlated types. When there are multiple, risk-neutral agents, the designer can use lotteries to improve her worst-case revenue.
}

\subsection{Three or More Agents \& First Moment}

{
	In this section, we consider the case of $J \geq 3$ symmetrical agents with the ambiguity set $\mathcal{P}_{\boldsymbol{\mu}}$, where $\mu_j = \mu$ for all $j \in [J]$. Under these conditions, the adversary cannot do better than choosing the Dirac distribution at $\boldsymbol{\mu}$, and
	the designer can choose arbitrary three agents to allocate via lotteries, upon which the optimal aggregate payment yields the highest achievable worst-case expected payoff, $\mu$. As the number of agents does not affect the maximum worst-case expected payoff, and the three-agent case is easier to illustrate, we discuss the case of $J=3$ in the rest of this section.
	
	As in the two-agent case, 
	we numerically recover a robustly optimal symmetric mechanism. We consider the same linear program from Section \ref{sec:2agents}, but for 3 agents, and reduce the cardinality of the discrete type to $40$, which is denoted by $\mathcal{T}^3$. The coefficients of the multi-dimensional polynomial, $\sum_{j \in [J]} \lambda_1^* \nu_j + \lambda_0^*$ should solve:
	\begin{equation*}
		3\lambda_1^* \mu + \lambda_0^* = \mu, \quad \& \quad 3\lambda_1^* + \lambda_0^* = 1.
	\end{equation*}
	Analogous to the two-agent case, these two equations ensure robust optimality and are solved by $\lambda_1= 1/3$ and $\lambda_0=0$. The linear program for three agents maintains the analogs of the same symmetry and monotonicity constraints as those in \ref{sec:2agents}. Moreover, we introduce two additional sets of constraints in the hopes of obtaining an easily interpretable solution. The first one enforces that $p_1(\boldsymbol{\nu})=0$ whenever $\nu_1 = 0$. The second one enforces that $p_1(\boldsymbol{\nu})$ is weakly increasing in $\nu_2$ when $\nu_1 = 1$ and $\nu_2 < 1$. Without these two constraints, the results are much more chaotic for the payment rule, whose aggregate is constrained to be a polynomial. 
	
	\begin{figure}[tbp]
		\centering\setlength{\belowcaptionskip}{-10pt}
		\begin{subfigure}[t]{0.48\textwidth}
			\centering
			\includegraphics[width=\textwidth, trim=0 10 0 0, clip]{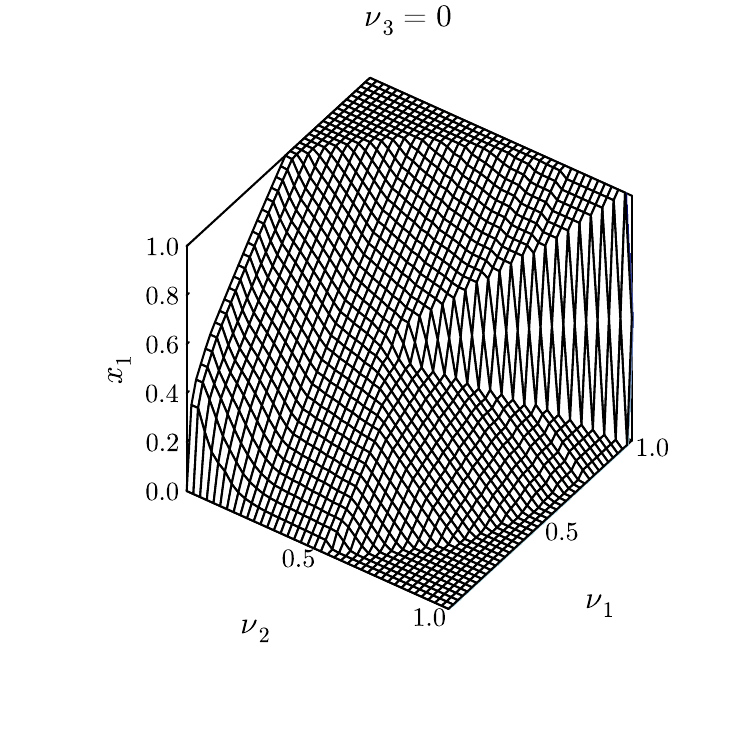}
			\label{fig:3agentAllocation-top_left}
		\end{subfigure}
		\hfill
		\begin{subfigure}[t]{0.48\textwidth}
			\centering
			\includegraphics[width=\textwidth, trim=0 10 0 0, clip]{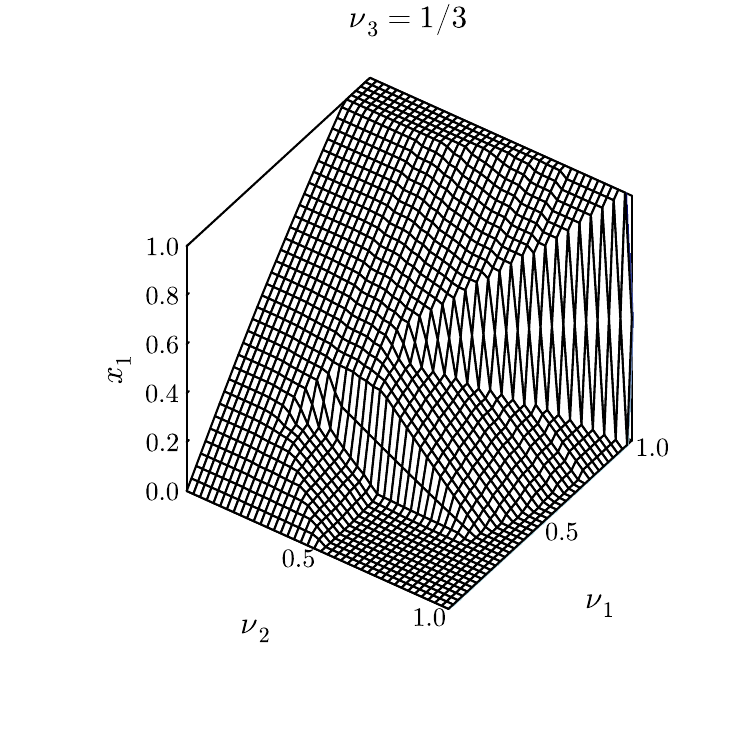}
			\label{fig:3agentAllocation-top_right}
		\end{subfigure}
		
		\vspace{-1.2cm} 
		
		\begin{subfigure}[t]{0.48\textwidth}
			\centering
			\includegraphics[width=\textwidth, trim=0 10 0 0, clip]{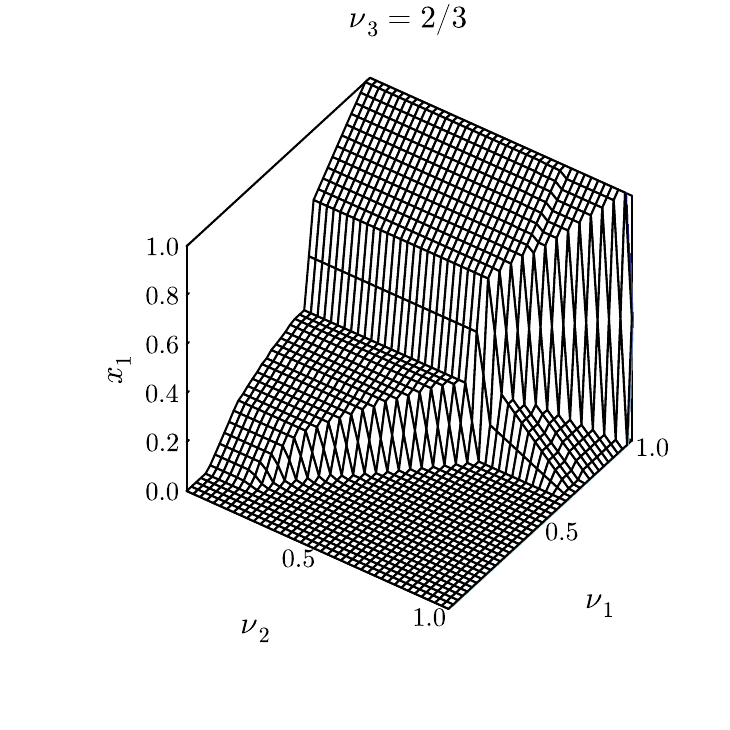}
			\label{fig:3agentAllocation-bottom_left}
		\end{subfigure}
		\hfill
		\begin{subfigure}[t]{0.48\textwidth}
			\centering
			\includegraphics[width=\textwidth, trim=0 10 0 0, clip]{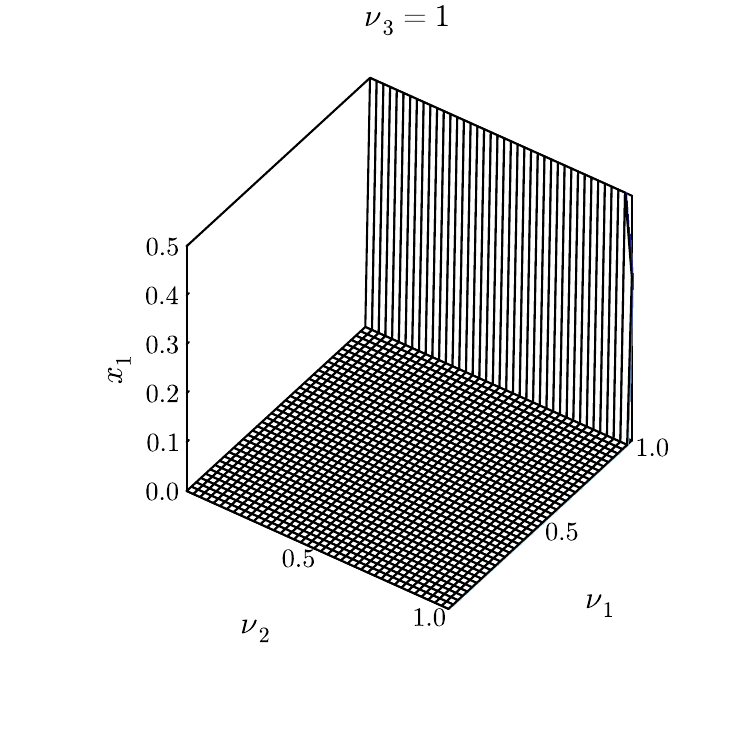}
			\label{fig:3agentAllocation-bottom_right}
		\end{subfigure}
		\vspace{-1cm} 
		\caption{The allocation rule of the robustly optimal symmetric mechanism illustrated for agent 1 and different $\nu_3$ values.}
		\label{fig:3agentAllocation}
	\end{figure}
	
	Figure \ref{fig:3agentAllocation} illustrates the allocation rule of a robustly optimal symmetric mechanism, which seems like a piece-wise linear function for most of the time. The complexity lies in identifying the nonlinear relation between the thresholds of the underlying piecewise function and the types of the other agents, $(\nu_2,\nu_3)$. We also see the discontinuity issue at play when $\nu_1 = \nu_2$ or $\nu_1 = \nu_3$.

	\begin{figure}[tbp]
		\centering
		\begin{subfigure}[t]{0.48\textwidth}
			\centering
			\includegraphics[width=\textwidth, trim=0 10 0 0, clip]{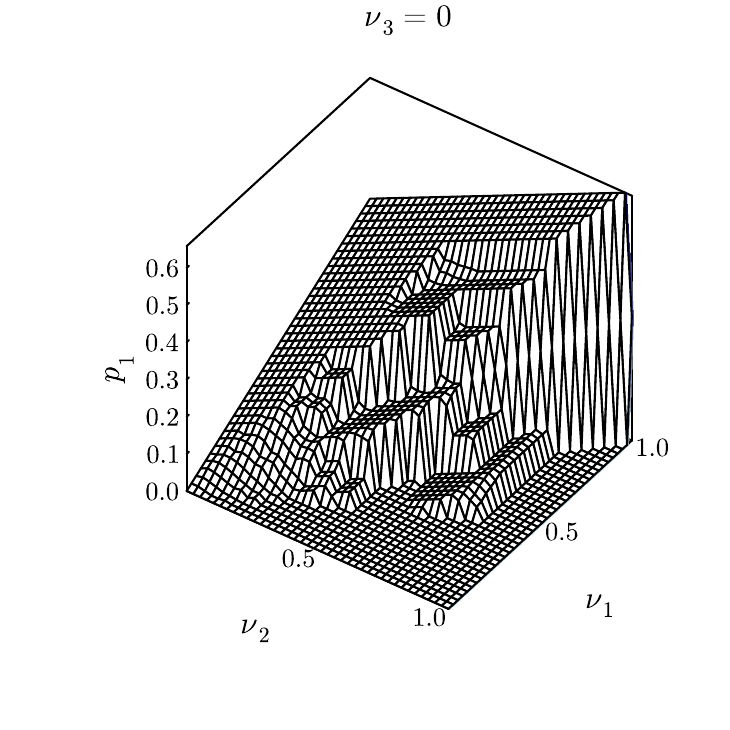}
			\label{fig:3agentPayment-top_left}
		\end{subfigure}
		\hfill
		\begin{subfigure}[t]{0.48\textwidth}
			\centering
			\includegraphics[width=\textwidth, trim=0 10 0 0, clip]{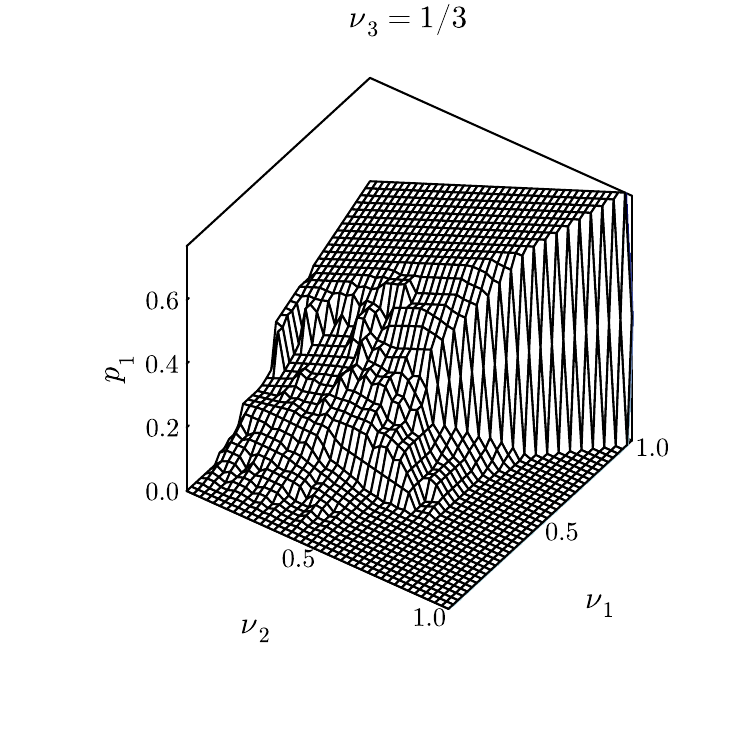}
			\label{fig:3agentPayment-top_right}
		\end{subfigure}
		
		\vspace{-1.3cm} 
		
		\begin{subfigure}[t]{0.48\textwidth}
			\centering
			\includegraphics[width=\textwidth, trim=0 10 0 0, clip]{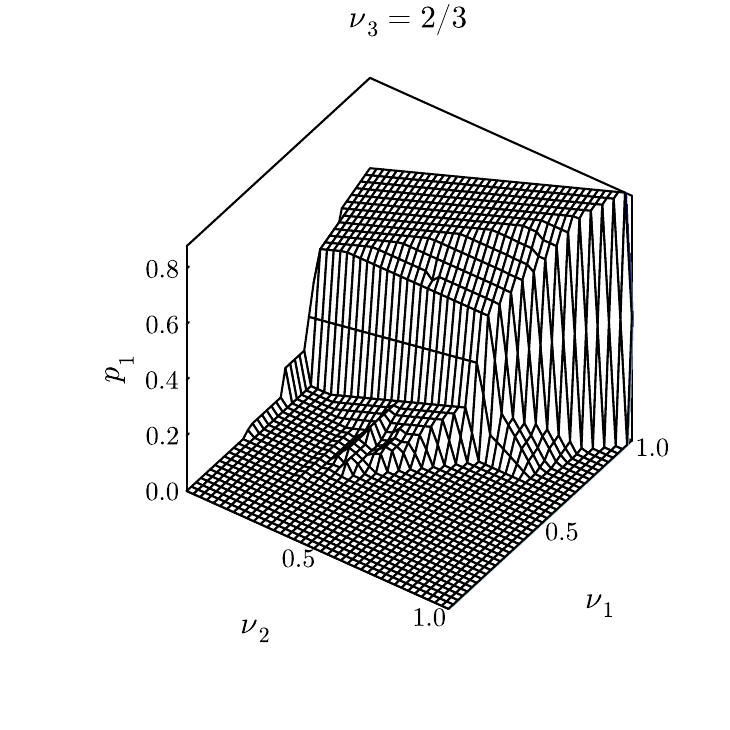}
			\label{fig:3agentPayment-bottom_left}
		\end{subfigure}
		\hfill
		\begin{subfigure}[t]{0.48\textwidth}
			\centering
			\includegraphics[width=\textwidth, trim=0 10 0 0, clip]{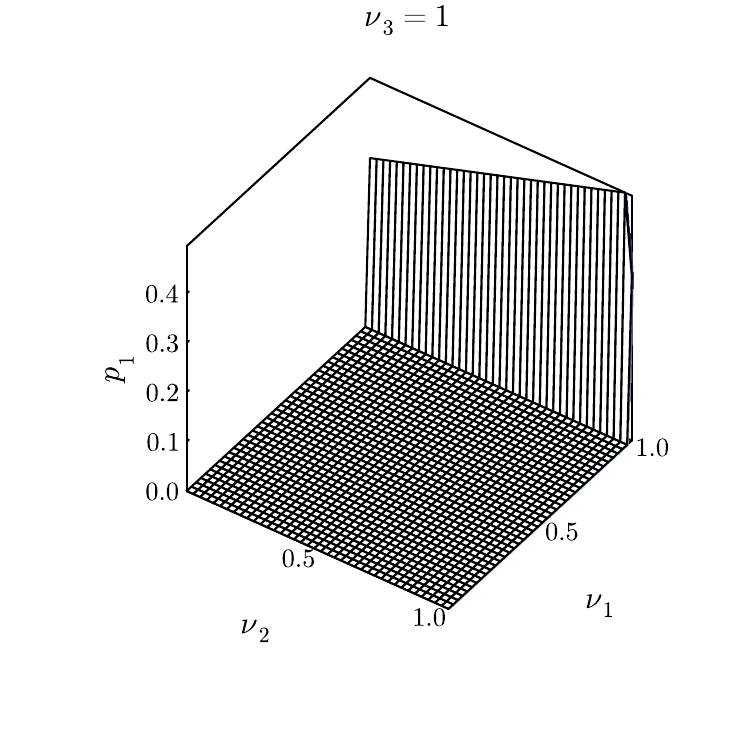}
			\label{fig:3agentPayment-bottom_right}
		\end{subfigure}
		\vspace{-1cm} 
		\caption{The payment rule (polynomial) of the robustly optimal symmetric mechanism illustrated for agent 1 and different $\nu_3$ values.}
		\label{fig:3agentPayment}
	\end{figure}
	
	In Figure \ref{fig:3agentPayment}, we see a much more complicated payment rule compared to the two-agent case. This payment rule is restricted to be symmetric, sum up to a polynomial, weakly increasing in $\nu_1$, weakly increasing in $\nu_2$ for $\nu_1 =1$ and $\nu_2 < 1$, and equal to zero whenever $\nu_1 = 0$. We observe that the payment rule, which simplifies the one-agent case, becomes increasingly complex as the number of agents increases. This is precisely the reason why our approach for characterizing the robustly optimal mechanism in the one-agent case cannot be extended to the multi-agent case. Recall that, in the one-agent case, the aggregate payment was only charged to a single agent, from which we recovered the robustly optimal allocation. In the multi-agent case, we must distribute the aggregate payment among more than one agent in a manner that satisfies the \eqref{DSIC} constraints for all agents, type profiles, and deviations. Such a procedure does not lead to a simple, robustly optimal mechanism, but rather complicates the analysis. We illustrate and comment on the maximal payment rule in Section \ref{sec: maximalPaymentSection}.
}

{From an optimization standpoint, the multi-agent problem differs qualitatively from the single-agent case. Robust optimality requires dominant-strategy feasibility for every realization, leading to a high-dimensional constraint system in which aggregate payment bounds are tractable but individual allocations are not. This creates an inherent gap between dual characterizations and implementable mechanisms.}

\section{Concluding Remarks} \label{sec: conclusion}

We study robust contract design with deferred inspection, motivated by applications in which a principal can observe realized outcomes ex post but lacks reliable distributional information ex ante. Building on the inspection framework of \citet{alaei2024optimal}, we relax the common prior assumption and adopt a moment-based distributionally robust approach.

Our main finding is that deferred inspection fundamentally changes the structure of robust contract design. In the single-agent setting with only first-moment information, we obtain a complete and tractable characterization of robustly optimal mechanisms. Robust optimality can be achieved by a simple contract with a concave allocation and a linear payment rule, yet robustness does not uniquely determine transfers. This flexibility allows the designer to choose among infinitely many contracts, including a Pareto robustly optimal one that attains the same worst-case revenue guarantee as all robustly optimal mechanisms while weakly dominating them in expected revenue for every admissible distribution and strictly improving performance away from the worst case. Numerical experiments show that these mechanisms perform favorably under distributional misspecification.

We also derive structural insights for multi-agent extensions. While moment-based ambiguity continues to imply a polynomial structure for aggregate payments, robustly optimal multi-agent mechanisms are substantially more complex. They must satisfy dominant-strategy incentive compatibility under ambiguity and typically rely on correlated worst-case distributions and randomized allocations. As a result, the simple and closed-form characterization obtained in the single-agent case does not extend to multi-agent settings; instead, numerical analysis is required to study their properties.

Finally, although our model assumes perfect verification, we conjecture that the robust optimality of linear payment rules extends to environments with noisy inspection, where the optimal contract would scale rewards to reflect the reduced quality of the verification signal.

Overall, our results highlight a sharp contrast between robust contract design with a single agent and robust multi-agent mechanism design under inspection. They suggest that deferred inspection enables simple and effective robust contracts in bilateral settings, while robustness in multi-agent environments comes at the cost of significant analytical and operational complexity. {Our analysis focuses on environments with costless inspection and a single inspection stage. Extending the framework to dynamic or costly inspection remains an open challenge.}

\bibliography{bib.bib}

@article{alaei2024optimal,
  title={Optimal auction design with deferred inspection and reward},
  author={Alaei, Saeed and Belloni, Alexandre and Makhdoumi, Ali and Malekian, Azarakhsh},
  journal={Operations Research},
  volume={72},
  number={6},
  pages={2413--2429},
  year={2024},
  publisher={INFORMS}
}

@article{iancu2014pareto,
  title={Pareto efficiency in robust optimization},
  author={Iancu, Dan A and Trichakis, Nikolaos},
  journal={Management Science},
  volume={60},
  number={1},
  pages={130--147},
  year={2014},
  publisher={INFORMS}
}

@misc{AmazonFulfillment,
    author = "Juozas Kaziukenas",
    title = "Fulfillment by Amazon Chokehold",
    url  = "www.marketplacepulse.com/articles/fulfillment-by-amazon-chokehold",
    note = "February 8, 2024",
    year={2022}
}

@article{myerson1983mechanism,
  title={Mechanism design by an informed principal},
  author={Myerson, Roger B},
  journal={Econometrica: Journal of the Econometric Society},
  pages={1767--1797},
  year={1983},
  publisher={JSTOR}
}

@article{hurwicz1973design,
  title={The design of mechanisms for resource allocation},
  author={Hurwicz, Leonid},
  journal={The American Economic Review},
  volume={63},
  number={2},
  pages={1--30},
  year={1973},
  publisher={JSTOR}
}

@article{ben2014optimal,
  title={Optimal allocation with costly verification},
  author={Ben-Porath, Elchanan and Dekel, Eddie and Lipman, Barton L},
  journal={American Economic Review},
  volume={104},
  number={12},
  pages={3779--3813},
  year={2014},
  publisher={American Economic Association 2014 Broadway, Suite 305, Nashville, TN 37203}
}

@inbook{Wilson_1987,
    place={Cambridge}, 
    series={Econometric Society Monographs}, 
    title={Game-theoretic analyses of trading processes}, 
    booktitle={Advances in Economic Theory: Fifth World Congress}, 
    publisher={Cambridge University Press}, 
    author={Wilson, Robert},   
    year={1987}, 
    pages={33–70}, 
    collection={Econometric Society Monographs}
}

@article{bergemann2005robust,
  title={Robust mechanism design},
  author={Bergemann, Dirk and Morris, Stephen},
  journal={Econometrica},
  pages={1771--1813},
  year={2005},
  publisher={JSTOR}
}

@article{hansen1985auctions,
  title={Auctions with contingent payments},
  author={Hansen, Robert G},
  journal={The American Economic Review},
  volume={75},
  number={4},
  pages={862--865},
  year={1985},
  publisher={JSTOR}
}

@article{skrzypacz2013auctions,
  title={Auctions with contingent payments—an overview},
  author={Skrzypacz, Andrzej},
  journal={International Journal of Industrial Organization},
  volume={31},
  number={5},
  pages={666--675},
  year={2013},
  publisher={Elsevier}
}

@article{townsend1979optimal,
  title={Optimal contracts and competitive markets with costly state verification},
  author={Townsend, Robert M},
  journal={Journal of Economic theory},
  volume={21},
  number={2},
  pages={265--293},
  year={1979},
  publisher={Elsevier}
}

@article{mylovanov2017optimal,
  title={Optimal allocation with ex post verification and limited penalties},
  author={Mylovanov, Tymofiy and Zapechelnyuk, Andriy},
  journal={American Economic Review},
  volume={107},
  number={9},
  pages={2666--2694},
  year={2017},
  publisher={American Economic Association 2014 Broadway, Suite 305, Nashville, TN 37203}
}

@article{li2020mechanism,
  title={Mechanism design with costly verification and limited punishments},
  author={Li, Yunan},
  journal={Journal of Economic Theory},
  volume={186},
  pages={105000},
  year={2020},
  publisher={Elsevier}
}

@article{carroll2019robustness,
  title={Robustness in mechanism design and contracting},
  author={Carroll, Gabriel},
  journal={Annual Review of Economics},
  volume={11},
  pages={139--166},
  year={2019},
  publisher={Annual Reviews}
}

@article{bandi2014optimal,
  title={Optimal design for multi-item auctions: A robust optimization approach},
  author={Bandi, Chaithanya and Bertsimas, Dimitris},
  journal={Mathematics of Operations Research},
  volume={39},
  number={4},
  pages={1012--1038},
  year={2014},
  publisher={INFORMS}
}

@article{carrasco2018optimal,
  title={Optimal selling mechanisms under moment conditions},
  author={Carrasco, Vinicius and Luz, Vitor Farinha and Kos, Nenad and Messner, Matthias and Monteiro, Paulo and Moreira, Humberto},
  journal={Journal of Economic Theory},
  volume={177},
  pages={245--279},
  year={2018},
  publisher={Elsevier}
}

@book{burnside1892theory,
  title={The theory of equations: with an introduction to the theory of binary algebraic forms},
  author={Burnside, William Snow and Panton, Arthur William},
  year={1892},
  publisher={Hodges, Figgis}
}

@article{nickalls1993new,
  title={A new approach to solving the cubic: Cardan’s solution revealed},
  author={Nickalls, Richard WD},
  journal={The Mathematical Gazette},
  volume={77},
  number={480},
  pages={354--359},
  year={1993},
  publisher={Cambridge University Press}
}

@article{myerson1981optimal,
  title={Optimal auction design},
  author={Myerson, Roger B},
  journal={Mathematics of operations research},
  volume={6},
  number={1},
  pages={58--73},
  year={1981},
  publisher={INFORMS}
}

@book{clarke2013functional,
  title={Functional analysis, calculus of variations and optimal control},
  author={Clarke, Francis},
  volume={264},
  year={2013},
  publisher={Springer}
}

@article{bayrak2025distributionally,
  title={Distributionally robust optimal allocation with costly verification},
  author={Bayrak, Halil Ibrahim and Ko{\c{c}}yi{\u{g}}it, {\c{C}}a{\u{g}}{\i}l and Kuhn, Daniel and P{\i}nar, Mustafa Celebi},
  journal={Operations Research},
  year={2025},
  publisher={INFORMS}
}

@inproceedings{belloni2025approximately,
  title={Approximately Optimal Online Mechanism for Multiunit Demand Buyers with Post-Allocation Inspection},
  author={Belloni, Alexandre and Li, Xuanjie and Makhdoumi, Ali},
  booktitle={Proceedings of the 26th ACM Conference on Economics and Computation},
  pages={354--354},
  year={2025}
}

@article{li2025mechanism,
  title={Mechanism Design with Post-Allocation Inspection and Exclusion Penalties},
  author={Li, Xuanjie and Makhdoumi, Ali and Peke{\v{c}}, Sa{\v{s}}a},
  journal={Available at SSRN 5139835},
  year={2025}
}

@article{chen2024screening,
  title={Screening with limited information: A dual perspective},
  author={Chen, Zhi and Hu, Zhenyu and Wang, Ruiqin},
  journal={Operations Research},
  volume={72},
  number={4},
  pages={1487--1504},
  year={2024},
  publisher={INFORMS}
}

@article{wang2024minimax,
  title={Minimax regret robust screening with moment information},
  author={Wang, Shixin and Liu, Shaoxuan and Zhang, Jiawei},
  journal={Manufacturing \& Service Operations Management},
  volume={26},
  number={3},
  pages={992--1012},
  year={2024},
  publisher={INFORMS}
}

@article{bachrach2022distributional,
  title={Distributional robustness: From pricing to auctions},
  author={Bachrach, Nir and Talgam-Cohen, Inbal},
  journal={arXiv preprint arXiv:2205.09008},
  year={2022}
}

\newpage
\appendix
\section{Appendix}

\subsection{Maximal payment rule in the multi-agent setting}\label{sec: maximalPaymentSection}

\begin{figure}
	\centering
	\includegraphics[scale=0.7, trim=0 45 0 18, clip]{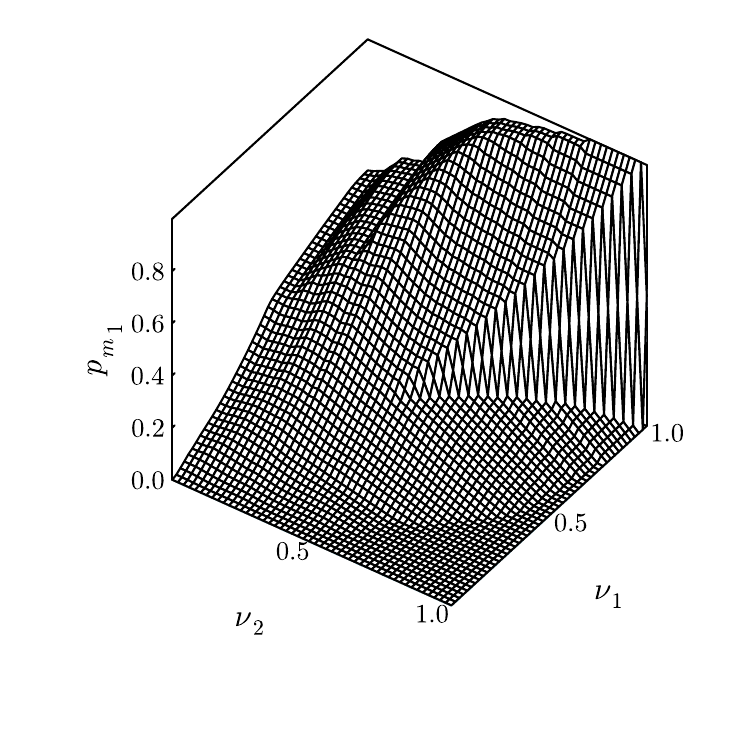}
	\caption{The individual payment rule of a symmetric mechanism that is robustly optimal for $\mu \in [\mu'',1]$ and has maximal aggregate payment.}
	\label{fig: 2agentsMaxPaymentRule}
\end{figure}

In Figure \ref{fig: 2agentsMaxPaymentRule}, we see that the maximal payment rule, ${p_m}_1$, only resolves the negative payment issue compared to $p_1$ in Figure \ref{fig: 2agentsPaymentRule}. Although ${p_m}_1$ seems to be more complex than $p_1$, it can be deduced directly from $x_1$ using part $(i)$ of Proposition \ref{prop: resultsAlaei}, which directly extends to the multi-agent case. Having said that, we are unable to formulate a robustly optimal $x_1$ (symmetric or asymmetric) in closed form.

\begin{figure}[htbp]
	\centering
	\begin{subfigure}[t]{0.48\textwidth}
		\centering
		\includegraphics[width=\textwidth, trim=0 10 0 0, clip]{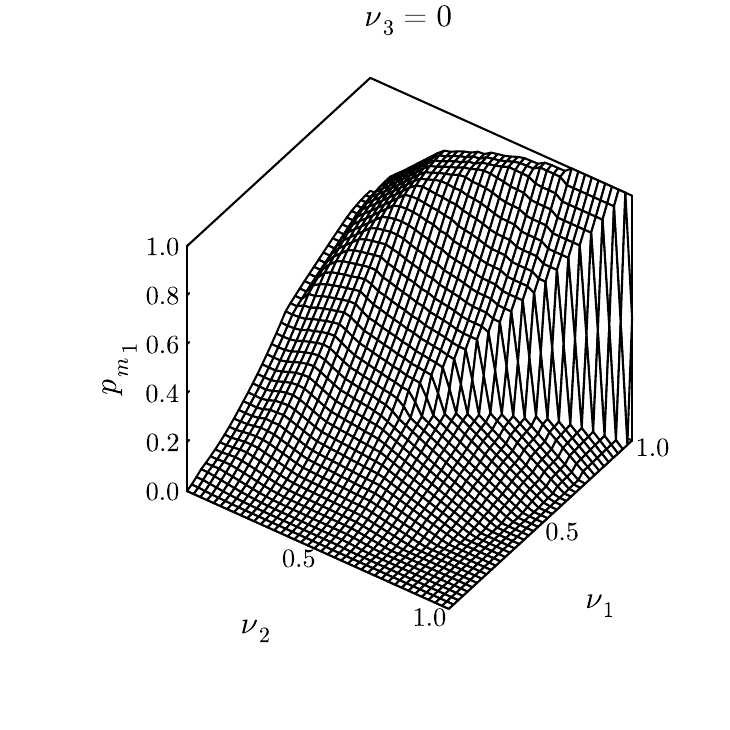}
		\label{fig:3agentPaymentMax-top_left}
	\end{subfigure}
	\hfill
	\begin{subfigure}[t]{0.48\textwidth}
		\centering
		\includegraphics[width=\textwidth, trim=0 10 0 0, clip]{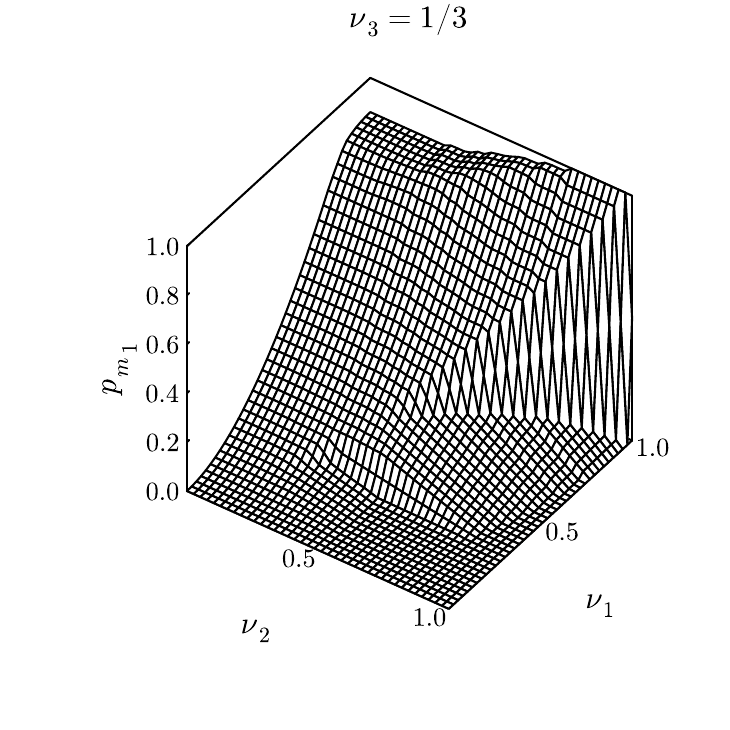}
		\label{fig:3agentPaymentMax-top_right}
	\end{subfigure}
	
	\vspace{-1.3cm} 
	
	\begin{subfigure}[t]{0.48\textwidth}
		\centering
		\includegraphics[width=\textwidth, trim=0 10 0 0, clip]{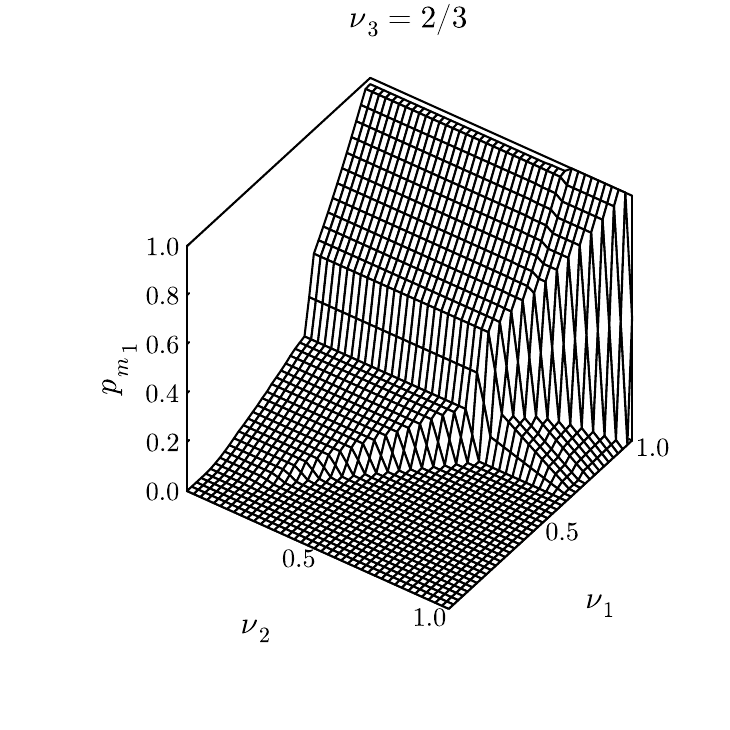}
		\label{fig:3agentPaymentMax-bottom_left}
	\end{subfigure}
	\hfill
	\begin{subfigure}[t]{0.48\textwidth}
		\centering
		\includegraphics[width=\textwidth, trim=0 10 0 0, clip]{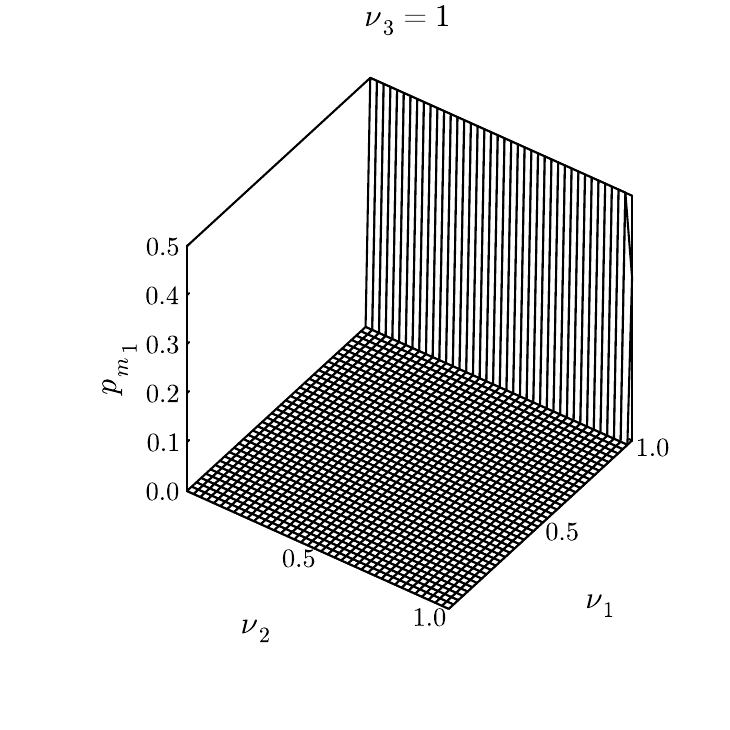}
		\label{fig:3agentPaymentMax-bottom_right}
	\end{subfigure}
	\vspace{-1cm} 
	\caption{The payment rule (maximal) of the robustly optimal symmetric mechanism illustrated for agent 1 and different $\nu_3$ values.}
	\label{fig:3agentPaymentMax}
\end{figure}

In Figure \ref{fig:3agentPaymentMax}, we see the maximal payment rule, which looks much nicer compared to the polynomial one. This is because $p_m$ can be directly recovered from $x$ given in Figure \ref{fig:3agentAllocation} through \eqref{DSIC} constraints. 
One may think that it can be easier to solve directly for a robustly optimal $x$ and then recover $p_m$. To do this, one must make sure that $x$ is feasible and leads to individual payments whose aggregate is lower bounded by a robustly optimal polynomial. Hence, the characterization of a multi-agent robustly optimal mechanism is tied to the relation between the individual payments and the polynomial that lower bounds their aggregate.

\subsection{Optimal \& approximation mechanisms for small $\mu$}\label{sec: smallMu}

In this section, we comment on the optimal worst-case payoff when there is one agent, the ambiguity set contains only the first moment, $\mu$, and the first moment is small, $\mu \in (0, \mu')$. Similar to the two-point distribution presented in Lemma \ref{lem:upperBounds}, the adversary can also use a three-point distribution. We illustrate the distributions in Lemma \ref{lem:upperBounds} and Lemma \ref{lem: SmallMuUpperBounds} in Figure \ref{fig: upperBounds}, which clearly shows that the three-point distribution leads to a smaller upper bound compared to that of the two-point mechanism when $\mu$ is small.

\begin{figure}
	\centering
	\includegraphics[scale=0.8]{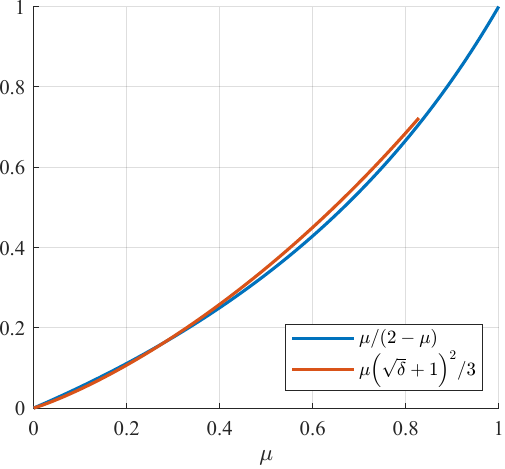}
	\caption{Upper bounds given in Lemma \ref{lem:upperBounds} and Lemma \ref{lem: SmallMuUpperBounds}.}
	\label{fig: upperBounds}
\end{figure}

\begin{lemma}\label{lem: SmallMuUpperBounds}
	For any $\mu \in (0,(15-3\sqrt{5})/10]$, the three-point distribution that assigns probability $\mu/3\underline{\nu}$ to type $\underline{\nu}= \delta(\sqrt{\delta}+1)$, probability $\mu/3\overline{\nu}$ to type $\overline{\nu}=\sqrt{\delta}(\sqrt{\delta}+1)$, and probability $\mu/3$ to type $1$ is contained in the ambiguity set $\mathcal{P}_{\mu}$, under which the seller's optimal revenue is $\frac{\mu}{3}(\sqrt{\delta}+1)^2$, where $\delta= \frac{ \mu}{3- \mu}$. Hence, $z^* \leq \frac{\mu}{3}(\sqrt{\delta}+1)^2$.
\end{lemma}
\noindent{\bf Proof}. The given distribution belongs to our ambiguity set if the following conditions hold:
\begin{equation*}
	\begin{aligned}
		(i) &\; \frac{ \mu}{3},\; \frac{ \mu}{3 \overline{\nu}},\; \frac{ \mu}{3 \underline{\nu}} \in [0,1],\\
		(ii) &\; \underline \nu,\; \overline{\nu} \in [0,1],\\
		(iii) &\; \frac{ \mu}{3} +\frac{ \mu}{3 \overline{\nu}}  + \frac{ \mu}{3 \underline{\nu}} = 1, \;\text{ and } \; \frac{ \mu}{3}(1) + \frac{ \mu}{3\overline{\nu}}  (\overline{\nu})+  \frac{ \mu}{3 \underline{\nu}}(\underline \nu) = \mu.
	\end{aligned}
\end{equation*}
It is easy to see that the types and the probabilities are non-negative as $\delta \geq 0$. These probabilities are also less than $1$ as they sum up to $1$:
\begin{equation*}
	\begin{aligned}
		\frac{ \mu}{3} + \frac{ \mu}{3 \overline{\nu}}  + \frac{ \mu}{3 \underline{\nu}} &= \frac{ \mu}{3}\bigg( 1 + \frac{1}{\sqrt{\delta}(\sqrt{\delta}+1)} + \frac{1}{{\delta}(\sqrt{\delta}+1)}\bigg),\\
		&= \frac{ \mu}{3}\bigg( 1 + \frac{\sqrt{\delta}+1}{{\delta}(\sqrt{\delta}+1)} \bigg),\\
		&= \frac{ \mu}{3}\bigg( 1 + \frac{3-
			\mu}{{ \mu}}\bigg) = 1.
	\end{aligned}
\end{equation*}
As the expectation of this distribution equals $ \mu$ by construction, the only remaining condition to check is $\underline \nu, \overline{\nu} \leq 1$. This is easy to see for type $\underline \nu = {\delta}(\sqrt{\delta}+1) = \mu (\sqrt{\delta}+1)/ (3- \mu) \leq \mu (2) / (3- \mu) < \mu$ for any $\mu \in (0,1)$. For type $\overline{\nu}$, the following inequality needs to hold:
\begin{equation*}
	\begin{aligned}
		\sqrt{\delta}(\sqrt{\delta}+1) &\leq 1,\\
		\sqrt{\delta} &\leq 1-\delta,\\
		\sqrt{\frac{ \mu}{3- \mu}} &\leq \frac{3-2 \mu}{3- \mu},\\ 
		\mu(3- \mu) &\leq (3-2 \mu)^2,\\
		0 &\leq 5  \mu^2 -15 \mu + 9.
	\end{aligned}
\end{equation*}
This condition holds only if $ \mu \leq (15-3\sqrt{5})/10$. Hence, for any $ \mu \in (0, (15-3\sqrt{5})/10]$, the conditions $(i)$, $(ii)$ and $(iii)$ are satisfied, and we have an admissible distribution.

Now, we solve the principal's problem under this three-point distribution. Note that limiting the adversary to pick a specific distribution from the ambiguity set would result in a higher expected revenue for the seller. Hence, the optimal mechanism under this distribution would lead to an upper bound for $z^*$.
Similar to the proof of Lemma \ref{lem:upperBounds}, we omit all irrelevant types, $\nu \in [0,1) \setminus \{\underline \nu, \overline{\nu}\}$, and some of the constraints to solve for the upper bound:
\begin{equation*}
	\begin{aligned}
		\max_{x,p} \;& \frac{ \mu}{3} p(1)+ \frac{ \mu}{3 \overline \nu}p(\overline \nu)+ \frac{ \mu}{3 \underline \nu}p(\underline \nu)  \nonumber\\
		\text{s.t.} \;&x(1)- p(1) \geq x(\overline{\nu}) - \overline{\nu},\\
		&\overline \nu x(\overline \nu)- p(\overline \nu) \geq \overline{\nu}x(\underline{\nu}) - \underline{\nu},\\
		&\underline{\nu} x(\underline{\nu})- p(\underline{\nu}) \geq 0, \\
		&x(\nu) \in [0,1] &\forall \nu \in \{1, \overline{\nu}, \underline \nu\}.
	\end{aligned}
\end{equation*}
In the optimal solution, all payment rules should be equal to their lowest upper bound:
\begin{equation*}
	\begin{aligned}
		\max_{x,p} \;& \frac{\mu}{3} \big( x(1)-x(\overline{\nu}) + \overline{\nu} \big) + \frac{ \mu}{3 \overline \nu} \big( \overline \nu (x(\overline \nu)- x(\underline{\nu})) + \underline{\nu}\big)+ \frac{ \mu}{3}x(\underline \nu)  \nonumber\\
		\text{s.t.} \;&x(\nu) \in [0,1] & \forall \nu \in \{1, \overline{\nu}, \underline \nu\}.
	\end{aligned}
\end{equation*}
Summing up the coefficients of the variables reduces the objective to the following term:
\begin{equation*}
	\frac{\mu}{3} \big( x(1) + \overline{\nu} + \frac{\underline{\nu}}{\overline{\nu}} \big).
\end{equation*}
To maximize this value, one needs to set $x(1) = 1$, which give the following upper bound:
\begin{equation*}
	\begin{aligned}
		\frac{\mu}{3} \big( 1 + \overline{\nu} + \frac{\underline{\nu}}{\overline{\nu}} \big) &= \frac{\mu}{3}(1 + \sqrt{\delta}(\sqrt{\delta}+1) + \sqrt{\delta})\\
		& = \frac{\mu}{3} (\sqrt{\delta} + 1)^2,
	\end{aligned}
\end{equation*}
completing our proof. \hfill \Halmos	

Next, we propose a mechanism tailored for the three-point worst-case distribution given in Lemma \ref{lem: SmallMuUpperBounds}. It is optimal when $\underline \mu \in [0.107, 0.25]$ and approximates the optimal solution when $ \mu \in (0,0.107)$. In the latter case, when $ \mu \in (0,0.107)$, the adversary can use a four-point discrete distribution that gives an upper bound for the worst-case expected payoff lower than those provided in Lemma \ref{lem:upperBounds} and Lemma \ref{lem: SmallMuUpperBounds}. Finding the optimal mechanism in this case necessitates repeating every step in this paper for worst-case distributions with four points, which would lengthen the paper without adding more insight. Hence, we do not pursue optimality for this case of extremely small $\mu$ and propose an approximation mechanism that performs close to optimality.

\begin{theorem}\label{thrm: threePointMechanism}
	If $ \mu \in [0.107,0.25]$, then the following mechanism is robustly optimal with a worst-case expected payoff of $\frac{ \mu}{3}(\sqrt{\delta}+1)^2$,
	\begin{equation*}\label{eq:linearPaymentMechanismSmallmu}
		\begin{aligned}
			&p^s(\nu) = \lambda_1 \nu+ \lambda_0 & \forall \nu \in [0,1],\\
			&x^s(\nu) = \begin{cases}
				-\frac{\lambda_0}{\underline \nu^2}\nu &\text{if } \nu \in [0, \underline \nu],\\
				\alpha+ \frac{\lambda_0}{\nu} &\text{if } \nu \in [\underline \nu, \nu^\circ],\\
				-\frac{\lambda_0}{ {\nu^\circ}^2}\nu + \lambda_1 + 2\frac{\lambda_0}{\nu^\circ} &\text{if } \nu \in [\nu^\circ,\nu^\star],\\
				\nu-\overline{\nu} +1 &\text{if } \nu \in 	[\nu^\star, \overline{\nu}],\\
				1 &\text{if } \nu \in [\overline{\nu},1],
			\end{cases}
		\end{aligned}
	\end{equation*}
	where $\delta= \frac{ \mu}{3- \mu}$, $\lambda_0 =-\delta\sqrt{\delta}(\sqrt{\delta}+1)$, $\lambda_1 = (\sqrt{\delta}+1)^2/3-\lambda_0/ \mu$, 
	$\underline \nu = -2\lambda_0/\lambda_1$, 
	$\nu^\circ = (-\lambda_0 - \sqrt{{\lambda_0}^2+\tau \lambda_0 \nu^\star})/\tau$, $\nu^\star = 2-\lambda_1 -2\sqrt{1-\overline{\nu}}$, $\tau = \lambda_1 -\nu^\star +\overline{\nu}-1$ and $\overline{\nu}= \lambda_1 + \lambda_0$. Moreover, $x^s$ is continuous and weakly increasing.
\end{theorem}
\noindent{\bf Proof of Theorem \ref{thrm: threePointMechanism}}. 
To prove that $(x^s,p^s)$ is robustly optimal, we will show that it is feasible and generates an expected payoff of $\frac{ \mu}{3}(\sqrt{\delta}+1)^2$ for $ \mu \in [0.107,0.25]$, attaining the upper bound given in Lemma \ref{lem: SmallMuUpperBounds}. That is, $(x^s,p^s)$ is feasible and satisfies $p^s(\underline \mu) = \lambda_1 \underline \mu +\lambda_0 =\frac{\underline \mu}{3}(\sqrt{\delta}+1)^2$. The latter follows trivially from the definitions of $\lambda_1$ and $\lambda_0$.

To show that $(x^s,p^s)$ is a feasible mechanism, we will prove that it is well-defined, satisfies \eqref{IC}, \eqref{IR} constraints, and its allocation is always between $0$ and $1$. To prove that it is well-defined, we must show that $\nu^\circ$ and $\nu^\star$ are well-defined real numbers. In particular, we need $1-\overline{\nu} \geq 0$ for a real-valued $\nu^\star$, and ${\lambda_0}^2+\tau \lambda_0 \nu^\star \geq 0$ and $\tau \neq 0$ for a well-defined and real valued $\nu^\circ$.
Since these are quite complex functions of $ \mu$, we refer to their plots given in Figures \ref{fig: discriminantForNuCirc} and \ref{fig: typesForThreePointMechanism}, which ensure that both requirements hold for $ \mu \in [0.107,0.25]$. From Figure \ref{fig: typesForThreePointMechanism}, it is also apparent that $0 \leq \underline \nu \leq \nu^\circ \leq \nu^\star \leq \overline{\nu} \leq 1$ holds. Note that $\tau$ is strictly positive for the interval $ \mu \in [0.107,0.25]$ and equals zero when $ \mu \approx 0.1067$. 

\begin{figure}
	\centering
	\begin{subfigure}[b]{0.42\textwidth}
		\centering
		\includegraphics[width=\textwidth]{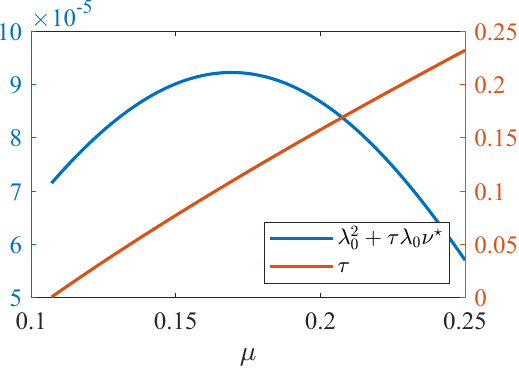}
		\caption{${\lambda_0}^2 +\tau \lambda_0 \nu^\star$ is non-negative, and $\tau$ is strictly positive.}
		\label{fig: discriminantForNuCirc}
	\end{subfigure}
	\hfill
	\begin{subfigure}[b]{0.42\textwidth}
		\centering
		\includegraphics[width=\textwidth]{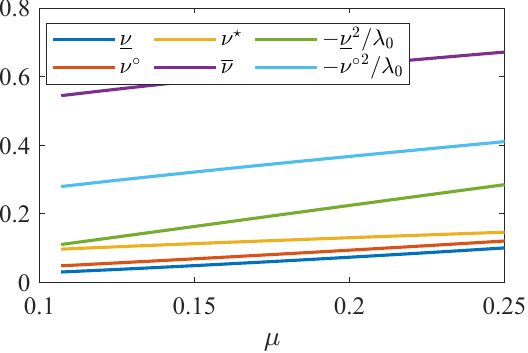}
		\caption{The types satisfy $0 \leq \underline \nu \leq \nu^\circ \leq \nu^\star \leq \overline{\nu} \leq 1$.}
		\label{fig: typesForThreePointMechanism}
	\end{subfigure}
	\caption{Two plots ensuring that $(x^s,p^s)$ is well-defined for $\mu \in [0.017,0.25]$.}
\end{figure}

Next, we prove some properties of $x^s$ that will help us in proving $x^s(\nu) \in [0,1]$ for any $\nu \in [0,1]$. First, we show that $x^s$ is continuous for $\nu \in [0,1]$, for which we only need to show that $x^s$ takes the same values at the corresponding endpoints of the intervals. The proof for $-\frac{\lambda_0}{\underline \nu^2} \underline \nu = \lambda_1 + \frac{\lambda_0}{\underline \nu}$ is presented in the proof of Theorem \ref{thrm: twoPointOptimalMechanism}. The equalities for $\nu^\circ$ and $\overline{\nu}$ follow directly from their definition and the definition of $x^s$. Hence, we only need to prove the continuity at the end point $\nu^\star$. We subtract the two allocation functions for the point $\nu^\star$ and show that it is equal to zero:
\begin{equation*}
	\begin{aligned}
		&-\frac{\lambda_0}{{\nu^\circ}^2} \nu^\star + \lambda_1 +2\frac{\lambda_0}{\nu^\circ} - (\nu^\star -\overline{\nu}+1) \\
		&\qquad = \big[ (\lambda_1 -\nu^\star +\overline{\nu}-1) {\nu^\circ}^2 +2\lambda_0\nu^\circ -\lambda_0 \nu^\star \big]/{\nu^\circ}^2.
	\end{aligned}
\end{equation*}
As $\nu^\circ$ is strictly positive for the interval $\mu \in [0.107,0.25)$ (see Figure \ref{fig: typesForThreePointMechanism}), the above function is well-defined and equals zero if the numerator is zero. Notice that the numerator is a quadratic function of $\nu^\circ$ that can be easily solved for $\nu^\circ$ values that make it equal to zero. Since its discriminant is ${\lambda_0}^2 + \tau \lambda_0 \nu^\star$, which is non-negative as depicted in Figure \ref{fig: discriminantForNuCirc}, it has real roots. One can check that one of the roots is equal to $(-\lambda_0 - \sqrt{{\lambda_0}^2+\tau \lambda_0 \nu^\star})/\tau$, which is how we define $\nu^\circ$ in Theorem \ref{thrm: threePointMechanism}. Hence, $x^s$ is also continuous $\nu^\star$.


Another useful property of $x^s$ is that it is weakly increasing over $\nu$, $i.e.$, its first derivative with respect to $\nu$ is non-negative. The first derivatives are $-\lambda_0/\underline \nu^2$ for interval $[0,\underline \nu]$, $-\lambda_0/\nu^2$ for $[\underline \nu,\nu^\circ]$, $-\lambda_0/ {\nu^\circ}^2$ for $[\nu^\circ,\nu^\star]$, $1$ for $[\nu^\star,\overline{\nu}]$, and $0$ for $[\overline{\nu},1]$, which are all non-negative as $\lambda_0 \leq 0$. Hence, as $x^s$ is weakly increasing and continuous for $\nu \in [0,1]$, and $x^s(0)=0$, and $x^s(1)=1$ by definition, we must have $x^s(\nu) \in [0,1]$ for all $\nu \in [0,1]$. 

Here, we also note two other properties of $x^s$ that will be helpful in proving \eqref{IC} constraints. First, the first derivative $x^s$ is non-increasing in $\nu$, because $-\lambda_0 /\underline \nu^2 \geq - \lambda_0/{\nu^\circ}^2 \geq 1$, where the first inequality follows from $\lambda_0 \leq 0$ and $\underline \nu \leq \nu^\circ \leq 1$, whereas the second inequality follows from $-{\nu^\circ}^2/\lambda_0 \leq 1$ as shown in Figure \ref{fig: typesForThreePointMechanism}. Second, the second derivative of $x^s$ is always non-positive, which makes $x^s$ a concave function. 

To prove that $(x^s,p^s)$ is individually rational, $i.e.$, it satisfies \eqref{IR}, we need to show that $x^s(\nu) \geq \lambda_1  +\lambda_0/\nu$ holds for all $\nu \in [0,1]$. It holds with equality for $\nu \in [\underline \nu, \nu^\circ]$ by definition. Notice that $\lambda_1  +\lambda_0/\nu$ is strictly concave as $\lambda_0 <0$. Then, the following inequalities are satisfied for any $\nu$:
\begin{equation}
	\begin{aligned}
		\lambda_1  +\frac{\lambda_0}{\nu} + \frac{-\lambda_0}{ \nu^2}(\nu'- \nu) &\geq \lambda_1  +\frac{\lambda_0}{\nu'} &\forall \nu' \in[0,1].
	\end{aligned}
\end{equation}
When $\nu = \underline \nu$, the left-hand side of the above inequality is equal to $x^s(\nu')$ for all $\nu' \in [0, \underline \nu]$ (see proof of Theorem \ref{thrm: twoPointOptimalMechanism}).
Next, we prove that, when $\nu = \nu^\circ$, the left-hand side is equal to $x^s(\nu')$ for $\nu' \in [\nu^\circ, {\nu}^\star]$:
\begin{equation*}
	\begin{aligned}
		\lambda_1  +\frac{\lambda_0}{\nu^\circ} + \frac{-\lambda_0}{{\nu^\circ}^2}(\nu'- \nu^\circ) &= -\frac{\lambda_0}{{\nu^\circ}^2}\nu' + \lambda_1 +2\frac{\lambda_0}{\nu^\circ},\\
		&\geq \lambda_1  +\frac{\lambda_0}{\nu'}.
	\end{aligned}
\end{equation*}
Hence, for $\nu \in [0,\nu^\star]$, satisfaction of \eqref{IR} follows from $\lambda_1+\lambda_0/\nu$ and its concavity.
For $\nu \in [\nu^\star, \overline{\nu}]$, the \eqref{IR} constraint is satisfied if the following is non-negative:
\begin{equation*}
	\begin{aligned}
		\nu - \overline{\nu} +1 -(\lambda_1\nu +\beta) &= \nu(1-\lambda_1) - \lambda_1 -2\lambda_0 +1,\\
		&=(1-\lambda_1)(\nu +1) -2 \lambda_0.
	\end{aligned}
\end{equation*}
As $\lambda_0 \leq 0$, this term is non-negative if $\lambda_1 \leq 1$. We next show that this is true for $ \mu \in [0.107,0.25)$:
\begin{equation*}
	\begin{aligned}
		\lambda_1 &= \frac{(\sqrt{\delta}+1)^2}{3} - \frac{\lambda_0}{ \mu} = (\sqrt{\delta}+1) \big( \frac{\sqrt{\delta} +1}{3} + \frac{\sqrt{\delta}}{3- \mu} \big),\\
		&= (\sqrt{\delta}+1) \big( \frac{1}{3} + \sqrt{\delta}(\frac{1}{3}+\frac{1}{3- \mu}) \big) = \frac{\sqrt{\delta}+1}{3} + (\frac{ \mu}{3- \mu}+\sqrt{\delta})(\frac{1}{3}+\frac{1}{3- \mu}),\\
		&\leq \frac{1/2+1}{3} + (\frac{1/4}{3- \mu}+\frac{1}{2})(\frac{1}{3}+\frac{1}{3- \mu}),\\
		&\leq \frac{1}{2} + (\frac{1/4}{11/4}+\frac{1}{2})(\frac{1}{3}+\frac{1}{11/4}) = \frac{1}{2} + \frac{13}{22} \cdot \frac{23}{33} \leq 1,
	\end{aligned}
\end{equation*}
where the inequalities follow from $\delta \leq \mu \leq 0.25$. Lastly, when $\nu \in [\overline{\nu},1]$, we can write \eqref{IR} as $\nu x^*(\nu) -(\lambda_1 \nu +\lambda_0) =\nu(1-\lambda_1)-\lambda_0 \geq 0$, which follows again from $\lambda_1 \leq 1$. Hence, $(x^s,p^s)$ satisfies the \eqref{IR} constraints.

To prove that $(x^s,p^s)$ is incentive compatible, $i.e.$, satisfies \eqref{IC}, we need to show the following for all $\nu \in[0,1]$:
\begin{equation*}
	\nu(x^s(\nu)-\lambda_1) -\lambda_0 + \min_{\hat \nu \in [0,\nu]} -\nu x^s(\hat \nu) + \hat \nu \geq 0.  
\end{equation*}
By construction, for any fixed $\nu$, the function $-\nu x^s(\hat \nu) + \hat \nu$ is convex as $x^s$ is concave, $i.e.$, $\text{d}x^s(\nu)/\text{d}{\nu}$ is decreasing over $\nu$ by definition. Hence, we can identify the minimizer by inspecting the derivative of $-\nu x^s(\hat \nu) + \hat \nu$ with respect to $\hat \nu$:
\begin{equation}\label{eq: best deviation}
	\begin{aligned}
		1 - \nu \frac{\text{d}x^s(\hat \nu)}{\text{d}\hat{\nu}}\Big\rvert_{\hat \nu \in [\underline \nu, 1]} &= \begin{cases}
			\nu\frac{\lambda_0}{\underline \nu^2} +1 &\text{if } \hat \nu \in [0,\underline \nu],\\
			\nu\frac{\lambda_0}{\hat \nu^2} +1 &\text{if } \hat \nu \in [\underline \nu, \nu^\circ],\\
			\nu\frac{\lambda_0}{{\nu^\circ}^2} +1 &\text{if } \hat \nu \in [\nu^\circ, \nu^\star],\\
			-\nu +1 &\text{if } \hat \nu \in [\nu^\star, \overline \nu],\\
			1 &\text{if } \hat \nu \in [\overline \nu,1].
		\end{cases}
	\end{aligned}
\end{equation}
If \eqref{eq: best deviation} is non-negative for all $\hat \nu \in [0, \nu]$, then the minimum is attained at $\hat \nu=0$, meaning that \eqref{IC} reduces to \eqref{IR}, which is already proven above. As we already showed the first derivative of $x^s$ is non-increasing in $\hat \nu$, for any $\nu \leq -\frac{\underline \nu^2}{\lambda_0}$,
\eqref{eq: best deviation} is non-negative for all $\hat \nu \in [0,1]$, so that \eqref{IC} is satisfied via \eqref{IR}. For $\nu > -\frac{\underline \nu^2}{\lambda_0}$, the minimum will be attained at some $\hat \nu \in [0, \nu]$, so that proving non-negativity at this point is enough to show that \eqref{IC} holds. Since $x^s$ is concave, the minimum will be attained when \eqref{eq: best deviation} equals zero or when it changes signs at an endpoint, which can happen in two different instances. First, when $\hat \nu \in [\underline \nu, \nu^\circ]$, we may have:
\begin{equation*}
	\begin{aligned}
		- \nu \frac{\text{d}x^s(\hat \nu)}{\text{d}\hat{\nu}}\Big\rvert_{\hat \nu \in [\underline \nu, \nu^\circ]} +1 = -\nu\frac{-\lambda_0}{\hat \nu^2} +1 &= 0, \quad \Rightarrow \quad \hat \nu &= \sqrt{-\lambda_0 \nu},
	\end{aligned}
\end{equation*}
which can be satisfied only for $\nu \in [-\underline \nu^2/\lambda_0, -{\nu^\circ}^2/\lambda_0]$. 
Second, \eqref{eq: best deviation} can change signs when $\hat \nu = \nu^\star$ and $\nu \lambda_0 / {\nu^\circ}^2 +1 \leq 0$, which can happen only if $\nu \geq -{\nu^\circ}^2/\lambda_0$.
Hence, for $\nu \in [-\underline \nu^2/\lambda_0, 1]$,
the minimum value of $-\nu x^s(\hat \nu) + \hat \nu$ is given by 
\begin{equation*}
	\begin{aligned}
		&\begin{cases}
			-\nu \lambda_1 + 2\sqrt{-\lambda_0 \nu} &\text{if } \nu \in [\frac{-\underline \nu^2}{\lambda_0},\frac{-{\nu^\circ}^2}{\lambda_0}],\\
			-\nu (\nu^\star-\overline{\nu} +1) + \nu^\star &\text{if } \nu \in [\frac{-{\nu^\circ}^2}{\lambda_0},1].
		\end{cases}
	\end{aligned}
\end{equation*}
First, we show that \eqref{IC} is satisfied for all $\nu \in [-{\nu^\circ}^2/\lambda_0,1]$, which reduces to the following inequality:
\begin{equation}
	\label{eq: ICnonnegativityForAll2}
	\begin{aligned}
		&\nu(x^s(\nu)-\lambda_1) -\lambda_0 + \min_{\hat \nu \in [0,\nu]} -\nu x^s(\hat \nu) + \hat \nu \\ 
		&\qquad = \nu(x^s(\nu)-\lambda_1) -\lambda_0 -\nu(\nu^\star-\overline{\nu}+1)+\nu^\star \geq 0.
	\end{aligned}
\end{equation}
Since $-{\nu^\circ}^2/\lambda_0 \in [\nu^\star,\overline{\nu}]$ (see Figure \ref{fig: typesForThreePointMechanism}), the left-hand side of \eqref{eq: ICnonnegativityForAll2} equals the following two functions that follow from the definition of $x^s(\nu)$:
\begin{equation*}
	\begin{aligned}
		\begin{cases}
			\nu(\nu-\nu^\star-\lambda_1) -\lambda_0 +\nu^\star &\text{if } \nu \in [-\frac{{\nu^\circ}^2}{\lambda_0}, \overline{\nu}],\\
			\nu(\overline{\nu}-\nu^\star-\lambda_1) -\lambda_0+\nu^\star &\text{if } \nu \in [\overline{\nu},1].
		\end{cases}
	\end{aligned}
\end{equation*}
The first function above is a quadratic convex function of $\nu$, whereas the second function is linear. To ensure that a linear function is non-negative for some interval, we can check its values at some endpoints beyond this interval. For $\nu =0$, it becomes $-\lambda_0 +\nu^\star$, which is non-negative since $\lambda_0 \leq 0$ by construction. For $\nu=1$, it equals zero since $\overline{\nu} = \lambda_1 +\lambda_0$ by construction. To ensure that a quadratic convex function is non-negative, we can check its minimum value.
Its minimum is attained when $\nu = (\nu^\star+\lambda_1)/2$ and equals:
\begin{equation*}
	\begin{aligned}
		\nu^\star-\big(\frac{\nu^\star+\lambda_1}{2}\big)^2-\lambda_0&= \nu^\star-(1-\sqrt{1-\overline{\nu}})^2-\lambda_0,\\
		&= \nu^\star-2 +\overline{\nu}+2\sqrt{1-\overline{\nu}} -\lambda_0,\\
		&= \nu^\star + \lambda_1-2+2\sqrt{1-\overline{\nu}}  = 0,
	\end{aligned}
\end{equation*}
where the equalities follow from the definitions of $\nu^\star$ and $\overline{\nu}$.
Lastly, we need to show that the following holds for all $\nu \in [-\underline \nu^2/\lambda_0, -{\nu^\circ}^2/\lambda_0]$:
\begingroup
\setlength{\abovedisplayskip}{2pt} 
\setlength{\belowdisplayskip}{2pt}
\begin{align}
	&\nu(x^s(\nu)-\lambda_1) -\lambda_0 + \min_{\hat \nu \in [0,\nu]} -\nu x^s(\hat \nu) + \hat \nu \hspace{0.5cm} \nonumber\\ 
	&\qquad = \nu(x^s(\nu)-2\lambda_1) -\lambda_0 +2\sqrt{-\lambda_0 \nu} \geq 0.
	\tag{\ref{eq: ICnonnegativityForAll}}
\end{align}
\endgroup
As it can be seen in Figure \ref{fig: typesForThreePointMechanism}, for any $ \mu \in [0.107,0.25]$, we have $\nu^\star \leq -\underline \nu^2/\lambda_0  \leq -{\nu^\circ}^2/\lambda_0 \leq \overline{\nu}$.
Hence, for any $\nu \in [-\underline \nu^2/\lambda_0, -{\nu^\circ}^2/\lambda_0]$ we also have $\nu \in [\nu^\star,\overline{\nu}]$, which means that we only need to consider the case $x^s(\nu) = \nu-\overline{\nu}+1$. Hence, \eqref{eq: ICnonnegativityForAll} reduces to the following:
\begin{equation}\label{eq: ICnonnegativityForAllReduced}
	\nu(\nu-\overline{\nu}+1-2\lambda_1)-\lambda_0+2\sqrt{-\lambda_0 \nu} \geq 0.
\end{equation}
{
	We let $f(\nu)$ denote the left-hand side of \eqref{eq: ICnonnegativityForAllReduced} and substitute the values of $\overline{\nu}$, $\lambda_1$, $\lambda_0$, and then the value of $\delta= \mu/ (3- \mu)$ to obtain a function of $\nu$ for any fixed $ \mu \in [0.107,0.25]$:
	\begin{equation*}
		\begin{aligned}
			f(\nu) &= \nu(\nu +1 -3\lambda_1 -\lambda_0) -\lambda_0+2\sqrt{-\lambda_0 \nu} \\
			&= \nu \big(\nu +1-(\sqrt{\delta}+1)^2 +\lambda_0\frac{3- \mu}{ \mu} \big) -\lambda_0+2\sqrt{-\lambda_0 \nu} \\
			&=  \nu \big(\nu +1 -(\sqrt{\delta}+1)^2 - \sqrt{\delta}(\sqrt{\delta}+1) \big) -\lambda_0+2\sqrt{-\lambda_0 \nu} \\
			&= \nu^2 -\nu(2\delta +3\sqrt{\delta} ) -\lambda_0 +2\sqrt{-\lambda_0 \nu}\\
			&=\nu^2 -\nu(2\delta +3\sqrt{\delta} ) + (\delta^2 + \delta \sqrt{\delta}) + 2\sqrt{\nu} \sqrt{\delta^2 + \delta \sqrt{\delta}}\\
			&=\nu^2 -\nu (\frac{2 \mu}{3-  \mu} + \frac{3 \sqrt{ \mu}}{\sqrt{3-  \mu}}) + 2\sqrt{\nu} \sqrt{\frac{ \mu^2}{(3- \mu)^2} +\frac{ \mu \sqrt{ \mu}}{(3-  \mu) \sqrt{3- \mu}}} \\
			&\qquad + (\frac{ \mu^2}{(3- \mu)^2} +\frac{ \mu \sqrt{ \mu}}{(3-  \mu) \sqrt{3-  \mu}})\\
			&=\Big( (3- \mu)^2 \nu^2 - \nu (3- \mu)\big[ 2  \mu +3 \sqrt{ \mu(3- \mu)} \big] \\
			&\qquad + 2 \sqrt{\nu}(3- \mu)\sqrt{ \mu^2 +  \mu \sqrt{ \mu(3- \mu)}} +  \mu^2 + \mu \sqrt{ \mu(3- \mu)} \Big)\frac{1}{(3- \mu)^2}.
		\end{aligned}
	\end{equation*}
	Then, to prove the non-negativity of $f(\nu)$ for any $ \mu \in [0.107,0.25]$ and any $\nu \in [-\underline \nu^2/\lambda_0, -{\nu^\circ}^2/\lambda_0]$, we just need to show that the numerator in the last line is non-negative, $i.e.$,
	\begin{equation*}
		\begin{aligned}
			&(3- \mu)^2 \nu^2 - \nu (3- \mu)\big[ 2  \mu +3 \sqrt{ \mu(3- \mu)} \big]  \\
			&\qquad + 2 \sqrt{\nu}(3- \mu)\sqrt{ \mu^2 +  \mu \sqrt{\underline \mu(3- \mu)}} +  \mu^2 + \mu \sqrt{ \mu(3- \mu)} \geq 0.
		\end{aligned}
	\end{equation*}
	We let $g(\nu)$ denote the numerator and solve for its stationary points:
	\begin{equation*}
		\begin{aligned}
			g'(\nu) = \;&2(3- \mu)^2 \nu -(3- \mu)\big[ 2 \mu +3 \sqrt{ \mu(3- \mu)} \big]\\
			&\qquad + \frac{(3- \mu)}{\sqrt{\nu}}\sqrt{ \mu^2 +  \mu \sqrt{ \mu(3- \mu)}} =0, \\
			\implies \quad
			&\nu \Big(2(3- \mu)\nu -\big[ 2  \mu +3 \sqrt{ \mu(3- \mu)} \big] \Big)^2 =  \mu^2 +  \mu \sqrt{ \mu(3- \mu)}, 
		\end{aligned}
	\end{equation*}
	which can be written as a cubic equation of $\nu$:
	\begin{equation*}
		\begin{aligned}
			&4(3- \mu)^2 \nu^3 -4 (3- \mu)\big[ 2  \mu +3 \sqrt{ \mu(3- \mu)} \big] \nu^2\\
			&\qquad  + \big[ 2  \mu +3 \sqrt{ \mu(3- \mu)} \big]^2 \nu - \mu^2 -  \mu \sqrt{ \mu(3- \mu)} =0. 
		\end{aligned}
	\end{equation*}
	We use Cardano's approach to find the real roots of the above equation (see \citet{burnside1892theory} and \citet{nickalls1993new}). Any cubic equation, $ax^3+bx^2+cx+d=0$ with parameters $(a,b,c,d)$, can be transformed into a depressed cubic equation $t^3 + pt+q =0$, where $t = x+ b/3a$, $p = (3ac - b^2)/3a^2$, and $q= (2b^3-9abc+27a^2d)/27a^3$. If the value of $\Delta =(q/2)^2 +(p/3)^3$ happens to be strictly positive, then there exists one real root, which is given by the following equation.
	\begin{equation*}
		t_{root} = \sqrt[3]{-\frac{q}{2} +\sqrt{\Delta}} + \sqrt[3]{-\frac{q}{2} -\sqrt{\Delta}}.
	\end{equation*}
	Since our cubic equation is quite complex, but continuous functions of $\nu$ and $ \mu$, we conclude the proof by evaluating the relevant values numerically. 
	
	\begin{figure}
		\centering
		\begin{subfigure}[t]{0.48\textwidth}
			\centering
			\includegraphics[width=\textwidth]{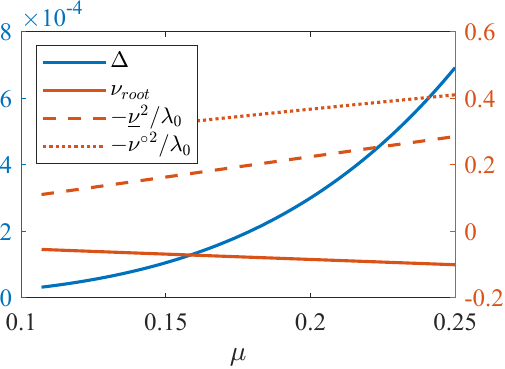}
			\caption{$\Delta$ is strictly positive, and $\nu_{root}$ does not belong to $[-\underline \nu^2/\lambda_0, -{\nu^\circ}^2/\lambda_0]$.}
			\label{fig: discriminantAndRoot}
		\end{subfigure}
		\hfill
		\begin{subfigure}[t]{0.48\textwidth}
			\centering
			\includegraphics[width=\textwidth]{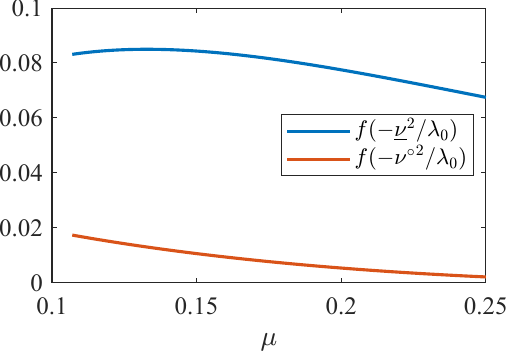}
			\caption{$f(\nu)$ evaluated at the values $-\underline \nu^2/\lambda_0$ and $-{\nu^\circ}^2/\lambda_0$.}
			\label{fig: functionAtBoundaryValues}
		\end{subfigure}
		\caption{Two plots ensuring that \eqref{eq: ICnonnegativityForAllReduced} holds for all $ \mu \in [0.107,0.25]$ and $\nu \in [-\underline \nu^2/\lambda_0, -{\nu^\circ}^2/\lambda_0]$ values.}
	\end{figure}
	
	In Figure \ref{fig: discriminantAndRoot}, we show that the value of $\Delta$ in our problem is strictly positive for any $ \mu \in [0.107,0.25]$, and the value of $\nu_{root} = t_{root} - b/3a$ is always outside of the interval $[-\underline \nu^2/\lambda_0, -{\nu^\circ}^2/\lambda_0]$. Hence, to prove the non-negativity of $f(\nu)$, we only need to consider its value evaluated at the boundaries of the interval $[-\underline \nu^2/\lambda_0, -{\nu^\circ}^2/\lambda_0]$. Figure \ref{fig: functionAtBoundaryValues} shows that $f(\nu)$ evaluated at the boundaries is always non-negative when $ \mu \in [0.107,0.25]$. 
}
\hfill \Halmos  \\

\begin{figure}
	\centering
	\includegraphics[scale=0.8]{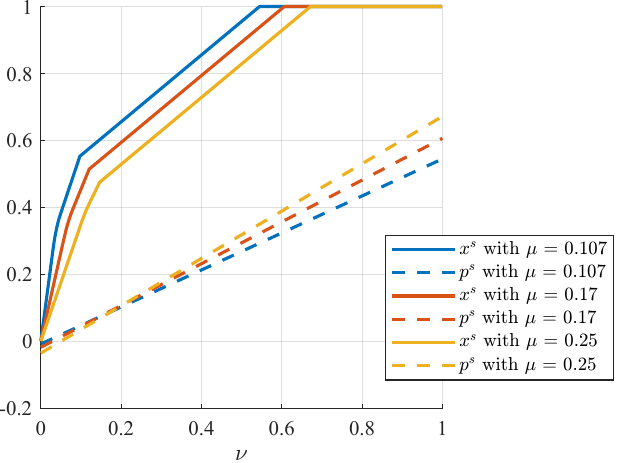}
	\caption{Mechanism $(x^s,p^s)$ from Theorem \ref{thrm: threePointMechanism} for different $ \mu$ values.}
	\label{fig: threePointMechanism}
\end{figure}

Figure \ref{fig: threePointMechanism} illustrates the robustly optimal mechanism $(x^s,p^s)$, which has a concave allocation rule and a linear payment rule.
It is only guaranteed to be feasible when $ \mu \in [0.107,0.25]$. We identified the bounds of this interval through numerical experiments as the mechanism is defined by high-order polynomials of $ \mu$.

Next, we also present the mechanism that extracts the maximal payment under the same allocation rule $x^s$. We will use this mechanism to approximate the optimal solution of the case $\mu \in (0,0.107)$.

\begin{theorem}\label{thrm: threePointMechanismMaximal}
	If $ \mu \in [0.107,0.25]$, then the following payment rule, $p_m^s$, extracts the maximum feasible payment from each type under $x^s$, and $(x^s,p_m^s)$ is Pareto robustly optimal:
	\begin{equation*}
		\begin{aligned}
			p_m^s(\nu) = &\begin{cases}
				\nu x^s(\nu) &\text{if } \nu \in [0,\nu'],\\
				\nu(\nu -\overline{\nu}+1 -\lambda_1)+2\sqrt{-\lambda_0 \nu} &\text{if } \nu \in [\nu',\nu''],\\
				\nu(\nu-\nu^\star)+\nu^\star &\text{if } \nu \in [\nu'', \overline{\nu}],\\
				\nu(\overline{\nu}-\nu^\star)+\nu^\star &\text{if } \nu \in [\overline{\nu},1],
			\end{cases}
		\end{aligned}
	\end{equation*}
	where $\nu' = -\underline \nu^2/\lambda_0$ and $\nu'' = -{\nu^\circ}^2/\lambda_0$.
\end{theorem}
{Any feasible mechanism that extracts strictly higher payments from some types must violate an incentive constraint that binds for $p_m^s$, and therefore cannot weakly dominate $(x^s,p_m^s)$ under all $\mathbb{P} \in\mathcal P_\mu$.}

\begin{figure}
	\centering
	\includegraphics[scale=0.8]{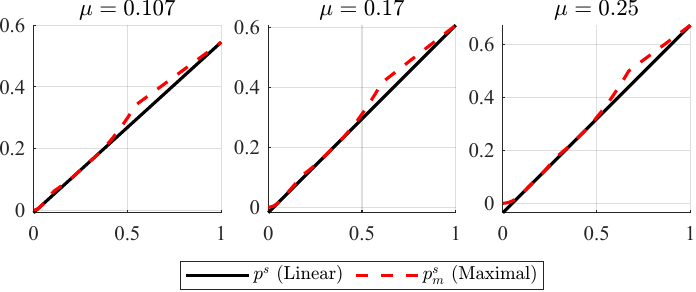}
	\caption{Plots of the linear and maximal payment rules for different $ \mu$ values.}
	\label{fig: threePointMechanismMaximal}
\end{figure}

We omit the proof of Theorem \ref{thrm: threePointMechanismMaximal} since it is merely an application of the arguments in the proof of Theorem \ref{thrm: twoPointMaximalPayment}. Figure \ref{fig: threePointMechanismMaximal} illustrates the maximal payment that is always bigger than or equal to the linear payment.

\begin{figure}
	\centering
	\includegraphics[scale=0.8]{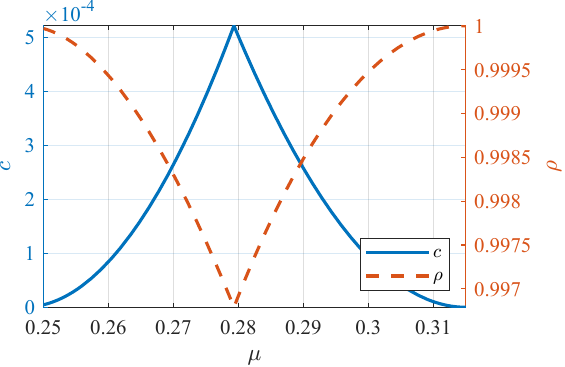}
	\caption{The relative, $\rho$, and absolute, $c$, performance guarantees of the mechanism $(x^*,p^*)$ defined for $\mu'$.}
	\label{fig: approximationMid}
\end{figure}

In the remainder of this section, we propose approximate solutions for small $\mu \in (0,0.107) \cup (0.25,\mu')$. We first consider the case $\mu \in (0.25,\mu')$.
In this case, the adversary can use a convex combination of the two-point and three-point distributions from Lemma \ref{lem:upperBounds} and Lemma \ref{lem: SmallMuUpperBounds} to get an upper bound that is less than those illustrated in Figure \ref{fig: upperBounds}. From our numerical experiments, we see that this is equivalent to finding the maximal convex function that is less than $ \mu/(2- \mu)$ and $ \mu(\sqrt{\delta}+1)^2/3$. To identify the robustly optimal mechanism, we first need to pin down this convex function, which requires solving a high-degree polynomial of $ \mu$, and use the linear programming approach from the proof of Lemma \ref{lem:upperBounds} to argue that this is indeed an upper bound. Instead of doing all these tedious calculations, we argue that the mechanism $(x^*,p^*)$ defined for $\underline \mu^*=\mu'$ demonstrates near-optimal performances when $ \mu \in (0.25,\mu')$. Figure \ref{fig: approximationMid} illustrates the performance metrics of $(x^*,p^*)$ relative to the upper bound $b = \min\{ \mu/(2- \mu), \mu(\sqrt{\delta}+1)^2/3\}$. Namely, $\rho = z/b$ and $c=b-z$, where $z$ is the worst-case expected payoff under $(x^*,p^*)$.
We know from Theorem \ref{thrm: twoPointOptimalMechanism} that this mechanism is optimal when $ \mu=\mu'$, and Figure \ref{fig: approximationMid} shows that it is nearly optimal when $ \mu=0.25$. In general, its relative performance, $\rho$, is always above $0.996$ when $ \mu \in (0.25,0.315)$.

We close this section by proposing an approximation mechanism for the case of $ \mu \in (0,0.107)$.
\begin{proposition} \label{prop: approximation}
	If $ \mu \in (0, 0.107)$, let $z^*$ be the optimal worst-case expected payoff, and let $z_{ \mu^*}$ be the worst-case expected payoff under the mechanism $(x^s,p_m^s)$ defined for $ \mu^* = 0.107$. The relative and absolute performance guarantees of $(x^s,p_m^s)$ are $z_{ \mu^*} \geq \rho z^*$ and $z_{ \mu^*} \geq z^*-c$ respectively, where $\rho = z_{ \mu^*}/b$, $c=b-z_{ \mu^*}$,
	\begin{equation*}
		\begin{aligned}
			z_{ \mu^*} =\begin{cases}
				\mu x^s( \mu) &\text{if }  \mu \in [0,\underline \nu_{ \mu^*}],\\
				\mu \lambda_1 + \lambda_0, &\text{if }  \mu \in [\underline \nu_{ \mu^*},  \mu^*],
			\end{cases}
		\end{aligned}
	\end{equation*}
	$b = \min\{ \mu/(2- \mu), \mu(\sqrt{\delta}+1)^2/3\}$, and $\underline \nu$, $\lambda_1$, and $\lambda_0$ are the parameters of $(x^s,p_m^s)$ as defined in Theorem \ref{thrm: threePointMechanism}.
\end{proposition}
\noindent{\bf Proof of Proposition \ref{prop: approximation}}. 
Given any $ \mu \in (0, 0.107)$, the relative and absolute performance guarantees follow by their definition when $z_{ \mu^*}$ is the worst-case expected payoff under $(x^s,p_m^s)$, and $b$ is an upper bound for $z^*$. By definition, $b$ is an upper bound of $z^*$, which is defined in Lemma \ref{lem:upperBounds}. To complete the proof, we must show that $z_{ \mu^*}$ is the worst-case expected payoff under $(x^s,p_m^s)$. By Proposition \ref{prop: equivalence of MDPs}, we can equivalently show that there exists a linear line that is below $p_m^s(\nu)$ for all $\nu \in [0,1]$, which is equal to $z_{ \mu^*}$ when $\nu = \mu$.    

Consider the payment made by type $\nu \in (0, 0.107)$ under the mechanism $(x^s,p_m^s)$ defined for the fixed $ \mu^* = 0.107$. It satisfies the following:
\begin{equation*}
	\begin{aligned}
		p_m^s(\nu) \geq \begin{cases}
			-\frac{\lambda_0}{\underline \nu^2} \nu^2 &\text{if } \nu \in [0,\underline \nu],\\
			\lambda_1 \nu +\lambda_0 &\text{if } \nu \in [\underline \nu,1],
		\end{cases}
	\end{aligned}
\end{equation*}
where the inequality follows from the definition of $p_m^s$ and from the fact that $p_m^s(\nu) \geq \lambda_1 \nu +\lambda_0$ for all $\nu \in [0,1]$. 
Then, by definition, there exists a linear line, $\lambda_1 \nu +\lambda_0$, that is below $p_m^s(\nu)$ for all $\nu \in [0,1]$ and equals to $z_{ \mu^*}$ when $\nu = \mu \in [\underline \nu,1]$. Next, notice that $-\frac{\lambda_0}{\underline \nu^2}\nu^2$ is a convex function of $\nu$ as $\lambda_0 \leq 0$, so that, for any $ \mu \in (0,\underline \nu]$, it satisfies:
\begin{equation}\label{eq: approximationLowerBound}
	-\frac{\lambda_0}{\underline \nu^2}\nu^2 \geq -\frac{\lambda_0}{\underline \nu^2} \mu^2+ (\nu- \mu)(-2\frac{\lambda_0}{\underline \nu^2} \mu) \quad \forall \nu \in [0,1],
\end{equation}
and the above right-hand side is a linear function of $\nu$. It is easy to see that this linear function reduces to $z_{ \mu^*}$ when $\nu = \mu$, and the left-hand side of \eqref{eq: approximationLowerBound} equals $p_m^s(\nu)$ by definition when $\nu = \mu \in (0,\underline \nu]$. We conclude the proof by showing that this linear function is also below $p_m^s(\nu)$ for $\nu \in [\underline \nu,1]$. This follows from the following:
\begin{equation*}
	\begin{aligned}
		-\frac{\lambda_0}{\underline \nu^2} \mu^2+ (\nu- \mu)(-2\frac{\lambda_0}{\underline \nu^2} \mu) &= -\frac{\lambda_0}{\underline \nu^2} \mu^2+ (\nu-\underline \nu+\underline \nu- \mu)(-2\frac{\lambda_0}{\underline \nu^2}  \mu),\\
		& \leq p_m^s(\underline \nu)+ (\nu-\underline \nu)(-2\frac{\lambda_0}{\underline \nu^2} \mu), \\
		& \leq p_m^s(\underline \nu)+ (\nu-\underline \nu)(-2\frac{\lambda_0}{\underline \nu^2}\underline \nu), \\
		& \leq \lambda_1 \nu +\lambda_0 \leq p_m^s(\nu),
	\end{aligned}
\end{equation*}
where the first inequality follows from \eqref{eq: approximationLowerBound} and the definition of $p_m^s$, the second inequality from $\nu \in [\underline \nu,1]$ and $-\lambda_0/\underline \nu^2 \geq 0$, the penultimate inequality from the definitions of $p_m^s(\underline \nu)$ and $\underline \nu$, and the last inequality from the fact that $p_m^s(\nu) \geq \lambda_1 \nu +\lambda_0$ for all $\nu \in [0,1]$. This completes the proof.
\hfill \Halmos  \\

\begin{figure}
	\centering
	\includegraphics[scale=0.8]{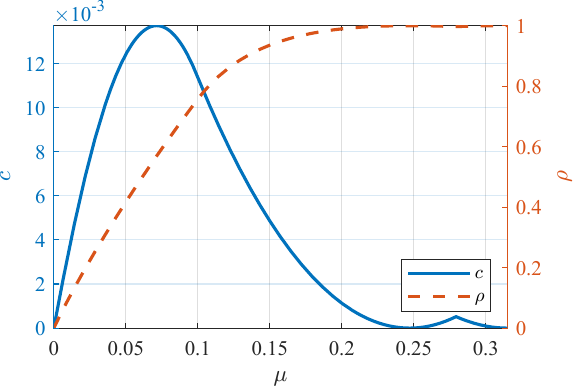}
	\caption{The relative, $\rho$, and absolute, $c$, performance guarantees of the mechanism $(x^*,p_m^*)$ defined for $0.107$.}
	\label{fig: approximation}
\end{figure}
Figure \ref{fig: approximation} shows the performance indicators of $(x^*,p_m^*)$. We see that it is near-optimal if $ \mu$ is close to $0.107$ and performs fairly well when it is not too small. As $ \mu$ approaches zero, both the relative and absolute performance guarantees tend to zero. This means that for the small values of $ \mu$, $(x^*,p_m^*)$ fails to perform well relative to the optimal worst-case expected payoff ($\rho \leq 0.5$ when $ \mu \leq 0.02$), whereas the distance to optimality will also be small ($c \leq 10^{-3}$ for all $ \mu \in [0,0.107]$). This is because the worst-case expected payoff guarantee is extremely small for small values of $ \mu$, even under the robustly optimal mechanism that we approximate.

\subsection{Relegated proofs}

\noindent{\bf Proof of Lemma \ref{lem:upperBounds}}.
{
	By definition, the two-point distribution assigns positive probabilities that add up to one. For feasibility, we only need to show that it has expectation $\mu$:
	\begin{equation*}
		\alpha(1)+ (1-\alpha)\nu'= \frac{ \mu}{2}+ (1-\frac{ \mu}{2})\frac{ \mu}{2- \mu} = \frac{ \mu}{2} + \frac{ \mu}{2} = \mu.
	\end{equation*}
	
	Next, we solve for the optimal mechanism under this two-point distribution. Notice that
	the objective reduces to the expected payment received from types $1$ and $\nu'$. Hence, all other types, $\nu \in [0,1) \setminus \{\nu'\}$, do not contribute to the seller's revenue. To avoid violating our constraints, we let
	\begin{equation*}
		\begin{aligned}
			&x(\nu) = \begin{cases}
				0 &\text{if } \nu < \nu',\\
				x(\nu') &\text{if } \nu \in (\nu',1),
			\end{cases} \\
			&p(\nu) = \begin{cases}
				0 &\text{if } \nu < \nu',\\
				p(\nu') &\text{if } \nu \in (\nu',1),
			\end{cases}
		\end{aligned}
	\end{equation*}
	for all $\nu \in [0,1) \setminus \{\nu'\}$. Through this reduction, we maintain feasibility and safely omit all variables and constraints for the types other than $1$ and $\nu'$. 
	We also omit the \eqref{IR} constraint of type 1, which only leads to an increased objective value. To find the given upper bound, we solve the following relaxed problem for the principal:
	\begin{equation*}
		\begin{aligned}
			z^*\leq \max_{x,p} \;& \alpha p(1)+ (1- \alpha)p(\nu') \nonumber\\
			\text{s.t.} \;&x(1)- p(1) \geq x({\nu}') - {\nu}',\\
			&\nu' x(\nu')- p(\nu') \geq 0, \\
			&x(\nu) \in [0,1] &\forall \nu \in \{\nu',1\}.
		\end{aligned}
	\end{equation*}
	As we maximize the expected payment, it is optimal to set all $p(\nu)$ to their upper bounds. This leads to:
	\begin{equation*}
		\begin{aligned}
			z^*\leq \max_{x,p} \;&\alpha \big(x(1)-x(\nu')+\nu'\big)+ (1- \alpha)\nu'x(\nu') \\
			\text{s.t.} \;&x(1),x(\nu') \in [0,1].
		\end{aligned}
	\end{equation*}
	We simplify the objective and see that it reduces to a function of $x(1)$ only:
	\begin{equation*}
		\begin{aligned}
			&\alpha\big(x(1) + \nu'\big) + \big((1-\alpha)\nu' -\alpha\big) x(\nu') = \alpha\big(x(1) + \nu'\big),
		\end{aligned}
	\end{equation*}
	where the equality follows from the fact that $\alpha = (1-\alpha)\nu' = \mu/2$. Then, it is optimal to set $x(1) = 1$, which gives the following upper bound for $z^*$:
	\begin{equation*}
		z^* \leq \alpha\big(1 + \nu'\big) = \frac{\mu}{2} \big( 1 + \frac{\mu}{2-\mu}\big) = \frac{\mu}{2-\mu}, 
	\end{equation*}
	which concludes our proof.
	\hfill \Halmos
}

\noindent{\bf Proof of Theorem \ref{thrm: twoPointOptimalMechanism}}.	For starters, for any $\mu \in (0,1)$, we prove two facts that will be useful throughout the proof. \begingroup
\setlength{\abovedisplayskip}{2pt} 
\setlength{\belowdisplayskip}{2pt}
\begin{align}
	\lambda_1  \mu +\lambda_0 &= \frac{2{z^*}^2}{\mu^2} \mu  + {z^*}^2 =\frac{2\mu}{(2- \mu)^2} - (\frac{ \mu}{2- \mu})^2, \nonumber\\
	&=\frac{ \mu}{(2- \mu)^2}(2- \mu) =\frac{ \mu}{2- \mu} = z^*, \label{eq: linear payment equals z*}\\
	-(1-\mu)^2 &\leq 0 \; \Rightarrow \mu(2-\mu) \leq 1 \;\Rightarrow \mu \leq \frac{1}{2-\mu}. \label{eq: useful mu relation}
\end{align}
\endgroup

We first show that $x^*$ is well-defined, $i.e.$, $0 \leq \underline \nu \leq \nu^\circ \leq \overline{\nu} \leq 1$ for any $ \mu \in (0,1)$. The non-negativity of $\underline \nu$ follows by definition, whereas the remaining three inequalities are proven below:
\begin{equation*}
	\begin{aligned}
		\nu^\circ &= z^* = \frac{ \mu}{2- \mu} \geq  \mu^2 = \underline \nu,\\
		\overline{\nu} &= \lambda_1 +\lambda_0 \geq \lambda_1  \mu +\lambda_0 = z^* = \nu^\circ,\\
		\overline{\nu} &= \lambda_1 + \lambda_0 = \frac{2{z^*}^2}{\mu^2} - {z^*}^2 =\frac{2-\mu^2}{(2-\mu)^2} \leq 1,
	\end{aligned}
\end{equation*}
where we make use of \eqref{eq: useful mu relation}, $z^* = \mu/(2-\mu)$, and $\mu \in(0,1)$. 

To prove that $(x^*,p^*)$ is robustly optimal, we will show that it is feasible and generates an expected payoff of $\mu/(2-\mu)$, attaining the upper bound given in Lemma \ref{lem:upperBounds}. The fact, $p^*( \mu) = \lambda_1  \mu +\lambda_0 = \mu/(2- \mu) = z^*$ follows from \eqref{eq: linear payment equals z*}. It only remains to show that $(x^*,p^*)$ is a feasible mechanism, $i.e.$, it satisfies \eqref{IC}, \eqref{IR} constraints, and $x^*$ is always between $0$ and $1$. Note that, if $x^*$ is weakly increasing and continuous for $\nu \in [0,1]$, as $x^*(0)=0$, and $x^*(1)=1$ by definition, we must have $x^*(\nu) \in [0,1]$ for all $\nu \in [0,1]$. Below, we prove that $x^*$ it is indeed weakly increasing and continuous. 

For continuity, we will show that $x^*$ takes the same values at the corresponding end points of the intervals. First, we show that the functions $(\frac{z^*}{ \mu^2})^2\nu$ and $\lambda_1 + \frac{\lambda_0}{\nu}$ evaluated at $\underline{\nu}=\mu^2$ have the same value:
\begin{equation*}
	(\frac{z^*}{ \mu^2})^2\underline \nu = \frac{{z^*}^2}{ \mu^2}= \frac{2{z^*}^2}{ \mu^2} - \frac{{z^*}^2}{ \mu^2}= \lambda_1 + \frac{\lambda_0}{\underline \nu}.
\end{equation*}
Second, the functions $\lambda_1 + \frac{\lambda_0}{\nu}$ and $\nu+\lambda_1-2z^*$ evaluated at $\nu^\circ = z^*$ should have the same value:
\begin{equation*}
	\lambda_1 + \frac{\lambda_0}{\nu^\circ} = \lambda_1 -z^* = \nu^\circ + \lambda_1 -2z^*.
\end{equation*}
Lastly, we show that $\nu+\lambda_1-2z^*$ evaluated at $\overline{\nu}=\lambda_1+\lambda_0$ equals $1$:
\begin{equation}\label{eq: one minus overlined nu}
	\begin{aligned}
		\overline \nu+\lambda_1-2z^* &= 2\lambda_1 + \lambda_0 -2z^* = (\frac{2{z^*}}{ \mu})^2 - {z^*}^2 - 2z^*\\
		&=(\frac{2{z^*}}{ \mu})^2 + 1 - (1+z^*)^2= 1, 
	\end{aligned}
\end{equation}
where the last equality follows after utilizing $z^*=\mu/(2-\mu)$. Hence, $x^*$ is continuous. 

To prove that $x^*$ is weakly increasing, it is enough to show that its first derivative is non-negative in each interval. The first derivatives are $\lambda_1/2\mu^2$ for the interval $[0, \underline \nu)$, $-\lambda_0/ \nu^2$ for $[\underline \nu, \nu^\circ)$, $1$ for $[\nu^\circ, \overline{\nu})$, and $0$ for $[\overline{\nu},1]$, which are all non-negative as $\lambda_1, -\lambda_0 \geq 0$, by definition. 

To prove that $(x^*,p^*)$ is individually rational, $i.e.$, it satisfies \eqref{IR}, we need to show $\nu x^*(\nu) \geq \lambda_1 \nu +\lambda_0$. 

Equivalently, $x^*(\nu) \geq \lambda_1  +\lambda_0/\nu$, which holds with equality for $\nu \in [\underline \nu, \nu^\circ]$. Let $g(\nu)= \lambda_1  +\lambda_0/\nu$. Notice that $g$ is strictly concave by definition, as $\lambda_0 <0$. Then, the following holds for any $\nu \in [0,1]$:
\begin{equation*}
	\begin{aligned}
		g(\nu) + \diff{g(\nu)}{\nu}(\nu'-\nu) & \geq g(\nu') &\forall \nu' \in[0,1],\\
		\lambda_1  +\frac{\lambda_0}{\nu} + \frac{-\lambda_0}{ \nu^2}(\nu'- \nu) &\geq \lambda_1  +\frac{\lambda_0}{\nu'} &\forall \nu' \in[0,1].
	\end{aligned}
\end{equation*}
Now we show that the left-hand side of the above inequality is equal to $x^*(\nu')$ for all $\nu' \in [0, \underline \nu]$ when $\nu= \underline{\nu}$, and it is equal to $x^*(\nu')$ for all $\nu' \in [\nu^\circ, \overline{\nu}]$ when $\nu = \nu^\circ$. First consider the case $\nu= \underline{\nu}$:
\begin{equation*}
	\begin{aligned}
		\lambda_1  +\frac{\lambda_0}{\underline \nu} + \frac{-\lambda_0}{ \underline \nu^2}(\nu'- \underline \nu) &= -\frac{\lambda_0}{\underline \nu^2}\nu' +\lambda_1 + 2\frac{\lambda_0}{\underline \nu},\\
		&= -\frac{\lambda_0}{\underline \nu^2}\nu' +2\frac{{z^*}^2}{\mu^2} - 2\frac{{z^*}^2}{\mu^2},\\
		&= x^*(\nu') \geq \lambda_1  +\frac{\lambda_0}{\nu'},
	\end{aligned}
\end{equation*}
where the last equality follows from the definition of $x^*$ for $\nu' \in [0, \underline \nu)$. Then, we consider the case $\nu = \nu^\circ$:
\begin{equation*}
	\begin{aligned}
		\lambda_1  +\frac{\lambda_0}{\nu^\circ} + \frac{-\lambda_0}{{\nu^\circ}^2}(\nu'- \nu^\circ) &= \nu' + \lambda_1 + 2\frac{\lambda_0}{\nu^\circ},\\
		&=\nu' + \lambda_1 + -2z^*,\\
		&= x^*(\nu') \geq \lambda_1  +\frac{\lambda_0}{\nu'},
	\end{aligned}
\end{equation*}
where the last equality follows from the definition of $x^*$ for $\nu' \in [\nu^\circ, \overline{\nu})$. {Finally, for any $\nu \in [\overline{\nu},1]$, \eqref{IR} can be written as $\nu x^*(\nu) -(\lambda_1 \nu +\lambda_0) =\nu(1-\lambda_1)- \lambda_0 \geq 0$, which happens to be true irrespective of the value of $\lambda_1$. If $(1-\lambda_1) \geq 0$, this is immediate. If $(1-\lambda_1) < 0$, on the other hand, we have the following for any $\nu \in [0,1]$:}
\begin{equation*}
	\nu(1-\lambda_1)-\lambda_0 \geq 1-\lambda_1-\lambda_0 = 1 - \overline{\nu} \geq 0. 
\end{equation*}

To prove that $(x^*,p^*)$ is incentive compatible, $i.e.$, satisfies \eqref{IC}, we need to show that $\nu(x^*(\nu)-x^*(\hat \nu)) +\hat \nu \geq \lambda_1 \nu + \lambda_0$ for any $\hat \nu \in [0,\nu]$ and $\nu \in[0,1]$. Equivalently, for all $\nu \in[0,1]$, we need to show:
\begin{equation}\label{eq: IC proof}
	\nu(x^*(\nu)-\lambda_1) -\lambda_0 + \min_{\hat \nu \in [0,\nu]} \{-\nu x^*(\hat \nu) + \hat \nu\} \geq 0.    
\end{equation}
Let $h(\nu,\hat \nu) = -\nu x^*(\hat \nu) + \hat \nu$ and consider its second derivative with respect to $\hat \nu$. That is, for any fixed $\nu \in [0,1]$, ${\text{d}^2h(\nu,\hat \nu)}/{\text{d}{\hat \nu}^2}$ equals $\lambda_0/{\hat \nu}^2$ if $\hat \nu \in [\underline{\nu},\nu^\circ)$, and $0$ otherwise. Hence, $h(\nu,\hat \nu)$ is convex over $\hat \nu$. To find its minimum, we inspect its first derivative with respect to $\hat \nu$.
As ${\text{d}h(\nu,\hat \nu)}/{\text{d}{\hat \nu}}$ is decreasing over $\hat \nu$ due to convexity, the derivative would be non-negative for all $\hat \nu \in [0, \nu]$, if it is non-negative at $\hat \nu =0$:
\begin{equation*}
	\begin{aligned}
		&\diff{h(\nu,\hat \nu)}{\hat \nu} \Bigr\rvert_{\hat \nu = 0} = -\nu\frac{\lambda_1}{2 \mu^2} +1 \geq 0, \quad \Rightarrow \quad
		\mu^2(2-\mu)^2 \geq \nu.
	\end{aligned}
\end{equation*}
Hence, for all $\nu \leq \mu^2(2-\mu)^2$, proving \eqref{IC} reduces to proving \eqref{IR}, which is already done above. For $\nu > \mu^2(2-\mu)^2$, the minimum will be attained at some $\hat \nu \in [0, \nu]$, so that proving non-negativity at this minimum is enough to show that \eqref{IC} holds. We consider the first-order condition:
\begin{equation*}
	\begin{aligned}
		&\diff{h(\nu,\hat \nu)}{\hat \nu} = -\nu\frac{-\lambda_0}{\hat \nu^2} +1 = 0, \quad \Rightarrow \quad \hat \nu = \sqrt{-\lambda_0 \nu} = z^* \sqrt{\nu}.
	\end{aligned}
\end{equation*}
Hence, for fixed $\nu$, the minimum value of $h(\nu,\hat \nu)$ is attained at $h(\nu,z^* \sqrt{\nu})$. Note that we need to prove \eqref{eq: IC proof} for all $\nu \in (\mu^2(2-\mu)^2,1]$, for which $z^* \sqrt{\nu} \in (\mu^2,z^*]=(\underline{\nu},\nu^\circ]$. Hence, \eqref{eq: IC proof} reduces to:
\begin{equation}\label{eq: ICnonnegativityForAll}
	\begin{aligned}
		&\nu(x^*(\nu)-\lambda_1) -\lambda_0 + h(\nu,z^* \sqrt{\nu}) \\ 
		&\qquad= \nu(x^*(\nu)-\lambda_1) -\lambda_0 -\nu\lambda_1- \frac{\lambda_0 \sqrt{\nu}}{z^*} + z^*\sqrt{\nu}, \\
		&\qquad =\nu(x^*(\nu)- (\frac{2z^*}{\mu})^2) +{z^*}^2 +2z^*\sqrt{\nu}\geq 0. 
	\end{aligned}
\end{equation}
We first show \eqref{eq: ICnonnegativityForAll} for $\nu \in [\overline{\nu},1]$, for which $x^*(\nu)=1$. This function is concave over $\nu$ so that it is enough to verify non-negativity at the endpoints. When $\nu=0$, \eqref{eq: ICnonnegativityForAll} reduces to ${z^*}^2 \geq 0$, which holds for any $z^* \in [0,1]$. When $\nu=1$, \eqref{eq: ICnonnegativityForAll} becomes:
\begin{equation*}
	\begin{aligned}
		&(1-(\frac{2z^*}{\mu})^2)+{z^*}^2 +2z^* = (z^*+1)^2 - (\frac{2z^*}{\mu})^2  \\
		&\qquad = (\frac{2}{2-\mu})^2 - (\frac{2}{2-\mu})^2 = 0.
	\end{aligned}
\end{equation*}
Lastly, we will prove \eqref{eq: ICnonnegativityForAll} for $\nu \in (\mu^2(2-\mu)^2, \overline{\nu})$ and conclude the proof. Note that, if $\mu^2(2-\mu)^2 \geq \nu^\circ = z^*$, we can focus on the allocation function for the interval $[\nu^\circ,\overline{\nu}]$. This follows if the following is true:
\begin{equation}\label{eq: ICtypesInequality}
	\mu^2(2-\mu)^2 - z^* \geq 0 
\end{equation}
which we will prove at the end of this proof for all $\mu \in [\mu',1]$. Assuming for now that \eqref{eq: ICtypesInequality} holds, we just need to consider the allocation function for the interval $[\nu^\circ, \overline{\nu})$, which is $x^*(\nu)=\nu+\lambda_1-2z^*$. Then, \eqref{eq: ICnonnegativityForAll} becomes:
\begin{equation*}
	\begin{aligned}
		&\nu(x^*(\nu)- 2 \lambda_1) +{z^*}^2 +2z^*\sqrt{\nu}\\
		&\qquad = \nu(\nu-2z^* - 2(\frac{z^*}{\mu})^2) +{z^*}^2 +2z^*\sqrt{\nu} \geq 0.
	\end{aligned}
\end{equation*}
Noting that $z^*/\mu = 1/(2-\mu) = (z^*+1)/2$, we can write the left-hand side of the above inequality as follows:
\begin{equation*}
	\begin{aligned}
		&\nu(\nu-2z^* - 2(\frac{z^*+1}{2})^2) +{z^*}^2 +2z^*\sqrt{\nu}\\
		&\qquad ={z^*}^2(1-\frac{\nu}{2}) + z^*(2\sqrt{\nu}-3\nu) + \nu^2-\frac{\nu}{2},
	\end{aligned}
\end{equation*}
which is, for any fixed $\nu$, a quadratic function of $z^*$. Note that the coefficient of ${z^*}^2$ is positive for any $\nu \in [0,1]$. Hence, any such function is non-negative if its determinant is negative, ensuring that \eqref{eq: ICnonnegativityForAll} holds. When it has a positive determinant, then \eqref{eq: ICnonnegativityForAll} may not be satisfied for some $z^*$. To avoid such cases and ensure feasibility, we require $z^*$ to be higher than the roots of the above quadratic function. We make a change of variable, $t=\sqrt{\nu}$ and write the determinant of the quadratic function:
\begin{equation*}
	\begin{aligned}
		\Delta &= (-3t^2+2t)^2 - 4(1-\frac{t^2}{2})(t^4-\frac{t^2}{2})\\
		&= t^2(-3t+2)^2-4t^2(1-\frac{t^2}{2}) (t^2-\frac{1}{2}) \\
		&=t^2(9t^2-12t+4) -4t^2(-\frac{t^4}{2}+ \frac{5t^2}{4}-\frac{1}{2})\\
		&= 2t^2(t^4+2t^2-6t+3).
	\end{aligned}
\end{equation*}
When $\Delta \geq 0$, the roots are:
\begin{equation*}
	z_{\pm} = \frac{(3t^2-2t) \pm t\sqrt{2(t^4+2t^2-6t+3)}}{2-t^2}.
\end{equation*}
In order to ensure that \eqref{eq: ICnonnegativityForAll} is satisfied, we define our mechanism only for $z^*$ that is bigger than both of these roots for any $t = \sqrt{\nu} \in [0,1]$. Hence, one can find the lower bound, $z'$, for which our mechanism is feasible, by solving a nonlinear optimization problem that maximizes the bigger root, $z_+$, subject to the constraint that $\Delta \geq 0$. This is exactly the model given in Theorem \ref{thrm: twoPointOptimalMechanism}.  As the function $z^* = \mu/(2-\mu)$ is a one-to-one correspondence for the domain $\mu \in [0,1]$, and its inverse function gives $\mu = 2z^*/(z^*+1)$, our mechanism $(x^*,p^*)$ should be feasible for any $\mu \in [\mu',1]$.

Finally, we prove that \eqref{eq: ICtypesInequality} holds for all $z^* \in [z',1]$, which can be written as:
\begin{equation*}
	\begin{aligned}
		\mu^2(2-\mu)^2 - z^* &= (\frac{2z^*}{z^*+1})^2(2-\frac{2z^*}{z^*+1})^2 - z^*\\
		&=z(\frac{16z^*}{(z^*+1)^4}-1).
	\end{aligned}
\end{equation*}
This equation is non-negative if $16z^* \geq (z^*+1)^4$. We make us of the fact that $(z^*+1)^4 = {z^*}^4 + {4z^*}^3 + {6z^*}^2 + {4z^*} + 1 \leq 11{z^*}^2 + {4z^*} + 1$ for $z^* \in [0,1]$ and focus on $z^*$ values that satisfy:
\begin{equation*}
	\begin{aligned}
		11{z^*}^2 + {4z^*} + 1 &\leq 16z^*,\\
		11{z^*}^2 - {12z^*} + 1&\leq 0.
	\end{aligned}
\end{equation*}
The above quadratic function has roots $\{\frac{1}{11},1\}$, so it is non-negative for all $z^* \in [\frac{1}{11},1]$. We also find a lower bound for $z'$ by using the solution $t=1/2$ in the given nonlinear program. It is straightforward to verify that this approach is feasible and yields an objective value of $(3\sqrt{2}-2)/14$, which is greater than $1/11$. Hence, we have $z' \geq 1/11$, and \eqref{eq: ICtypesInequality} is satisfied for any $z^* \in [z',1]$.
\hfill \Halmos  \\

\noindent{\bf Proof of Theorem \ref{thrm: twoPointMaximalPayment}}. 
Leveraging Proposition \ref{prop: resultsAlaei},
we just need to show that the following holds for all $\nu \in [0,1]$:
\begin{equation*}
	p_m^*(\nu) = \inf_{\hat \nu \in [0,\nu]} \{ \nu(x^*(\nu)-x^*(\hat \nu)) + \hat \nu \},
\end{equation*}
which ensures that both \eqref{IC} and \eqref{IR} are satisfied under $x^*$ while extracting the maximum payment from each type. But first, we show that $\nu^\circ \leq \nu^\star \leq \overline{\nu}$ for any $\mu \in [\mu',1]$.  The first inequality, $\nu^\circ \leq \nu^\star =\mu^2(2-\mu)^2$, is already proven in the proof of Theorem \ref{thrm: twoPointOptimalMechanism} (See equation \eqref{eq: ICtypesInequality}).
To prove $\nu^\star=\mu^2(2-\mu)^2 \leq \overline{\nu} =\lambda_1+\lambda_0$, we examine their difference:
\begin{equation*}
	\begin{aligned}
		\overline{\nu}-\nu^\star &= 2(\frac{z^*}{\mu})^2 -{z^*}^2 - \mu^2(2-\mu)^2\\
		&= \frac{2-\mu^2}{(2-\mu)^2} - \mu^2(2-\mu)^2 \\
		&= 1- \frac{2(1-\mu)^2}{(2-\mu)^2} -[1-(1-\mu)^2]^2\\
		&= (1-\mu)^2[- \frac{2}{(2-\mu)^2} +2 -(1-\mu)^2]\\
		&= (1-\mu)^2[\frac{2(1-\mu)(3-\mu)}{(2-\mu)^2} -(1-\mu)^2]\\
		&= (1-\mu)^3[\frac{2(3-\mu) -(1-\mu)(2-\mu)^2}{(2-\mu)^2}],
	\end{aligned}
\end{equation*}
which is non-negative if $2(3-\mu) -(1-\mu)(2-\mu)^2 \geq 0$ for all $\mu \in [0,1]$. That is equivalent to:
\begin{equation*}
	\mu^3-5\mu^2+6\mu+2 = \mu(\mu-3)(\mu-2)+2,
\end{equation*}
which is non-negative for all $\mu \in [0,1]$.

We now show that $p_m^*$ extracts the maximum payment from each type. From Proposition \ref{prop: resultsAlaei}, we know that the maximum payment is derived from the tightest \eqref{IC} constraint, and from the proof of Theorem \ref{thrm: twoPointOptimalMechanism}, we know which \eqref{IC} constraints bind for each type (see the discussion after \eqref{eq: IC proof}). That is, for each $\nu \in [0,1]$:
\begin{equation*}
	\begin{aligned}
		&\nu x^*(\nu) + \min_{\hat \nu \in [0,\nu]} \{-\nu x^*(\hat \nu) +\hat \nu\}\\
		&\qquad =\begin{cases}
			\nu x^*(\nu) &\text{if } \nu \leq \nu^\star,\\
			\nu x^*(\nu) -\nu x^*(z^* \sqrt{\nu}) +z^*\sqrt{\nu} &\text{if } \nu > \nu^\star.
		\end{cases}
	\end{aligned}
\end{equation*}
Hence, for any $\nu \leq \nu^\star$, the maximum payment function is:
\begin{equation*}
	\begin{aligned}
		p_m^*(\nu)= \begin{cases}
			\frac{\lambda_1}{2 \mu^2} \nu^2 &\text{if } \nu \in [0, \underline \nu),\\
			\lambda_1 \nu +\lambda_0 &\text{if } \nu \in [\underline \nu, \nu^\circ),\\
			\nu(\nu+\lambda_1 -z^*) &\text{if } \nu \in [\nu^\circ,\nu^\star].
		\end{cases}
	\end{aligned}
\end{equation*}
For $\nu \in (\nu^\star,1]$, we have $z^*\sqrt{\nu} \in (\mu^2, z^*] = (\underline{\nu},\nu^\circ]$. Hence, the maximum payment for any $\nu \in (\nu^\star,1]$ is equal to:
\begin{equation*}
	\begin{aligned}
		&\nu x^*(\nu) -\nu (\lambda_1 -\frac{z^*}{\sqrt{\nu}}) +z^*\sqrt{\nu} \\
		&\qquad = \nu (x^*(\nu) -\lambda_1) +2z^*\sqrt{\nu} \\
		&\qquad = \begin{cases}
			\nu^2 -2z^*(v-\sqrt{\nu}) &\text{if } \nu \in [\nu^*, 	\overline \nu],\\
			\nu (1 -\lambda_1)+2z^*\sqrt{\nu} &\text{if } \nu \in [\overline \nu, 1].
		\end{cases}
	\end{aligned}
\end{equation*}
Moreover, robust optimality of $(x^*,p_m^*)$ follows from robust optimality of $(x^*,p^*)$ as we have $p_m^*(\nu) =\min_{\hat \nu \in [0,\nu]} \{ \nu(x^*(\nu)-x^*(\hat{\nu}))+\hat{\nu} \} \geq p^*(\nu)$ for all $\nu \in [0,1]$ by construction.

{Now, we prove that $(x^*,p_m^*)$ is Pareto robustly optimal for any $\mu \in [\mu',1]$. Consider a feasible mechanism $(x',p')$ that weakly Pareto robustly dominates $(x^*,p_m^*)$. Then, we must have:
	\begin{equation}\label{eq: PRO definition for mu}
		\mathbb{E}_{\mathbb{P}}[p'(\nu)] \geq \mathbb{E}_{\mathbb{P}}[p_m^*(\nu)], \quad \forall \mathbb{P} \in \mathcal{P}_{\mu}.
	\end{equation}
	We use admissible distributions to show that any such mechanism should be identical to $(x^*,p_m^*)$. That is, $(x^*,p_m^*)$ cannot be strictly dominated by any feasible mechanism. Hence, it is Pareto robustly optimal.
	
	\textbf{Step 1.} $p'(\nu) \geq p_m^*(\nu)$ for all $\nu \in [\mu,1]$. Consider any $\nu \in [\mu,1]$. The two-point distribution that assigns probability $\alpha =\mu/\nu$ to type $\nu$, and probability $1-\alpha$ to type $0$ is contained in $\mathcal{P}_{\mu}$. Then, \eqref{eq: PRO definition for mu} can be satisfied only if $p'(\nu) \geq p_m^*(\nu)$ for all $\nu \in [\mu,1]$.
	
	\textbf{Step 2.} $x'(\nu) \leq x^*(\nu)$ for all $\nu \in [\nu^\circ,\overline{\nu})$. From Step 1 and \eqref{IC} constraints, the following should hold for any $\nu$:
	\begin{equation*}
		p_m^*(1) \leq p'(1) \leq x'(1)-x'(\nu) +\nu.
	\end{equation*}
	As $p_m^*(1) = \lambda_1+\lambda_0$ and $x'(1) \leq 1$, we get:
	\begin{equation*}
		x'(\nu) \leq 1+\nu - \lambda_1-\lambda_0 = \nu + \lambda_1 -2z^* = x^*(\nu),
	\end{equation*}
	where the first equality follows from \eqref{eq: one minus overlined nu}, and the last equality follows from the definition of $x^*(\nu)$ when $\nu \in [\nu^\circ,\overline{\nu})$.
	
	\textbf{Step 3.} $(x',p')(\nu) = (x^*,p_m^*)(\nu)$ for all $\nu \in [\nu^\circ,{\nu}^\star]$. We first show that $\mu < \nu^\star = \mu^2(2-\mu)^2$ for all $\mu \in [\mu',1]$:
	\begin{equation*}
		\begin{aligned}
			\mu^2(2-\mu)^2 -\mu &= \mu[\mu(2-\mu)^2 - 1]\\
			&= \mu(\mu-1)(\mu^2-3\mu+1).
		\end{aligned}
	\end{equation*}
	As the quadratic function above is negative for all $\mu \in [(3-\sqrt{5})/2, (3+\sqrt{5})/2]$, and $\mu' < (3-\sqrt{5})/2$, the above function is strictly positive for all $\mu \in [\mu',1)$. Then, as $\nu^\circ = z^* < \mu < \nu^\star$ for any $\mu \in [\mu',1]$, there exists a unique two-point distribution in $\mathcal{P}_\mu$ that puts probability masses to $\nu^\circ$ and $\nu^\star$. Under this distribution, $\mathbb{E}_{\mathbb{P}}[p'(\nu)]$ can be written as $\alpha p'(\nu^\circ) + (1-\alpha) p'(\nu^\star)$ for some $\alpha \in (0,1)$, and it has the following upper bound due to \eqref{IR} constraints: 
	\begin{equation*}
		\begin{aligned}
			\alpha p'(\nu^\circ) + (1-\alpha) p'(\nu^\star) &\leq \alpha \nu^\circ x'(\nu^\circ) + (1-\alpha)\nu^\star x'(\nu^\star)\\
			&\leq \alpha \nu^\circ x^*(\nu^\circ) + (1-\alpha)\nu^\star x^*(\nu^\star)\\
			&\leq \alpha p_m^*(\nu^\circ) + (1-\alpha) p_m^*(\nu^\star) = \mathbb{E}_{\mathbb{P}}[p_m^*(\nu)],
		\end{aligned}
	\end{equation*}
	where the second inequality follows due to Step 2, the third from the definition of $(x^*,p_m^*)$. Then, $(x',p')$ can weakly Pareto robustly dominate $(x^*,p_m^*)$ only if $(x',p')(\nu) = (x^*,p_m^*)(\nu)$ for all $\nu \in \{\nu^\circ,\nu^\star\}$. One can extend this result to $\nu \in [\nu^\circ,\nu^\star]$ by considering the two-point distribution that puts probability masses to points $\varepsilon \nu^\circ +(1-\varepsilon)\mu$ and $\varepsilon \nu^\star + (1-\varepsilon)\mu$ for any $\varepsilon \in [0,1]$.
	
	\textbf{Step 4.} $(x',p')(1) = (x^*,p_m^*)(1)$. We can write the following due to \eqref{IC}:
	\begin{equation*}
		\begin{aligned}
			p'(1) &\leq [x'(1) - x'(\nu^\circ)] + \nu^\circ\\
			&\leq 1 - x^*(\nu^\circ) +\nu^\circ = p_m^*(1), 
		\end{aligned}
	\end{equation*}
	where the second inequality follows from Step 3. Also, there exists a unique two-point distribution in $\mathcal{P}_\mu$ that puts probability masses to $1$ and $0$. Then, $(x',p')$ can weakly Pareto robustly dominate $(x^*,p_m^*)$ only if $(x',p')(1) = (x^*,p_m^*)(1)$.
	
	\textbf{Step 5.} $(x',p')(\nu) \geq (x^*,p_m^*)(\nu)$ for all $\nu \in [0,\mu)$. For any $\nu \in [0,\mu)$, there exists a unique two-point distribution in $\mathcal{P}_\mu$ that puts probability masses to $1$ and $\nu$. Then, as we have $p'(1)=p_m^*(1)$ from Step 4, we must also have $p'(\nu) \geq p_m^*(\nu)$ for all $\nu \in [0,\mu)$ for a weakly Pareto robustly dominating $(x',p')$. Then, \eqref{IR} and the definition of $(x^*,p_m^*)$ implies:
	\begin{equation*}
		p_m^*(\nu) = \nu x^*(\nu) \leq p'(\nu) \leq \nu x'(\nu),
	\end{equation*}
	$i.e.$, $x'(\nu) \geq x^*(\nu)$ for all $\nu \in [0,\mu)$.		
	
	\textbf{Step 6.} $(x',p')(\nu) = (x^*,p_m^*)(\nu)$ for all $\nu \in [\overline{\nu},1)$. For any $\nu \in [\overline{\nu},1)$, the definition of $(x^*,p_m^*)$ implies:
	\begin{equation*}
		\begin{aligned}
			p_m^*(\nu) &= \nu[1 - x^*(z^*\sqrt{\nu})] + z^*\sqrt{\nu}\\
			&\geq \nu[1 - x'(z^*\sqrt{\nu})] + z^*\sqrt{\nu} \geq p'(\nu),
		\end{aligned}
	\end{equation*}
	where the inequality follows from Step 5, as for any $\nu \in [\overline{\nu},1)$, we have $z^*\sqrt{{\nu}} \in [0,\mu)$. Then, Step 1 can be satisfied only if $p'(\nu) = p_m^*(\nu)$ for all $\nu \in [\overline{\nu},1)$. This is only possible if $x'(\nu) = x^*(\nu) =1$ for all $\nu \in [\overline{\nu},1)$.
	
	\textbf{Step 7.} $(x',p')(\nu) = (x^*,p_m^*)(\nu)$ for all $\nu \in ({\nu}^\star,\overline{\nu})$. For any $\nu \in ({\nu}^\star,\overline{\nu})$, the definition of $(x^*,p_m^*)$ implies:
	\begin{equation*}
		\begin{aligned}
			p_m^*(\nu) &= \nu[x^*(\nu) - x^*(z^*\sqrt{\nu})] + z^*\sqrt{\nu}\\
			&\geq \nu[x'(\nu) - x'(z^*\sqrt{\nu})] + z^*\sqrt{\nu} \geq p'(\nu),
		\end{aligned}
	\end{equation*}
	where the inequality follows from Steps 2 and 5, as for any $\nu \in ({\nu}^\star,\overline{\nu})$, we have $z^*\sqrt{{\nu}} \in [0,\mu)$. Then, as in the previous step, the mechanisms should coincide for all $\nu \in ({\nu}^\star,\overline{\nu})$.
	
	\textbf{Step 8.} $(x',p')(\nu) = (x^*,p_m^*)(\nu)$ for all $\nu \in [\underline{\nu},\nu^\circ)$. For any $\nu \in [\underline{\nu},\nu^\circ)$, the definition of $(x^*,p_m^*)$ implies:
	\begin{equation*}
		\begin{aligned}
			p_m^*(\frac{\nu^2}{{z^*}^2}) &= \frac{\nu^2}{{z^*}^2}[x^*(\frac{\nu^2}{{z^*}^2}) - x^*(\nu)] + {\nu}\\
			&\geq \frac{\nu^2}{{z^*}^2}[x'(\frac{\nu^2}{{z^*}^2}) - x'(\nu)] + {\nu} \geq 	p'(\frac{\nu^2}{{z^*}^2}),
		\end{aligned}
	\end{equation*}
	where the inequality follows from Steps 5, 6 and 7, as for any $\nu \in [\underline{\nu},\nu^\circ)$, we have ${\nu}^2/{z^*}^2 \in [\nu^\star,1)$. Due to Steps 6 and 7, we have $p_m^*({\nu^2}/{{z^*}^2}) = p'({\nu^2}/{{z^*}^2})$, which implies $x'(\nu) = x^*(\nu)$ for all $\nu \in [\underline{\nu},\nu^\circ)$ in the above inequality. Then, Step 5 can be satisfied only if we also have $p'(\nu) = p_m^*(\nu)$ for all $\nu \in [\underline{\nu},\nu^\circ)$.
	
	\textbf{Step 9.} $(x',p')(\nu) = (x^*,p_m^*)(\nu)$ for all $\nu \in [0,\underline{\nu})$. For any $\nu \in [0,\underline{\nu})$, the definition of $(x^*,p_m^*)$ implies:
	\begin{equation*}
		\begin{aligned}
			p_m^*(\nu^\star) &= \nu^\star[x^*(\nu^\star) - x^*(\nu)] + {\nu}\\
			&\geq \nu^\star[x'(\nu^\star) - x'(\nu)] + {\nu} \geq p'(\nu^\star),
		\end{aligned}
	\end{equation*}
	where the inequality follows from Steps 5 and 7. Due to Step 7, we have $p_m^*(\nu^\star) = p'(\nu^\star)$, which implies $x'(\nu) = x^*(\nu)$ for all $\nu \in [0,\underline{\nu})$. Then, Step 5 can be satisfied only if we also have $p'(\nu) = p_m^*(\nu)$ for all $\nu \in [0,\underline{\nu})$.
}
\hfill \Halmos  \\

{
	\noindent{\bf Proof of Lemma \ref{lem: polynomialPayment-multiAgent}}. Note that this proof closely follows the proof of Lemma 2 from \cite{carrasco2018optimal}. It is included for completeness, as \cite{carrasco2018optimal} do not state that their Lemma 2 also holds in the multi-agent setting.
	
	Fix any $p(.)$ that is bounded and satisfy $\min_{\mathbb{P} \in \mathcal{P}_{\mathbf{K}}} \int_{[0,1]^J} \sum_{j \in [J]} p_j(\boldsymbol{\nu}) d \mathbb{P}(\boldsymbol{\nu}) \geq 0$, and let $\mathbb{P}^* \in \mathcal{P}_{\mathbf{K}}$ denote the minimizer. If $\int_{[0,1]^J} \sum_{j \in [J]} p_j(\boldsymbol{\nu}) d \mathbb{P}(\boldsymbol{\nu}) = 0$, then the result is trivial. Otherwise, we solve the adversary's problem to deduce the given results.
	
	Let $D = \{\mathbb{P} \in \mathcal{P}_0([0,1]^J): \; \int_{[0,1]^J} \nu_j^N d\mathbb{P}(\boldsymbol{\nu}) < \infty, \; \forall j \in [J]\}$ and $C = \{r\mathbb{P} \in \mathcal{P}_0([0,1]^J): \; r \geq 0, \; \mathbb{P} \in D\}$. Then, the adversary's solution, $\mathbb{P}^*$ solves the following:
	\begin{equation*}
		\begin{aligned}
			\min_{\mathbb{P} \in C} \;&\int_{[0,1]^J} \sum_{j \in [J]} p_j(\boldsymbol{\nu}) d\mathbb{P}(\boldsymbol{\nu})\\
			\text{s.t.} \; & \int_{[0,1]^J} \nu_j^i d\mathbb{P}(\boldsymbol{\nu}) - K_{ij} =0 &\forall i \in [N], \forall j \in [J],\\
			& \int_{[0,1]^J} d\mathbb{P}(\boldsymbol{\nu}) = 1.
		\end{aligned}
	\end{equation*}
	Define $h_{ij}(\mathbb{P}) = \int_{[0,1]^J} \nu_j^i d\mathbb{P}(\boldsymbol{\nu}) - K_{ij}$ for all $\mathbb{P} \in C$, $i \in [N]$, and $j \in [J]$, and $h_0(\mathbb{P}) = \int_{[0,1]^J} d\mathbb{P}(\boldsymbol{\nu}) - 1$.
	Due to \cite{clarke2013functional} (see Theorem 9.4, page 182), there exists nontrivial $\eta$, $\theta_0$, and $\{\theta_{ij}\}_{i \in [N], j \in [J]}$ such that $\eta \in \{0,1\}$, and for all $\mathbb{P} \in C$:
	\begin{equation}\label{eq: nontrivialKKT}
		\begin{aligned}
			&\eta \int_{[0,1]^J} \sum_{j \in [J]} p_j(\boldsymbol{\nu}) d\mathbb{P}(\boldsymbol{\nu}) + \sum_{i \in [N]} \sum_{j \in [J]} \theta_{ij} h_{ij}(\mathbb{P}) \\
			&\qquad + \theta_0 h_0(\mathbb{P}) \geq \eta \int_{[0,1]^J} \sum_{j \in [J]} p_j(\boldsymbol{\nu}) d\mathbb{P}^*(\boldsymbol{\nu}).
		\end{aligned}
	\end{equation}
	Suppose that $\eta = 0$. Due to Assumption \ref{assumption: N moments}, there exists $\varepsilon$ such that $\mathcal{P}_{\hat{\mathbf{K}}} \neq \emptyset$ for $\hat{K}_{ij} = K_{ij} - \varepsilon \text{sign}(\theta_{ij})/2$. Then, for any $\mathbb{P}' \in \mathcal{P}_{\hat{\mathbf{K}}} \subseteq C$:
	\begin{equation*}
		\begin{aligned}
			\sum_{i \in [N]} \sum_{j \in [J]} \theta_{ij} h_{ij}(\mathbb{P}') + \theta_0 h_0(\mathbb{P}') = \frac{\varepsilon}{2}\Big( - \sum_{i \in [N]} \theta_{ij} \text{sign}(\theta_{ij})\Big) < 0,
		\end{aligned}
	\end{equation*}
	which contradicts \eqref{eq: nontrivialKKT}. Hence, we must have $\eta = 1$.
	
	As $\int_{[0,1]^J} \sum_{j \in [J]} p_j(\boldsymbol{\nu}) d \mathbb{P}(\boldsymbol{\nu}) > 0$ and $0 \in C$, we cannot have all zero $\{\theta_{ij}\}_{i \in [N], j \in [J]}$ an $\theta_0$. As $\mathbb{P}^* \in \mathcal{P}_{\mathbf{K}}$, we can rewrite \eqref{eq: nontrivialKKT}, for all $\mathbb{P} \in C$, as follows:
	\begin{equation}\label{eq: minimalityKKT}
		\begin{aligned}
			&\int_{[0,1]^J} \Big( \sum_{j =1}^J p_j(\boldsymbol{\nu}) + \sum_{i =1}^N \sum_{j =1}^J \theta_{ij} \nu_j^i +\theta_0\Big) d\mathbb{P}(\boldsymbol{\nu}) \\
			&\qquad \geq \int_{[0,1]^J} \Big( \sum_{j =1}^J p_j(\boldsymbol{\nu}) + \sum_{i =1}^N \sum_{j =1}^J \theta_{ij} \nu_j^i +\theta_0\Big) d\mathbb{P}^*(\boldsymbol{\nu}).
		\end{aligned}
	\end{equation}
	If we have $\sum_{j \in [J]} p_j(\boldsymbol{\nu}) + \sum_{i \in [N]} \sum_{j \in [J]} \theta_{ij} \nu_j^i +\theta_0< 0$ for some $\boldsymbol{\nu} \geq 0$, then one can choose $r \delta_{\boldsymbol{\nu}}$ and let $r \to \infty$ to contradict \eqref{eq: minimalityKKT}. Hence, one must have
	\begin{equation*}
		\sum_{j \in [J]} p_j(\boldsymbol{\nu}) + \sum_{i \in [N]} \sum_{j \in [J]} \theta_{ij} \nu_j^i + \theta_0 \geq 0,
	\end{equation*}
	for all $\boldsymbol{\nu} \in [0,1]^J$. Moreover, $0 \in C$ implies that 
	\begin{equation*}
		\int_{[0,1]^J} \Big( \sum_{j =1}^J p_j(\boldsymbol{\nu}) + \sum_{i =1}^N \sum_{j =1}^J \theta_{ij} \nu_j^i +\theta_0\Big) d\mathbb{P}^*(\boldsymbol{\nu}) = 0.
	\end{equation*}
	That is, the set $\{\boldsymbol{\nu} \in [0,1]^J: \; \sum_{j \in [J]} p_j(\boldsymbol{\nu}) + \sum_{i \in [N]} \sum_{j \in [J]} \theta_{ij} \nu_j^i +\theta_0>0\}$ has null $\mathbb{P}$ measure. Defining $\lambda_{ij} = -\theta_{ij}$ and $\lambda_0 = \theta_0$ establishes part $(i)$. Besides, as we have $\sum_{j \in [J]} p_j(\boldsymbol{\nu}) + \sum_{i \in [N]} \sum_{j \in [J]} \theta_{ij} \nu_j^i +\theta_0 = 0$ $\mathbb{P}^*$ almost surely, we also establish $(iii)$.
	
	Part $(ii)$ follows from $(i)$ and \eqref{EPIR} constraints. At scenario $\boldsymbol{\nu}_0 = (0,\ldots,0)$, \eqref{EPIR} enforces $p_j(\boldsymbol{\nu}_0) \leq 0$ 
	for all $j \in [J]$. Hence, due to $(i)$, we have $0 \geq \sum_{j \in [J]} p_j(\boldsymbol{\nu}_0) \geq \lambda_0.$
	\hfill \Halmos  \\
}

{
	\noindent{\bf Proof of Proposition \ref{prop: equivalence of MDPs-multiAgent}}. We first prove the only-if part. Let $(x^*,p^*)$ and $\mathbb{P}^*$ be a solution of corresponding optimization problems in \eqref{eq: MDPmultiAgent} and hence constitute Nash equilibrium of the zero-sum game. Then, $p^*$ satisfies the second set of inequalities given in Proposition \ref{prop: equivalence of MDPs-multiAgent} due to \eqref{IC} constraints. Also, from Lemma \ref{lem: polynomialPayment}, there exists $\lambda_0^*$ and $\boldsymbol{\lambda}^* = \{\lambda_{ij}^*\}_{i \in [N], j\in [J]}$ such that $\sum_{j \in [J]} p_j^*(\boldsymbol{\nu}) \geq \sum_{i \in [N]} \sum_{j \in [J]} \lambda_{ij}^* \nu_j^i$ for all $\boldsymbol{\nu} \in [0,1]^J$, and
	\begin{equation}\label{eq: polynomial optimality}
		z^* = \inf_{\mathbb{P} \in \mathcal{P}_{\mathbf{K}}} \int_{[0,1]^J} \sum_{j =1}^J p_j^*(\boldsymbol{\nu}) d\mathbb{P}(\boldsymbol{\nu}) = \sum_{i =1}^N \sum_{j =1}^J \lambda_{ij}^* K_i.
	\end{equation}
	Notice that restricting the payment rules to be polynomials of degree $N$ would only decrease the optimal objective value. Hence, \eqref{eq: MDPmultiAgent} gives an upper bound for \eqref{eq: MDP lambda-multiAgent}. Moreover, if $(x^*,p^*)$ is feasible in \eqref{eq: MDPmultiAgent}, and $\sum_{j \in [J]} p_j^*(\boldsymbol{\nu}) \geq \sum_{i \in [N]} \sum_{j \in [J]} \lambda_{ij}^* \nu_j^i$ for all $\boldsymbol{\nu} \in [0,1]^J$, then $(x^*,\lambda_0^*,\boldsymbol{\lambda}^*)$ should also be feasible in \eqref{eq: MDPmultiAgent} and consequently in \eqref{eq: MDP lambda-multiAgent}. Then, due to \eqref{eq: polynomial optimality}, we can conclude that $(x^*,\boldsymbol{\lambda}^*)$ solves \eqref{eq: MDP lambda-multiAgent}.
	
	For the other direction, let $(x^*,\lambda_0^*,\boldsymbol{\lambda}^*)$ be a solution of \eqref{eq: MDP lambda-multiAgent} and let $p^*$ satisfy the set of inequalities given in Proposition \ref{prop: equivalence of MDPs-multiAgent}. Then, the second set of inequalities ensure that $(x^*,p^*)$ is feasible in \eqref{eq: MDPmultiAgent}. Assume to the contrary that $(x^*,p^*)$ does not solve \eqref{eq: MDPmultiAgent}, $i.e.$, there exists a mechanism $(\hat x, \hat p)$ that yield a strictly higher worst-case expected revenue:
	\begin{equation*}
		\begin{aligned}
			z^* =\inf_{\mathbb{P} \in \mathcal{P}_{\mathbf{K}}} \int_{[0,1]^J} \sum_{j \in [J]}\hat p_j(\boldsymbol{\nu}) d\mathbb{P}(\boldsymbol{\nu})
			> \inf_{\mathbb{P} \in \mathcal{P}_{\mathbf{K}}} \int_{[0,1]^J} \sum_{j \in [J]} p_j^*(\boldsymbol{\nu}) d\mathbb{P}(\boldsymbol{\nu}).
		\end{aligned}
	\end{equation*}
	Then, the results of Lemma \ref{lem: polynomialPayment} applied for $\hat p$ enables us to replicate the arguments of the only-if part of this proof to show that there exists $\hat{\lambda}_0$ and $\hat{\boldsymbol{\lambda}}$ feasible in \eqref{eq: MDP lambda-multiAgent} and generate a revenue of $z^*$, which is strictly higher than that of $(x^*,\lambda_0^*,\boldsymbol{\lambda}^*)$. This is a contradiction.
	\hfill \Halmos  \\
}

\noindent{\bf Proof of Lemma \ref{lem: upperbounds-2agents}}.
{
	Let $\mathbb{P}^*$ denote the distribution defined in Lemma \ref{lem: upperbounds-2agents}. We first show that $\mathbb{P}^*$ is contained in $\mathcal{P}_{\mu}$. That is, all assigned probabilities are nonnegative, sum up to one, and the expected value is $\mu$ for each agent. The first two conditions hold thanks to the constraints \eqref{eq: 2agentDist-equal to one} and \eqref{eq: 2agentDist-nonnegativity}:
	\begin{equation*}
		\beta/r + 2 \beta + \beta r + 2\beta \nu^* = \beta (2 \nu^* + r + 1/r + 2) = 1.
	\end{equation*}
	For the expected value, we first note that $\mathbb{E}_{\mathbb{P}^*}[\nu] = \beta (r{\nu^*}^2 + (2r +3) \nu^*)$ by definition. Then, $\mathbb{E}_{\mathbb{P}^*}[\nu]$ is fixed to $\mu$ due to the constraint \eqref{eq: 2agentDist-expectation}:
	\begin{equation*}
		\begin{aligned}
			& r{\nu^*}^2 + (2r +3 -2\mu) \nu^* - \mu(r + 1/r + 2) = 0,\\
			&\implies r{\nu^*}^2 + (2r +3) \nu^* = \mu (2\nu^* + r + 1/r + 2),\\
			&\implies \beta (r{\nu^*}^2 + (2r +3) \nu^*) = \mu.
		\end{aligned}
	\end{equation*}
	
	Let $r, \nu^*, \beta$ solve the nonlinear model in Lemma \ref{lem: upperbounds-2agents} for some fixed $\mu \in (0,1)$ and lead to the objective value of $f(\mu)$. Next, we consider the mechanism design problem under the unique prior $\mathbb{P}^*$ in order to derive the upper bound $f(\nu)$, which is the optimal objective value of the given nonlinear program. First, we omit all scenarios other than those assigned a positive probability by $\mathbb{P}^*$ together with all the corresponding variables and constraints. As they do not appear in the objective, this action results in a relaxation of the original mechanism design model. We further omit some \eqref{IR} and \eqref{IC} constraints on the remaining scenarios. Specifically, we maximize
	\begin{equation*}
		\begin{aligned}
			\max_{x \geq 0,p} \; \sum_{j \in J} \Bigg[ &\beta /r p_j(r\nu^*,r\nu^*) + \beta r p_j(\nu^*,\nu^*)  \\
			&+ \beta \Big(p_j(\nu^*,r \nu^*) +p_j(r\nu^*,\nu^*) \Big)\\
			&+ \beta \nu^* \Big( p_j(1, r \nu^*) + p_j(r \nu^*, 1)\Big) \Bigg],
		\end{aligned}
	\end{equation*}
	subject to the feasibility constraints $\sum_{j \in J} x_j(\nu_j,\nu_{-j}) \leq 1$ for all $\nu_j, \nu_{-j} \in \{r \nu^*,\nu^*,1\}$, \eqref{IR} constraints
	\begin{equation*}
		p_j(r \nu^*,\nu_{-j}) \leq r \nu^* x_j(r \nu^*,\nu_{-j}),
	\end{equation*}
	for all $\nu_{-j} \in \{r \nu^*,\nu^*,1\}$, and $j \in J$,
	and \eqref{IC} constraints
	\begin{equation*}
		\begin{aligned}
			&p_j(\nu^*,\nu_{-j}) \leq \nu^* \big( x_j(\nu^*,\nu_{-j}) - x_j(r\nu^*,\nu_{-j}) \big) + r \nu^*, \\
			&p_j(1,r \nu^*) \leq \big( x_j(1,r \nu^*) - x_j(\nu^*,r \nu^*) \big) + \nu^*,
		\end{aligned}
	\end{equation*}
	for all $\nu_{-j} \in \{r \nu^*,\nu^*\}$, and $j \in J$. Note that, in the above objective function $p_j(1,r \nu^*)$ denotes the payment received from agent $j$ when he reports type $1$, and the other agent reports $r \nu^*$. As each payment variable above has only one upper bound, it is optimal to substitute them in the objective function:
	\begin{equation*}
		\begin{aligned}
			\max_{x \geq 0} \; \sum_{j \in J} \Bigg[ &\beta \nu^* x_j(r \nu^*,r \nu^*) \\
			&+ \beta r \Big( \nu^* \big( x_j(\nu^*,\nu^*) - x_j(r\nu^*,\nu^*) \big) + r \nu^* \Big)  \\
			&+ \beta \Big( \nu^* \big( x_j(\nu^*,r \nu^*) - x_j(r\nu^*,r \nu^*) \big) + r \nu^* \Big)\\
			& + \beta r \nu^* x_j(r \nu^*,\nu^*)+ \beta r {\nu^*}^2 x_j(r \nu^*,1) \\
			&+ \beta \nu^* \Big( \big( x_j(1,r \nu^*) - x_j(\nu^*,r \nu^*) \big) + \nu^*\Big) \Bigg],
		\end{aligned}
	\end{equation*}
	subject to $\sum_{j \in J} x_j(\nu_j,\nu_{-j}) \leq 1$ for all $\nu_j, \nu_{-j} \in \{r \nu^*,\nu^*,1\}$. Rewriting the objective will lead to
	\begin{equation*}
		\begin{aligned}
			\sum_{j \in J} \Bigg[ &\beta r^2 \nu^* + \beta r \nu^* + \beta {\nu^*}^2 + \beta r \nu^* x_j(\nu^*,\nu^*) \\
			&+ \beta r {\nu^*}^2 x_j(r \nu^*,1) + \beta \nu^* x_j(1,r \nu^*) \Bigg].
		\end{aligned}
	\end{equation*}
	As the agents are symmetric, it is optimal to omit the subscript $j$ and solve for a symmetric solution. Then, the allocation constraints become $2 x(\nu^*,\nu^*) \leq 1$, and $x(r\nu^*,1) + x(1, r \nu^*) \leq 1$. As $\beta \nu^* \geq \beta r {\nu^*}^2$, it is optimal to set $x(\nu^*,\nu^*) = 1/2$ and $x^*(1, r \nu^*) = 1$. Then, the optimal solution is:
	\begin{equation*}
			2 \Big( \beta r^2 \nu^* + \beta r \nu^* + \beta {\nu^*}^2 + \beta r \nu^* /2 + \beta \nu^* \Big) = 2\beta \nu^* \Big( r^2 + 3r/2 + \nu^* + 1\Big),
	\end{equation*}
	which is exactly the objective function of the given nonlinear program. Hence, the optimal objective value, $f(\mu)$, represents an upper bound for $z^*$, the optimal value of our robust mechanism model.
	
	Finally, as the Dirac distribution at point $(\mu,\mu)$ is contained in $\mathcal{P}_{\mu}$, and the maximum payoff under such a distribution is equal to $\mu$ due to the allocation and \eqref{IR} constraints, we conclude that $z^*$ is upper bounded by $\min\{f(\mu), \mu\}$.
	\hfill \Halmos  \\
}

\end{document}